\documentclass[floatfix,aps,prb,onecolumn,showpacs,nofootinbib,superscriptaddress]{revtex4-2}
               
\usepackage{libertine}
\usepackage[libertine,slantedGreek,
            bigdelims,
            vvarbb
            ]{newtxmath}
\usepackage{graphicx}
\usepackage{dcolumn}
\usepackage{bm}
\usepackage{color}
\usepackage{soul}

\usepackage{mathtools}
\usepackage{bbm}

\usepackage{tikz}
\usepackage{lipsum} 

\usepackage{natbib}
\renewcommand\harvardurl[1]{\textbf{URL:} \url{#1}}

\renewcommand{\vec}[1]{\mathbf{#1}}

\usepackage{hyperref}

\usepackage{xparse}
\newcommand{\dagg}{\mkern 2mu \dagger}
\ExplSyntaxOn
\NewDocumentCommand{\mref}{m}{\quinn_mref:n {#1}}
\seq_new:N \l_quinn_mref_seq
\cs_new:Npn \quinn_mref:n #1
 {
  \seq_set_split:Nnn \l_quinn_mref_seq { , } { #1 }
  \seq_pop_right:NN \l_quinn_mref_seq \l_tmpa_tl
  ( 
  \seq_map_inline:Nn \l_quinn_mref_seq
    { \ref{##1},\nobreakspace } 
  \exp_args:NV \ref \l_tmpa_tl 
  ) 
 }
\ExplSyntaxOff

\newcommand{\overbar}[1]{\mkern 1.5mu\overline{\mkern-1.5mu#1\mkern-1.5mu}\mkern 1.5mu}

\DeclareMathOperator*{\SumInt}{%
\mathchoice%
  {\ooalign{$\displaystyle\sum$\cr\hidewidth$\displaystyle\int$\hidewidth\cr}}
  {\ooalign{\raisebox{.14\height}{\scalebox{.7}{$\textstyle\sum$}}\cr\hidewidth$\textstyle\int$\hidewidth\cr}}
  {\ooalign{\raisebox{.2\height}{\scalebox{.6}{$\scriptstyle\sum$}}\cr$\scriptstyle\int$\cr}}
  {\ooalign{\raisebox{.2\height}{\scalebox{.6}{$\scriptstyle\sum$}}\cr$\scriptstyle\int$\cr}}
}

\newcommand{\mbb}[1]{\ensuremath{\mathbb{#1}}}

\usepackage{braket}

\begin{document}
\title{Information thermodynamics for Markov jump processes coupled to underdamped diffusion: \\ Application to nanoelectromechanics}

\newcommand\unilu{\affiliation{Complex Systems and Statistical Mechanics, Department of Physics and Materials Science, University of Luxembourg, L-1511 Luxembourg City, Luxembourg}}
\newcommand\uniarg{\affiliation{Universidad de Buenos Aires, Facultad de Ciencias Exactas y Naturales, Departamento de Física, Buenos Aires, Argentina}}

\author{Ashwin Gopal}
\email{ashwin.gopal@uni.lu}
\unilu 

\author{Nahuel Freitas}
\email{nahuel.freitas@gmail.com}
\uniarg 

\author{Massimiliano Esposito}
\email{massimiliano.esposito@uni.lu}
\unilu

\date{\today}

\begin{abstract}
We extend the principles of information thermodynamics to study energy and information exchanges between coupled systems composed of one part undergoing a Markov jump process and another underdamped diffusion. We derive integral fluctuation theorems for the partial entropy production of each subsystem and analyze two distinct regimes. First, when the inertial dynamics is slow compared to the discrete-state transitions, we show that the steady-state energy and information flows vanish at the leading order in an adiabatic approximation, if the underdamped subsystem is governed purely by conservative forces. To capture the non-zero contributions, we consistently derive dynamical equations valid to higher order. Second, in the limit of infinite mass, the underdamped dynamics becomes a deterministic Hamiltonian dynamics driving the jump processes, we capture the next-order correction beyond this limit. We apply our framework to study self-oscillations in the single-electron shuttle - a nanoelectromechanical system (NEMS) - from a measurement-feedback perspective. We find that energy flows dominate over information flows in the self-oscillating regime, and study the efficiency with which this NEMS converts electrical work into mechanical oscillations. 
\end{abstract}

\maketitle

\section{Introduction}

Thermodynamic principles have been crucial for developing macroscopic engines that extract useful work from various nonequilibrium resources, and optimizing their designs. Such machines, usually operating at steady-state, consume a constant input of free energy which is converted from one form to another to carry out the desired function. The notion of information as a thermodynamic resource was first proposed by James Clerk Maxwell through his renowned thought experiment, the Maxwell demon \cite{maruyama2009colloquium, leff1990maxwell, leff2002maxwell}. 
The apparent violation of the second law in Maxwell demons is caused by ignoring the entropic transfers between the demon and the system which modify the second law of the system while leaving its first law unaffected. This thought experiment revealed a deep connection between information and physical dissipation \cite{szilard1929entropieverminderung, landauer1961irreversibility, bennett1982thermodynamics}. \par
Over the past two decades, stochastic thermodynamics has emerged as a natural framework for incorporating concepts from information theory - such as Shannon entropy and mutual information - into its formalism. Maxwell demons are analyzed as measurement-feedback systems, where the demon acts as a controller coupled to the system. It detects microscopic fluctuations, increasing the mutual information between itself and the system, and later utilizes this information to perform feedback on the system. Together, the system and the controller obey the second law, but the mutual information exchanged among them modifies their own local second laws. The controller produces mutual information that is then used as a resource by the system to perform useful tasks. This formalism has been extended beyond Maxwell demon setups to encompass more general coupled systems that exchange both energy and information between the system and the controller. This broader framework is commonly referred to as information thermodynamics \cite{sagawa2012thermodynamics,parrondo2015thermodynamics, ehrich2023energy,parrondo2023information}.\par

Initially developed for systems with non-autonomous measurement-feedback protocols \cite{PhysRevLett.102.250602}, this framework was later extended to autonomous classical systems described by discrete Markov jump (MJ) processes - where transitions between discrete states occur randomly over time \cite{horowitz2014thermodynamics,barato2014efficiency} — and continuous overdamped diffusion processes, which neglect inertial effects \cite{allahverdyan2009thermodynamic,horowitz2015multipartite}. Extensions have also been made to underdamped diffusion (UD), taking into account both position and velocity variables \cite{horowitz2014second,herpich2020effective}, and autonomous quantum systems \cite{strasberg2017quantum,ptaszynski2019thermodynamics}. Moreover, the local second law in these systems has been generalized to fluctuation theorems that account for mutual information flows \cite{sagawa2010generalized,rosinberg2016continuous,shiraishi2015fluctuation,prech2024quantum}.\par


This framework has proven particularly powerful in characterizing biological and synthetic nanomachines that transduce both energy and information to perform useful work \cite{horowitz2014thermodynamics, leighton2024flow,sanchez2019autonomous,amano2022insights,saha2021maximizing,parrondo2023information}. These theoretical developments have been further supported by experimental realizations of Maxwell demons in diverse single-molecule and single-electron setups \cite{serreli2007molecular,raizen2009comprehensive,toyabe2010experimental,koski2014experimental,koski2015chip,camati2016experimental,chida2017power,masuyama2018information,ribezzi2019large,saha2021maximizing,ciliberto2017experiments}, demonstrating the possibility to use information about microscopic states of the system to drive traditionally impossible processes. The macroscopic limit of the framework was also analyzed and applied to a variety of systems including CMOS transistors \cite{freitas2022maxwell,freitas2023information}, chemical reaction networks \cite{penocchio2022information,bilancioni2023chemical} and Potts models \cite{ptaszynski2024dissipation}.
%

\par


Recently, a new class of autonomous systems has emerged in which discrete MJ dynamics is intrinsically coupled with UD processes. Notable examples include nanomechanical resonators interacting with electronic (photonic) transport phenomena \cite{lassagne2009coupling, steele2009strong, okazaki2016gate, ekinci2005nanoelectromechanical, vigneau2022ultrastrong, wen2020coherent}. These nanoelectromechanical (NEM) systems, characterized by high quality factors and resonant frequencies, have significant applications in information processing \cite{lahaye2009nanomechanical, reed2017faithful}, ultra-sensitive sensing \cite{knobel2003nanometre, moser2013ultrasensitive}, and the development of autonomous heat engines exploiting self-oscillations \cite{tonekaboni2018autonomous, strasberg2021autonomous}. Similarly, CMOS clock circuits typically involve an inductive (mechanical) oscillating element in a feedback loop with a CMOS based amplifier to sustain the oscillations \cite{vittoz2010low, thommen1999improved, siniscalchi2020ultra}. Understanding the stochastic nature of these clocks at ultra-low power operation also requires mixed noise modeling. Here, the Johnson-Nyquist noise in mechanical components or RLC follows UD dynamics, whereas the thermal shot noise in CMOS transistors of the amplifiers requires Markov jump modeling \cite{freitas2020stochastic, freitas2021stochastic, gopal2024thermodynamic}. Operating at voltage scales near the thermal voltage $k_B T/q_e$ \cite{siniscalchi2020ultra, steele2009strong, vigneau2022ultrastrong, parrondo2023information}, their dynamics is strongly influenced by stochastic fluctuations from the environment. Consequently, information (quantified as correlated fluctuations) can be an important thermodynamic resource in such systems.\par


Despite system-specific studies using the framework of stochastic thermodynamics \cite{wachtler2019stochastic, strasberg2021autonomous, wachtler2019proposal} and one recent study including information flows \cite{parrondo2023information}, an information thermodynamic framework for general systems with jump-underdamped diffusive dynamics is still lacking. Furthermore, as there is no obvious way to relate an underdamped process with a jump process, an extention of information thermodynamics to such systems requires a tailored approach. 

Motivated by these challenges, in this article, we extend the information thermodynamics framework to such autonomous systems composed of a subsystem with Poisson MJ dynamics coupled to a subsystem described by UD dynamics. We also systematically derive closed dynamical equations in two limiting cases, the first assuming timescale separation and the second assuming a heavy (large mass) underdamped subsystem.\par

The paper is structured as follows: In Sec.~\ref{sec: General_model}, we describe the stochastic dynamics of our composite system, involving discrete jump dynamics coupled to continuous inertial dynamics, in a thermodynamically consistent fashion. In Sec.~\ref{sec: Thermo_composite}, we establish the thermodynamic laws, as we obtain the balance equations for energy and entropy for the composite system, both at the trajectory and the average level. We also identify the information flows between subsystems leading to local second laws and derive integrated fluctuation theorems (IFTs) for the partial entropy production of the individual subsystems. In Sec.~\ref{sec: TS}, we derive dynamical equations beyond the adiabatic approximation \cite{esposito2012stochastic, vigneau2022ultrastrong}, assuming that the UD dynamics is slower compared to the fast MJ dynamics. We also derive the lowest-order contributions to the energy and information flows between the subsystems. In Sec.~\ref{sec: MF_approx}, we derive dynamical equations considering the large mass limit of the underdamped subsystem and its equivalence to the meanfield approximation. In Sec.~\ref{sec: Electron_shuttle}, we apply our theoretical framework to a single-electron shuttle which displays a spontaneous transition to coherent oscillations \cite{wachtler2019stochastic, gorelik1998shuttle}.

\section{General Model}
\label{sec: General_model}
We consider a composite system composed of two subsystems, namely $\mbb{X}$ and $\mbb{Y}$. The subsystem $\mbb{X}$ is described by a continuous state space, $\Vec{\Gamma}\equiv (x, v)$, representing its position and velocity, respectively. The subsystem $\mbb{Y}$ has discrete coarse-grained states $y$. Here, the discrete state $y$ can represent a vector of discrete states $y \equiv (n_1, n_2, ..., n_N)$. In different contexts, the components of this vector have specific interpretations: for instance, in chemical reaction networks, $n_i$ represents the number of molecules of the $i^{\text{th}}$ chemical species \cite{rao2016nonequilibrium}, whereas in electronic circuits, $n_i$ denotes the number of charges in the $i^{\text{th}}$ conductor \cite{freitas2021stochastic}. This representation is general and can be applied to any system where the state can be described as a vector of discrete quantities. \par
The total energy of the system can be decomposed into contributions from the individual subsystems and an interaction term:
\begin{eqnarray}
    \hat{E}(\Vec{\Gamma}, y) = \frac{1}{2}m v^2 + \hat{U}(x,y) = \frac{1}{2}m v^2 + \hat{U}_{\mbb{X}}(x)  + \hat{U}_{\mbb{Y}}(y) + \hat{U}_{\mbb{XY}}(x, y),
    \label{eqn: total_energy}
\end{eqnarray}
where $\hat{U}(x, y)$ represents the potential energy that depends on both subsystems, and is further decomposed into the potential energy $\hat{U}_{\mathbb{X}}(x)$ of individual subsystems $\mathbb{X}(\mbb{Y})$, and the interaction energy $\hat{U}_{\mathbb{XY}}(x, y)$ between $\mathbb{X}$ and $\mathbb{Y}$.\par

The bare internal energies of the individual subsystems$\mbb{X}$ and $\mbb{Y}$ are then given by:
\begin{eqnarray}
    \hat{E}_{\mbb{X}}(\Vec{\Gamma}) =  \frac{1}{2}m v^2 + \hat{U}_{\mbb{X}}(x), \quad\quad
    \hat{E}_{\mbb{Y}}(y) = \hat{U}_{\mbb{Y}}(y).
\end{eqnarray}



\subsection{Langevin Dynamics}
At the trajectory level, the stochastic dynamics of the composite system can be described by coupled Langevin dynamics. The stochastic dynamics for the subsystem $\mbb{X}$ is modeled by underdamped diffusive (UD) dynamics. The position $x_t$ and the velocity $v_t$ of subsystem $\mbb{X}$ satisfy the following Langevin dynamics:
\begin{equation}
\begin{aligned}  
    dx_t &= v_t dt,\\
    m\: dv_t &=\left[ -\partial_x  \hat{U}(x_t,y_t) - \gamma v_t + g(x_t, y_t) \right] dt+ \sqrt{2\gamma \beta_{\mbb{X}}^{-1}}\,.\,dB_t,
    \label{eqn: Langevin_under}  
\end{aligned}
    \end{equation}
where $g(x, y)$ is a non-conservative force, $\gamma$ is the damping coefficient and $dB_t$ is the thermal Gaussian noise with zero mean and autocorrelation function $\langle dB_t dB_{t'}\rangle= dt \:\delta(t-t')$.\par
Similarly, the Langevin dynamics in the discrete state space is modeled as Markov jump (MJ) processes. The corresponding stochastic trajectory can be obtained by solving the following equations:
\begin{eqnarray}
    dy_t = \sum_{\rho, y'}(y'-y_t)\:d\mathcal{N}_\rho^{y',y_t}(x_t),\label{eqn: Langevin_MJP}
\end{eqnarray}
where $d\mathcal{N}_\rho^{y',y}(x)\in \{0,1\}$ is a Poisson random variable with mean $\langle d\mathcal{N}_\rho^{y',y}(x) \rangle = \lambda_\rho^{y',y}(x)dt$ starting from $y$ and ending at $y'$. Note that here we assume that the jump rates $\lambda_\rho^{y',y}(x)$ only depend on the position of the underdamped particle, as observed in specific example systems of \cite{wachtler2019proposal,gopal2024thermodynamic,pietzonka2022classical}. Moreover, we consider the ideal case with only one jump occurring at any time.
Since the jumps between the states are mediated by the reservoir $\rho$ of the discrete subsystem $\mbb{Y}$, each transition $y \to y'$ needs to have a reverse transition $y' \to y$ with it corresponding rates $\lambda_\rho^{y,y'}(x)$, to satisfy the  microscopic reversibility. In addition, these bidirectional processes also satisfy the local detailed balance condition (LDB) for thermodynamic consistency. The LDB associated with the process, $(y' \xrightarrow{\rho} y)$, for any given state $\Vec{\Gamma}$ of $\mbb{X}$, is given as
\begin{flalign}
    \log\frac{ \lambda_\rho^{y,y'}(x)}{\lambda_\rho^{y',y}(x)}  = -\beta_{\mbb{Y}}\left\{\left[\hat{\phi}_{\mbb{Y}}(\Vec{\Gamma}, y) - \hat{\phi}_{\mbb{Y}}(\Vec{\Gamma}, y')\right] - w^{y' \to y}_\rho(x)\right\},
    \label{eqn: LDB_condition}
\end{flalign}
where $\hat{\phi}_{\mbb{Y}}(\Vec{\Gamma}, y) = \hat{E}(\Vec{\Gamma}, y) - T_{\mbb{Y}} s_i(y)$ is the Helmholtz free energy of state $(\Vec{\Gamma}, y)$ associated with the thermal reservoir at the inverse temperature $\beta_{\mbb{Y}} = (k_{B} T_{\mbb{Y}})^{-1}$, and $s_i(y)$ is the internal entropy for each discrete state $y$. We thus assumed that the sole contribution to the internal entropy changes is due to the discrete state. The reservoir $\rho$ also performs non-conservative work $w^{y' \to y}_\rho(x) = - w^{y \to y'}_\rho(x)$ in the transition $y' \to y$. Examples of non-conservative work by the reservoir $\rho$ include the work done by its chemical or electrical potential \cite{rao2018conservation, freitas2021stochastic}. Therefore, the transition rates, in general, can be written as $\lambda_\rho^{y,y'}(x) = \Lambda^{y,y'}(x) \exp\left({\frac{-\beta_{\mbb{Y}}\left\{\left[\hat{\phi}_{\mbb{Y}}(\Vec{\Gamma}, y) - \hat{\phi}_{\mbb{Y}}(\Vec{\Gamma}, y')\right] - w^{y' \to y}_\rho(x)\right\}}{2}}\right)$, where $ \Lambda^{y,y'}(x) = \sqrt{\lambda_\rho^{y,y'}\lambda_\rho^{y',y}}$ is the symmetric (kinetic) part of the rates, which may also depend on the position $x$. 

\begin{figure}[h!]
    \centering
    \includegraphics[trim=10 0 10 0, clip, width=0.85\textwidth]{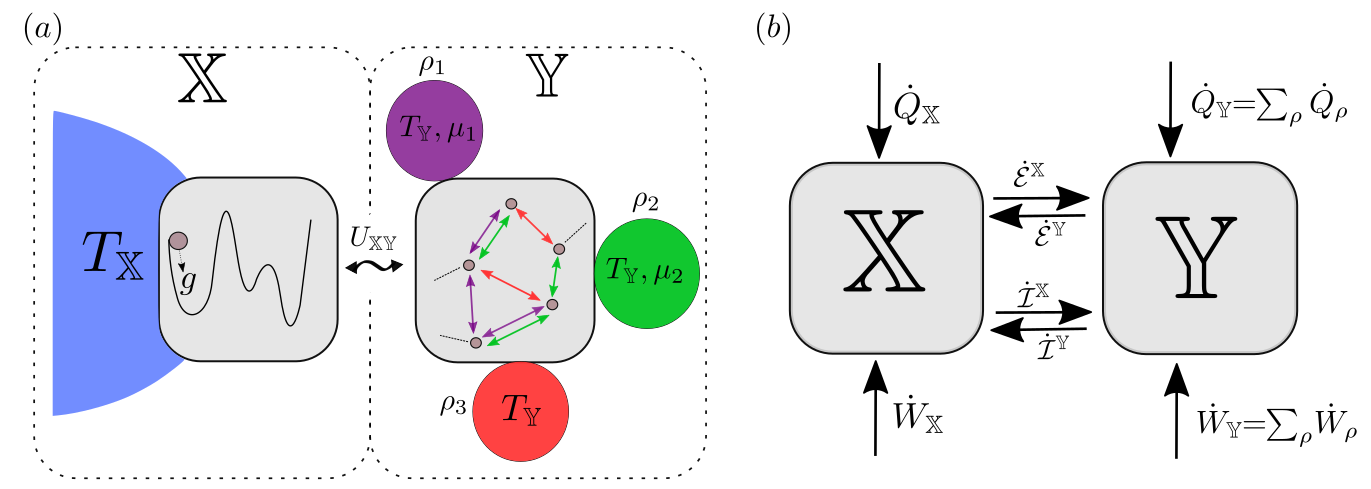}
   \caption{(a): Schematic diagram of the composite system and its associated reservoirs. The system is composed of subsystem $\mbb{X}$ connected to a thermal reservoir at temperature $T_\mbb{X}$, and subsystem $\mbb{Y}$ interacting with generalized reservoirs $\rho$ at temperature $T_\mbb{Y}$. (b): The thermodynamic picture of the composite system involving energy and information flows between the subsystems at the average description. Arrows indicate the convention for positive directions.}
    \label{fig: setup_figure}
\end{figure}

\subsection{Master Equation}
At the ensemble level, the dynamics of the probability density $P_t(\Vec{\Gamma},y)$ is described through a master equation, given as:
\begin{eqnarray}
    d_t P_t(\Vec{\Gamma},y) &=& \sum_\rho\sum_{y'} J_\rho^{y,y'}(\Vec{\Gamma}) - \nabla.\;\Vec{J}_{\Vec{\Gamma}}^y,
    \label{eqn: master_equation_joint}
\end{eqnarray}
where $J_\rho^{y,y'}$ is the probability current due to the jumps associated with $\rho$ in MJ dynamics, and $\Vec{J}_{\Vec{\Gamma}}^y$ is probability current associated with the UD dynamics. For the Langevin dynamics described by Eq.~\eqref{eqn: Langevin_under} and Eq.~\eqref{eqn: Langevin_MJP}, the above quantities are given as
\begin{eqnarray}
    J_\rho^{y,y'} &=& \lambda_\rho^{y,y'}(x) P_t(\Vec{\Gamma},y') -  \lambda_\rho^{y',y}(x) P_t(\Vec{\Gamma},y), \label{eqn: prob_current_MJP}\\
    \Vec{J}_{\Vec{\Gamma}}^y &=& \left( \Vec{\mathcal{L}}^{\rm{det}} + \Vec{\mathcal{L}}^{\rm{diss}}\right) P_t(\Vec{\Gamma},y) ,\label{eqn: prob_current_under}
\end{eqnarray}
where the underdamped probability current is further split into deterministic and dissipative contributions, such that:
\begin{eqnarray}
    \Vec{\mathcal{L}}^{\rm{det}} \equiv \begin{bmatrix}
        v \\
        \frac{1}{m}\left[-\partial_x \hat{U}(x,y)+g(x,y)\right]
    \end{bmatrix} ,\;\;\; \Vec{\mathcal{L}}^{\rm{diss}} \equiv \begin{bmatrix}
                                        0\\
                                       -\frac{\gamma}{m}\left( v+(m\beta_{\mbb{X}})^{-1} \partial_v \log P_t(\Vec{\Gamma},y)\right)
                                        \end{bmatrix}.\label{eqn: prob_current_operator_joint}
\end{eqnarray}
For the ease of notation in the following section, we will identify the independent force $f_{\mbb{X}} \equiv -\partial_x \hat{U}_{\mbb{X}}(x)$ due to the bare potential $\hat{U}_{\mbb{X}}$ and the coupled forces depending on the state of $\mbb{Y}$ as, $f_{\mbb{XY}} \equiv \left[-\partial_x \hat{U}_{\mbb{XY}}(x,y)+g(x,y)\right]$. \par

\section{Thermodynamics of the composite system}
\label{sec: Thermo_composite}
\subsection{1st Law }
\label{subsec: COmposite_1stlaw}
The first law of thermodynamics describes the balance of energy flows in the system. At the trajectory level, the change in total energy $\hat{E}(\Vec{\Gamma}, y)$ (Eq.~\eqref{eqn: total_energy}) for a given $(\Vec{\Gamma}_t,y_t)$ can be written as
\begin{eqnarray}
   d\hat{E}(\Vec{\Gamma}_t, y_t) = mv_t\circ dv_t + \partial_x \hat{U}(\Vec{\Gamma}_t, y_t) \circ dx_t + \partial_y \hat{E}(\Vec{\Gamma}_t,y_t) \circ dy_t ,
\end{eqnarray}
where $\circ$ denotes Stratonovich calculus. For the Poisson random variable $d\mathcal{N}_\rho^{y',y}$, the Stratonovich calculus can be converted to the Ito convention as $\partial_y \hat{E}(\Vec{\Gamma}_t,y_t) \circ dy_t = \sum_\rho \sum_{y'}\left[ \hat{E}(\Vec{\Gamma}_t, y') - \hat{E}(\Vec{\Gamma}_t, y_t)\right].\:dN^{y',y_t}_\rho(\Vec{\Gamma}_t)$. For ease of notation, we denote the thermodynamic quantities at the trajectory level by $\hat{o} \equiv \hat{o}(\Vec{\Gamma_t},y_t,t)$, corresponding to a given state $(\Vec{\Gamma}_t,y_t)$ at time $t$. Substituting $dv_t$ from Eq.~\eqref{eqn: Langevin_under}, we can rewrite the energy change in the Ito convention as
\begin{flalign}
    d\hat{E} &= - \gamma [ v^2_t + (\beta_{\mbb{X}} m)^{-1}]dt -  \sqrt{2\gamma\beta_{\mbb{X}}^{-1}} v_t \:.\: dB_t + g(x_t,y_t)\:.\:dx_t + \sum_{\rho,y'} \left[ \hat{E}(\Vec{\Gamma}_t, y') - \hat{E}(\Vec{\Gamma}_t, y_t)\right].\: dN^{y',y_t}_\rho(\Vec{\Gamma}_t)\nonumber\\
    &= \delta \hat{Q}_{\mbb{X}}  + \delta \hat{W}_{\mbb{X}}  + \sum_\rho \delta \hat{Q}_\rho + \sum_\rho \delta \hat{W}_\rho.
    \label{eqn: first_law_traj}
\end{flalign}
In the last step, we distinguish the different energy flows into the system at the trajectory level.  The work done on the system by the reservoir $\rho$ is $\delta \hat{W}_\rho = \sum_{y'}w^{y_t \to y'}_\rho(x_t).\: dN^{y',y_t}_\rho(x_t,t)$. The heat transferred from the reservoir $\rho$ to the system is then given as $\delta \hat{Q}_\rho = \sum_{y'} \left\{\left[ \hat{E}(\Vec{\Gamma}, y') - \hat{E}(\Vec{\Gamma}, y_t)\right] - w^{y_t \to y'}_\rho(x_t) \right\}\:.\: dN^{y',y_t}_\rho(x_t,t) $. Similarly, the heat transferred into the system from the reservoir $\mbb{X}$ corresponding to the UD dynamics as $\delta \hat{Q}_{\mbb{X}} = -\gamma v_t^2 dt + v_t\circ \sqrt{2\gamma\beta_{\mbb{X}}^{-1}}dB_t = - \gamma [ v^2_t - (\beta_{\mbb{X}}m)^{-1}]dt +  \sqrt{2\gamma\beta_{\mbb{X}}^{-1}} v_t \:.\: dB_t$. Finally, the work done by the non-conservative force can be computed as $ \delta \hat{W}_{\mbb{X}}  = g(x_t,y_t)\:.\:dx_t$. \par

\par
At the average level, the changes in energy of the system can be computed by taking the ensemble average of Eq.~\eqref{eqn: first_law_traj}, leading to:
\begin{eqnarray}
     d_t E  = \underbrace{\vphantom{ \sum_\rho }\Dot{Q}_{\mbb{X}}  + \Dot{W}_{\mbb{X}}}_{ d_t E_{\mbb{X}}  + \Dot{\mathcal{E}}^{\mbb{X}}}  + \underbrace{\sum_\rho \left[\Dot{Q}_\rho + \Dot{W}_\rho\right]}_{ d_t E_{\mbb{Y}}  + \Dot{\mathcal{E}}^{\mbb{Y}}} .
    \label{eqn: first_law_average}
\end{eqnarray}
From here on, the observables without the hat notation are ensemble averages, calculated over the instantaneous probability distribution, $P_t(\Vec{\Gamma},y)$. Specifically,  $\mathcal{O}\equiv\langle \hat{\mathcal{O}}\rangle_t = \SumInt \, \hat{\mathcal{O}}(\Vec{\Gamma},y) P_t(\Vec{\Gamma},y)$, where we introduce the shorthand notation $\SumInt \equiv\sum_y \int d\Vec{\Gamma}$. The average heat $\Dot{Q}_{\mbb{X}}$ and the average work $\Dot{W}_{\mbb{X}}$ entering the system from the reservoir $\mbb{X}$ is given as,
\begin{eqnarray}
    \Dot{Q}_{\mbb{X}} &=& - \gamma\left[\langle v^2\rangle -\frac{1}{m\beta_{\mbb{X}}}\right] , \label{eqn: heat_mech}\\
     \Dot{W}_{\mbb{X}} &=& \SumInt P_t(\Vec{\Gamma},y) \:v\: g(x, y) \label{eqn: work_mech}.
\end{eqnarray}
Similarly, the average heat $\Dot{Q}_\rho$ and the average work $\Dot{W}_\rho$ entering from reservoir $\rho$ are,
\begin{eqnarray}
    \Dot{Q}_\rho &=&    \sum_{y>y'}\int d\Vec{\Gamma}\: J^{y,y'}_\rho(\Vec{\Gamma},t)\left\{\left[U(x, y) - U(x, y')\right] - w^{y' \to y}_\rho(x)\right\},\\
    \Dot{W}_\rho &=&  \sum_{y>y'}\int d\Vec{\Gamma}\: J^{y,y'}_\rho(\Vec{\Gamma},t) \: w_\rho^{y'\to y}(x).
\end{eqnarray}

In addition to these external exchanges, energy is also exchanged internally between subsystems as a result of energetic interactions $\hat{U}){\mbb{XY}}$. The energy flow $\Dot{\mathcal{E}}^{\mbb{X}}$ represents the rate of increase in the interaction energy $\hat{U}_{\mbb{XY}}(x,y)$ caused solely by the UD dynamics of $\mbb{X}$.  Hence, $\Dot{\mathcal{E}}^{\mbb{X}}$ denotes the energy flowing from $\mbb{X}$ to $\mbb{Y}$. Similarly, $\Dot{\mathcal{E}}^{\mbb{Y}}$ is identified as the contribution to $\hat{U}_{\mbb{XY}}(x,y)$ exclusively due to the MJ dynamics in $\mbb{X}$, representing the energy transfer from $\mbb{Y}$ to $\mbb{X}$, 
\begin{eqnarray}  
    \Dot{\mathcal{E}}^{\mbb{X}} &=& \SumInt P_t(\Vec{\Gamma},y) \,v\,\partial_x \hat{U}_{\mbb{XY}}(x,y),\label{eqn: E_x_to_y}\\
    \Dot{\mathcal{E}}^{\mbb{Y}}&=&\sum_{\rho, y>y'}\int d\Vec{\Gamma}\: J^{y,y'}_\rho(\Vec{\Gamma})\left[\hat{U}_{\mbb{XY}}(x, y) - \hat{U}_{\mbb{XY}}(x, y')\right].
    \label{eqn: E_y_to_x}
\end{eqnarray}
In the steady-state, the average energies remains constant ($ d_t E = d_t E_{\mbb{X}}= d_t E_{\mbb{Y}} =0) $, leading to energy flow balancing each other, i.e. $\Dot{\mathcal{E}}^{\mbb{Y}}= - \Dot{\mathcal{E}}^{\mbb{X}}$. For instance, in the steady-state, a positive energy flow due to the dynamics in subsystem $\mbb{Y}$ ($\dot{\mathcal{E}}^{\mbb{Y}}>0$) implies that the work done by the reservoirs $\rho$ dominates over the corresponding heat exchanged with it,  i.e. $\sum_{\rho}\Dot{W}_{\rho} \ge -\sum_{\rho} \Dot{Q}_{\rho} > 0 $. In this case, this excess energy flows from $\mbb{Y}$ to the subsystem $\mbb{X}$, where it can be used to extract work such that $ -\Dot{W}_{\mbb{X}} \ge \Dot{Q}_{\mbb{X}} $. 

\subsection{2nd Law}
\label{sec: 2ndlaw_composite}
In this section, we will focus on entropic changes and their implications on thermodynamics. At the trajectory level, the nonequilibrium system entropy at any given time $t$ given as:
\begin{eqnarray}
    \hat{s} = s_i(y_t) - k_B\log P_t(\Vec{\Gamma}_t,y_t),
\end{eqnarray}
which accounts for the internal entropy $s_i(y)$ and the uncertainty $- k_B\log P_t(\Vec{\Gamma},y)$ at a given state. This quantity depends on both the given trajectory and also on the whole ensemble due to its dependence on the instantaneous probability density $P_t(\Vec{\Gamma}_t,y_t)$. The irreversible aspects of a trajectory are quantified by the entropy production, which is the sum of changes in the entropy of the system $\hat{s}(t)$ and the entropy associated with the reservoirs. For equilibrated baths, the change in reservoir entropy is given by the heat exchanged divided by its temperature, $-\delta \hat{Q}/T$. Hence, the instantaneous entropy production at any given instant t, is given as,
\begin{eqnarray}
    \delta \hat{\sigma}_{\rm{tot}} &=& d\hat{s}  - \frac{\delta \hat{Q}_{\mbb{X}} }{T_\mbb{X}} - \sum_\rho\frac{\delta \hat{Q}_{\rho}}{T_\mbb{Y}}\\
    &=& \underbrace{d\hat{s}_{\mbb{X}}-\frac{\delta \hat{Q}_{\mbb{X}}}{T_\mbb{X}}}_{\delta \hat{\sigma}_{\mbb{X}}} +  \underbrace{d\hat{s}_{\mbb{Y}}-\frac{\sum_\rho \delta \hat{Q}_{\rho}}{T_\mbb{Y}}}_{\delta \hat{\sigma}_{\mbb{Y}}}. \label{eqn: traj_entropy_prod} 
\end{eqnarray}
In the above expression, the instantaneous entropy production $\delta \hat{\sigma}_{\rm{tot}}$ is divided to partial contributions, $\delta \hat{\sigma}_{\mbb{X}}$ and $\delta \hat{\sigma}_{\mbb{Y}}$, due to the dynamics in subsystems $\mbb{X}$ and $\mbb{Y}$ respectively. This quantity satisfies an integrated fluctuation theorem, as shown in Sec.~\ref{sec: IFT}. Similarly, we also identify $d\hat{s}_{\mbb{X}(\mbb{Y})}$ as the time-variation of the system entropy $\hat{s}$ exclusively due to the $\mbb{X}(\mbb{Y})$ dynamics, given as,
\begin{eqnarray}
        d\hat{s}_{\mbb{X}} &= & - k_B \nabla\log P_t(\Vec{\Gamma}_t,y_t) \:.\: d\Vec{\Gamma}_t + k_B\:\frac{\nabla\:.\:\Vec{J}^y_{\Vec{\Gamma}}(t)}{P_t(\Vec{\Gamma}_t,y_t)}dt, \\
     d\hat{s}_{\mbb{Y}} &=& \left[s_i(y')- s_i(y_t) - k_B\log\frac{P_{t}(\Vec{\Gamma}_{t},y')}{P_{t}(\Vec{\Gamma}_{t},y_t)} \right]\:.\: dN^{y',y_t}_\rho(x_t,t) - k_B \left[\sum_{\rho,y'}\frac{J^{y_t,y'}_\rho(\Vec{\Gamma}_t,t)}{P_t(\Vec{\Gamma}_t,y_t)}\right]dt. \label{eqn: instant_entropy change}
\end{eqnarray}
Now moving to the average description, the nonequilibrium system entropy of the composite system, is given as:
\begin{eqnarray}
    S(t) =  \SumInt P_t(\Vec{\Gamma}, y)[s_i(y)-k_B\log P_t(\Vec{\Gamma}, y)].
\end{eqnarray}
The average entropy production rate $\Dot{\sigma}_{\rm{tot}}$ can be obtained by taking the average over Eq.~\eqref{eqn: instant_entropy change}, leading to:
\begin{eqnarray}
    \Dot{\sigma}_{\rm{tot}} &=& d_t S -\frac{\Dot{Q}_{\mbb{X}}}{T_{\mbb{X}}} - \sum_\rho \frac{\Dot{Q}_{\rho}}{T_{\mbb{Y}}} 
    \label{eqn: sigma_tot_ss_1}\\
     &=& \underbrace{\vphantom{ \sum_\rho }\frac{\gamma}{T_{\mbb{X}}}  \SumInt P_t(\Vec{\Gamma}, y)\left[v+(m\beta_{\mbb{X}})^{-1}\partial_v \log P_t(\Vec{\Gamma}, y)\right]^2}_{\Dot{\sigma}_{\mbb{X}} \ge 0} + \underbrace{\sum_\rho \int d\Vec{\Gamma} \sum_{y>y'} J^{y,y'}_\rho(\Vec{\Gamma},t) \log\frac{ \lambda_\rho^{y,y'}(x)P_t(\Vec{\Gamma}, y')}{\lambda_\rho^{y',y}(x) P_t(\Vec{\Gamma}, y)}}_{\Dot{\sigma}_{\mbb{Y}}\ge 0} \ge 0. \label{eqn: sigma_tot_ss_3}
\end{eqnarray}
This quantity is zero only when the system is at equilibrium, such that the dynamics is detailed balanced with zero irreversible fluxes. In the average description, the partial entropy productions $\Dot{\sigma}_{\mbb{X}/\mbb{Y}}$ due to individual subsystems $\mbb{X}/\mbb{Y}$ are themselves non-negative. Combining it with the first law (Eq.~\eqref{eqn: first_law_average}), we can rewrite the second law as follows:
\begin{eqnarray}
    \Dot{\sigma}_{\rm{tot}} = \left[d_t S - \beta_{\mbb{Y}}\; d_t E \right]  + \beta_{\mbb{Y}}\left[\Dot{Q}_{\mbb{X}}\Big(1-\frac{\beta_{\mbb{X}}}{\beta_{\mbb{Y}}}\Big) + \Dot{W}_{\rm{tot}} \right] \ge 0,
    \label{eqn: sigma_tot_ss_2}
\end{eqnarray}
where the total work done on the system is, $\Dot{W}_{\rm{tot}} = \Dot{W}_{\mbb{X}} + \sum_\rho \Dot{W}_\rho \equiv \Dot{W}_{\mbb{X}} + \Dot{W}_{\mbb{Y}} $. In the steady-state, since the entropy and the average energy do not change, i.e. $d_t S= d_t E = 0$, the second law imposes the following bound on its working:
\begin{eqnarray}
    \Dot{Q}_{\mbb{X}} \Big(1-\frac{\beta_{\mbb{X}}}{\beta_{\mbb{Y}}}\Big) + \Dot{W}_{\rm{tot}} \ge 0. \label{eqn: second_law_ss}
\end{eqnarray}
Hence, for non-isothermal conditions with $\beta_{\mbb{X}} > \beta_{\mbb{Y}}$, we recover the Carnot efficiency for converting heat into work, i.e $-\Dot{W}_{\rm{tot}}/\Dot{Q}_{\mbb{X}} \le [1-(\beta_{\mbb{X}}/\beta_{\mbb{Y}})]$. In an isothermal setting ($\beta_{\mbb{X}}= \beta_{\mbb{Y}})$, the expression is simplified and the efficiency of a work transducer is upper bounded by 1, i.e. $\eta_T = -\Dot{W}_{\mbb{X}}/\Dot{W}_{\mbb{Y}} \le 1$. 

\subsection{Mutual Information and information flows}
\label{subsec: infoflows}
In addition to the internal energy flows, there are also continuous information flows between the subsystems that create and / or destroy correlations between the two subsystems. The measure of correlations at any time $t$ can be tracked by mutual information. At the trajectory level, the instantaneous mutual information  $\hat{i}_t$, is given as:
\begin{eqnarray}
    \hat{i}(\Gamma_t:y_t) = k_B\log \frac{\mathbb{P}_t(\Gamma_t|y_t)}{P_t^{\mbb{X}}(\Gamma_t)} =  k_B\log \frac{\mathbb{P}_t(y_t|\Gamma_t)}{P_t^{\mbb{Y}}(y_t)}.
\end{eqnarray}
This quantity measures the distance between the conditional distribution and its corresponding marginal distribution and satisfies the following equality $\hat{i} = \hat{s}_{\mbb{X}}^{\text{marg}}(t) + \hat{s}_{\mbb{Y}}^{\text{marg}}(t) - \hat{s}(t) $. Here, $\hat{s}_{\mbb{X}}^{\text{marg}} = -k_B\log{P_t^{\mbb{X}}(\Vec{\Gamma})}$ is the marginalized system entropy for the subsystem $\mbb{X}$, and $\hat{s}_{\mbb{Y}}^{\text{marg}} = [s_i(y)-k_B \log{P_t^{\mbb{Y}}(y)}]$ is the marginalized system entropy for the subsystem $\mbb{Y}$. Similar to the other thermodynamic quantities, it time variation can also be divided into two contributions due to the dynamics in $\mbb{X}$ and $\mbb{Y}$, given as:
\begin{eqnarray}
    d\hat{i}^{\mbb{X}}&=& d\hat{s}_{\mbb{X}}^{\text{marg}}(t) - d\hat{s}_{\mbb{X}}(t) \nonumber\\
    &=&  k_B \left[\nabla\log P_t(\Vec{\Gamma}_t,y_t) - \nabla\log P^{\mbb{X}}_t(\Vec{\Gamma}_t) \right] \:.\: d\Vec{\Gamma}_t + k_B\:\frac{\partial_v \left[f_{\mbb{XY}}(x_t,y_t) P_t(\Vec{\Gamma}_t,y_t)\right]}{P_t(\Vec{\Gamma}_t,y_t)}- k_B\:\frac{\partial_v \left[\langle f_{\mbb{XY}}(\Gamma_t) \rangle_y P^{\mbb{X}}_t(\Vec{\Gamma}_t)\right]}{P^{\mbb{X}}_t(\Vec{\Gamma}_t)}dt, \\
    d\hat{i}^{\mbb{Y}}&=& d\hat{s}_{\mbb{Y}}^{\text{marg}}(t) - d\hat{s}_{\mbb{Y}}(t)\\
    &=& \left[\hat{i}_t(\Gamma_t:y') -\hat{i}(\Gamma_t:y) \right].dN_{\rho}^{y',y}(x_t,t) + k_B\sum_{\rho,y'}\frac{J^{y_t,y'}_\rho(\Vec{\Gamma}_t,t)}{P_t(\Vec{\Gamma}_t,y_t)}- k_B\sum_{\rho,y'}\int d\Vec{\Gamma}\frac{J^{y_t,y'}_\rho(\Vec{\Gamma},t)}{P^{\mbb{Y}}_t(y_t)},
\end{eqnarray}
where $\langle f_{\mbb{XY}}(\Gamma_t) \rangle_y = \sum_y f_{\mbb{XY}}(x_t,y) \mbb{P}_t(y|\Gamma_t) $ is the average interactive force that the underdamped particle experiences. \par

In the average description, the mutual information becomes the Kullback–Leibler divergence between the joint probability density $P_t(\Vec{\Gamma},y)$ and the product of marginal densities $P_t^{\mbb{X}}(\Vec{\Gamma})\otimes P_t^{\text{e}}(y)$, given as
\begin{eqnarray}
   I(t) \equiv I(\Vec{\Gamma}:y;t) = k_B\SumInt P_t(\Vec{\Gamma},y)\log\frac{P_t(\Vec{\Gamma},y)}{P^{\mbb{{X}}}_t(\Vec{\Gamma})P^{\mbb{Y}}_t(y)}.
   \label{eqn: mutual_information}
\end{eqnarray}
This quantity is positive (as $\log P \le P-1$) and is the non-additive contribution to the Shannon entropy $S(t)$ of the composite system, i.e. $I(t) = S_{\mbb{X}}^{\text{marg}}(t) + S_{\mbb{Y}}^{\text{marg}}(t) - S(t)$, where $S_{\mbb{X}(\mbb{Y})}^{\text{marg}} = \langle \hat{s}_{\mbb{X}(\mbb{Y})}^{\text{marg}}\rangle$ are the marginal Shannon entropy. By differentiating Eq.~\eqref{eqn: mutual_information} with respect to time, we identify the information flows $\Dot{\mathcal{I}}^{\mbb{X}(\mbb{Y})}$ due to the dynamics in the subsystem $\mbb{X}(\mbb{Y})$:  
\begin{flalign}
    \Dot{\mathcal{I}}^{\mbb{X}} &=  \frac{k_B}{m}\int d\Vec{\Gamma} \; P_t^{\mbb{X}}(\Vec{\Gamma}) \sum_y \left\{f_{\mbb{XY}}(x,y)\partial_v \mathbb{P}_t(y|\Vec{\Gamma}) -\left(\frac{\gamma}{\beta m}\right) \mathbb{P}_t(y|\Vec{\Gamma})\left[\partial_v\log \mathbb{P}_t(y|\Vec{\Gamma})\right]^2\right\}, \label{eqn: Idot_etom}\\
     \Dot{\mathcal{I}}^{\mbb{Y}} &= k_B\sum_\rho \sum_{y\ge y'} \int d\Vec{\Gamma} \;J_\rho^{y,y'}(\Vec{\Gamma})\log\frac{\mathbb{P}_t(\Vec{\Gamma}|y)}{\mathbb{P}_t(\Vec{\Gamma}|y')}\label{eqn: Idot_mtoe},
\end{flalign}
which satisfies $d_t I = \Dot{\mathcal{I}}^{\mbb{X}} + \Dot{\mathcal{I}}^{\mbb{Y}}$. The information flow $\Dot{\mathcal{I}}^{\mbb{Y}(\mbb{X})}$ is the rate of increase in mutual information $I(t)$ exclusively due to the dynamics of $\mbb{Y}(\mbb{X})$. It provides a qualitative insight into the measurement-feedback mechanism in such autonomous systems. For example, a positive $ \Dot{\mathcal{I}}^{\mbb{X}}$ implies that the UD dynamics in $\mbb{X}$ at a fixed state $y$ increases correlations on average and is equivalent to a measurement process. Therefore, such dynamics can be interpreted as the subsystem $\mbb{Y}$ transmitting information to the subsystem $\mbb{X}$ or the subsystem $\mbb{X}$ learning about the subsystem $\mbb{Y}$. In general, the magnitudes of the information flows at a given time $t$ may differ, that is, $|\Dot{\mathcal{I}}^{\mbb{X}}| \neq |\Dot{\mathcal{I}}^{\mbb{Y}}|$. However, in the stationary state, the mutual information converges to a constant, such that $d_t I = \Dot{\mathcal{I}}^{\mbb{X}} + \Dot{\mathcal{I}}^{\mbb{Y}}  = 0 $, similar to other time-independent observables, such as energy.
\par
Combining changes in marginal entropy $d_t S_{\mbb{X}(\mbb{Y})}^{\text{marg}}$ with information flow $\Dot{I}^{\mbb{X}(\mbb{Y})}$ leads to local second laws, given as:
\begin{equation}
    \begin{aligned}
        \Dot{\sigma}_{\mbb{Y}} &= \underbrace{\left[d_t S_{\mbb{Y}}^{\text{marg}} - \sum_\rho\frac{\Dot{Q}_\rho}{T_{\mbb{Y}}}\right]}_{\Dot{\Sigma}_{\mbb{Y}}} - \Dot{\mathcal{I}}^{\mbb{Y}} =  \Dot{\Sigma}_{\mbb{Y}} - \Dot{\mathcal{I}}^{\mbb{Y}}  \ge 0,\\
        \Dot{\sigma}_{\mbb{X}} &= \underbrace{\left[d_t S_{\mbb{X}}^{\text{marg}} - \frac{\Dot{Q}_{\mbb{X}}}{T_{\mbb{X}}} \right]}_{\Dot{\Sigma}_{\mbb{X}}}- \Dot{\mathcal{I}}^{\mbb{X}} =  \Dot{\Sigma}_{\mbb{X}} - \Dot{\mathcal{I}}^{\mbb{X}} \ge 0 .
    \end{aligned}
    \label{eqn: local_2nd_law}
\end{equation}
Only keeping track of an individual subsystem $S_{\mbb{X/Y}}^{\text{marg}}$ along with changes in associated reservoirs $\Dot{Q}_\rho/T_\rho$ leads to the estimation of marginal entropy production $\Dot{\Sigma}_{\mbb{X}(\mbb{Y})}$, which can violate the second law (Appendix.~\ref{appsec: marginal}). The positive quantity satisfying the local second law is the partial entropy production $\Dot{\sigma}_{\mbb{Y}/\mbb{X}}$, capturing the entropy produced by the local dynamics in subsystem $\mbb{Y}/\mbb{X}$ when the other subsystem $\mbb{X}/\mbb{Y}$ remains fixed. Information flows represent the non-additive component of partial entropy production, reflecting changes in correlations between subsystems. Hence, we find that bipartite information thermodynamics \cite{horowitz2014thermodynamics, herpich2020effective, parrondo2023information} naturally extends to these systems. \par

The information flow $\Dot{\mathcal{I}}^{\mbb{X}}$ due to $\mbb{X}$ can be further divided as $\Dot{\mathcal{I}}^{\mbb{X}} = \Dot{\mathcal{I}}^{\mbb{X}}_S +\Dot{\mathcal{I}}^{\mbb{X}}_F$. We identify these two terms as the entropic $\Dot{\mathcal{I}}^{\mbb{X}}_S$ and the force contribution $\Dot{\mathcal{I}}^{\mbb{X}}_F$, given as
\begin{eqnarray}
    \Dot{\mathcal{I}}^{\mbb{X}}_S &=& -\frac{\gamma k_B}{T_{\mbb{X}} m^2}\int d\Vec{\Gamma} \; P_t^{\mbb{X}}(\Vec{\Gamma})\sum_y \mathbb{P}_t(y|\Vec{\Gamma})\left[\partial_v\log \mathbb{P}_t(y|\Vec{\Gamma})\right]^2 \leq 0,\\
    \Dot{\mathcal{I}}^{\mbb{X}}_F &=& \frac{k_B}{m}\int d\Vec{\Gamma} \; P_t^{\mbb{X}}(\Vec{\Gamma})  \sum_y f_{\mbb{XY}}(x, y) \:  \partial_v \mathbb{P}_t(y|\Vec{\Gamma}).
\end{eqnarray}

Therefore, in steady state operation, when the interacting force$f_{\mathbb{XY}} = 0$, the subsystem $\mbb{Y}$ will always behave as a measurement device, as $\Dot{\mathcal{I}}^{\mbb{X}}=\Dot{\mathcal{I}}^{\mbb{X}}_S \le 0$. In such a setting, the UD dynamics in $\mathbb{X}$ becomes completely independent of the subsystem $\mathbb{Y}$ and relaxes to equilibrium. However, the MJ process in $\mathbb{Y}$ still depends on the state of $\mbb{X}$ through the jump rates, thus performing the measurement ($\Dot{\mathcal{I}}^{\mathbb{Y}} \geq 0$) on the state of $\mbb{X}$. This gained information cannot be used due to the lack of a feedback mechanism in the subsystem $\mbb{X}$ ($f_{\mathbb{XY}} = 0$). Hence, the subsystem $\mbb{Y}$ acts solely as a sensor. This scenario applies to a recent toy model \cite{pietzonka2022classical} and its electronic implementation \cite{gopal2024thermodynamic} of classical escapement clocks, where the UD dynamics drives the MJ process without any back-action. Conversely, the subsystem $\mbb{X}$ can be a measurement device (i.e. with $\Dot{\mathcal{I}}^{\mbb{X}} > 0$) only with an interacting force $f_{\mathbb{XY}}(x,y)\neq0$.  \par


\subsection{Integral Fluctuation theorem for partial entropy production}
\label{sec: IFT}
At the average level, the local second laws (Eq.~\eqref{eqn: local_2nd_law}) impose restrictions on thermodynamic flows in these autonomous systems. In this section, we generalize the fluctuation relations for partially masked dynamics, derived earlier for bipartite MJ processes \cite{shiraishi2015fluctuation} and bipartite UD dynamics \cite{rosinberg2016continuous}, to composite systems with coupled MJ-UD dynamics.  This generalization establishes fundamental bounds on fluctuations in the information processing capabilities of these systems.
Specifically, we show the existence of integral fluctuation theorems for the partial entropy production satisfying the dynamics in Eq.~\eqref{eqn: master_equation_joint}, given as:
\begin{eqnarray}
    \langle e^{-\hat{\sigma}_{\mbb{Y}}(t)/k_B} \rangle = \langle e^{-[\hat{\Sigma}_{\mbb{Y}}(t) - \Delta{\hat{I}}^{\mbb{Y}}(t)]/k_B} \rangle= 1, \hspace{1.5cm} 
     \langle e^{-\hat{\sigma}_{\mbb{X}}(t)/k_B} \rangle = \langle e^{-[\hat{\Sigma}_{\mbb{X}}(t) - \Delta{\hat{I}}^{\mbb{X}}(t)]/k_B} \rangle= 1,
     \label{eqn: IFT_main}
\end{eqnarray}
where $\hat{o}_{\mbb{X}(\mbb{Y})}(t) = \int_0^t \delta\hat{o}_{\mbb{X}(\mbb{Y})}(\tau)$ represents the time-integrated quantities along individual trajectories, corresponding to the partial entropy production $\delta\hat{\sigma}_{\mbb{X}(\mbb{Y})} = d\hat{s}_{\mbb{X}(\mbb{Y})}-\frac{\delta \hat{Q}_{\mbb{X}(\mbb{Y})}}{T_{\mbb{X}(\mbb{Y})}}$ (Eq.~\eqref{eqn: instant_entropy change}) and marginal entropy production $\delta\hat{\sigma}_{\mbb{X}(\mbb{Y})} = d\hat{s}^{\text{marg}}_{\mbb{X}(\mbb{Y})}-\frac{\delta \hat{Q}_{\mbb{X}(\mbb{Y})}}{T_{\mbb{X}(\mbb{Y})}}$ (see Sec.~\ref{subsec: infoflows}). Here, $\Delta{\hat{I}}^{\mbb{X}(\mbb{Y})}(t) = \int_0^t d\hat{I}^{\mbb{X}(\mbb{Y})}(\tau)$ is the mutual information accumulated due to the dynamics of the UD (MJ) in the individual subsystem $\mbb{X}(\mbb{Y})$. In Appendix.~\ref{appsec: IFT_proof}, we provide a detailed proof of the above generalized fluctuation theorems for such composite systems with mixed dynamics.

\section{Timescale separation}
\label{sec: TS}
In this section, we will consider the case when the jump dynamics in the subsystem $\mbb{Y}$ is much faster than the inertial dynamics in the subsystem $\mbb{X}$. For the dynamics described by Eq.~\eqref{eqn: master_equation_joint}, we can associate the time scale of the dynamics in subsystem $\mbb{Y}$ as $\tau_{\mbb{Y}}$ and a timescale due to the potential $\hat{U}_{\mbb{X}}$ in the subsystem $\mbb{X}$ as $\tau_{\mbb{X}}$. This allows us to define a dimensionless parameter $ \epsilon = \tau_{\mbb{Y}}/\tau_{\mbb{X}} $. Using the characteristic length scale $l_\mbb{X}$ of the dynamics in subsystem $\mbb{X}$, we can adimensionalize the master equation using the following variable transformation:
\begin{equation}
    \begin{aligned}
        \Tilde{t} &= t/\tau_{\mbb{Y}}, \quad \Tilde{x} = x/l_\mbb{X}, \quad \Tilde{v} = v\frac{\tau_{\mbb{X}}}{l_\mbb{X}},\\
        \quad \Tilde{U}(\Tilde{x},y) = \frac{\tau_{\mbb{X}}^2}{m \,l_\mbb{X}^2} U(x,y), \quad \Tilde{g}(\Tilde{x},y) &= \frac{\tau_{\mbb{X}}^2}{m \,l_\mbb{X}} g(x,y), \quad \Tilde{\gamma}=\frac{\tau_{\mbb{X}}}{m}\gamma,\quad \Tilde{\beta}_{\mbb{X}(\mbb{Y})}=\frac{ml_\mbb{X}^2}{\tau_{\mbb{X}}^2}\beta_{\mbb{X}(\mbb{Y})},\quad \Tilde{\lambda}^{y,y'}_\rho(\Tilde{x}) = \tau_{\mbb{Y}}\, \lambda_\rho^{y,y'}(x).
    \end{aligned}
    \label{eqn: rescaling}
    \end{equation}
Dropping the tilde notation, the adimensional master equation is given as
\begin{equation}
    d_t P_t(\Vec{\Gamma},y) = \sum_{\rho,y'}\lambda_\rho^{y,y'}(x) P_t(\Vec{\Gamma},y') + \epsilon \underbrace{\Big[-v\partial_x  - [f_{\mbb{X}} (x) + f_{\mbb{XY}} (x,y)]\partial_v + \gamma \partial_v(v+ \beta_{\mbb{X}}^{-1}\partial_v) \Big]}_{\Vec{\mathcal{L}}= \Vec{\mathcal{L}}^{\rm{det}} + \Vec{\mathcal{L}}^{\rm{diss}}} P_t(\Vec{\Gamma},y) ,
    \label{eqn: TS_master_eqn}
\end{equation}
where we have defined the escape rate $ \lambda_\rho^{y,y}(x)\equiv - \sum_{\rho,y'}\lambda_\rho^{y',y}(x)$ for state $y$. The adimensional forces are divided to identify the purely underdamped contributions $f_{\mbb{X}} (x) = -\partial_x \hat{U}_{\mbb{X}}(x)$ and the interacting ones $f_{\mbb{XY}} (x,y) = -\partial_x \hat{U}_{\mbb{XY}}(x,y) + g(x,y)$. To see the changes due to interaction terms, we further divide the Fokker-Planck operator $\Vec{\mathcal{L}} = \Vec{\mathcal{L}}^{\rm{det}} + \Vec{\mathcal{L}}^{\rm{diss}}= \Vec{\mathcal{L}}_{\mbb{X}} + \Vec{\mathcal{L}}_{\rm{int}}$, such that
\begin{eqnarray}
    \Vec{\mathcal{L}}_{\mbb{X}} = -v\partial_x  - f_{\mbb{X}} (x)\partial_v + \gamma \partial_v(v+ \beta_{\mbb{X}}^{-1}\partial_v), \hspace{0.8cm}\text{and}\hspace{0.8cm} \Vec{\mathcal{L}}_{\rm{int}} = - f_{\mbb{XY}} (x,y)  \partial_v.
\end{eqnarray}
Assuming a MJ dynamics to be much faster than the UD dynamics, i.e. $\epsilon \ll 1$, we can do a perturbative expansion of both the conditional $\mathbb{P}_t(y|\Vec{\Gamma})$ and marginal distribution $P_t^{\mbb{X}}(\Vec{\Gamma})$ in $\epsilon$, as
\begin{eqnarray}
    \mathbb{P}_t(y|\Vec{\Gamma}) &=& \mathbb{P}_t^{(0)}(y|\Vec{\Gamma}) + \epsilon \mathbb{P}_t^{(1)}(y|\Vec{\Gamma}) + \epsilon^2 \mathbb{P}_t^{(2)}(y|\Vec{\Gamma}) + \mathcal{O}(\epsilon^3)\nonumber \\
    P_t^{\mbb{X}}(\Vec{\Gamma}) &=& P_t^{\mbb{X},0}  +\epsilon P_t^{\mbb{X},1}(\Vec{\Gamma}) + \epsilon^2 P_t^{\mbb{X},1}(\Vec{\Gamma})+ \mathcal{O}(\epsilon^3),  \label{eqn: expansion}
\end{eqnarray}
such that we get back $P_t(\Vec{\Gamma}, y) = \mathbb{P}_t(y|\Vec{\Gamma})P_t^{\mbb{X}}(\Vec{\Gamma}).$
Normalization of the probabilities implies that $\sum_{y} \mathbb{P}_t^{(0)}(y|\Vec{\Gamma}) =1$, $\int dx \,dv P_t^{\mbb{X},0}(\Vec{\Gamma}) =1$, $\sum_{y} \mathbb{P}_t^{(n)}(y|\Vec{\Gamma}) = 0$  and $\int dx \,dv P_t^{(n)}(\Vec{\Gamma}) = 0$ for all $n\ge 1$. Due to the slower changes in UD dynamics of the $\mbb{X}$, the thermodynamic quantities will have a similar consequence, and the (adimensional) first and second law satisfies:
\begin{eqnarray}
     d_t E  &=& \epsilon\left[\Dot{Q}_{\mbb{X}} + \Dot{W}_{\mbb{X}}\right]  + \sum_\rho \Dot{Q}_\rho + \sum_\rho \Dot{W}_\rho \\
    &=& \epsilon \left[d_t E_{\mbb{X}}  + \Dot{\mathcal{E}}^{\mbb{X}}\right] + d_t E_{\mbb{Y}}  + \Dot{\mathcal{E}}^{\mbb{Y}},\\
    \Dot{\sigma}_{\rm{tot}}  &=& \epsilon\left[d_t S_{\mbb{X}} - \beta_{\mbb{X}} \Dot{Q}_{\mbb{X}}\right]+ \left[d_t S_{\mbb{Y}} -\beta_{\mbb{Y}}\sum_\rho\Dot{Q}_\rho\right] = \epsilon\Dot{\sigma}_{\mbb{X}} + \Dot{\sigma}_{\mbb{Y}} \ge 0 ,\\
    d_t I &=& \epsilon \left[ \Dot{\mathcal{I}}^{\mbb{X}}_S + \Dot{\mathcal{I}}^{\mbb{X}}_F \right] + \Dot{\mathcal{I}}^{\mbb{Y}},
\end{eqnarray}
where $d_t S_{\mbb{X}(\mbb{Y})}$ is the time-variation of the Shannon entropy $S$ exclusively due to $\mbb{X}(\mbb{Y})$ dynamics (Appendix.~\ref{appsec: 2nd_law_simplification}). Putting Eq.\ref{eqn: expansion} back into the master equation, we can systematically derive dynamical equations governing the marginal and conditional probabilities at different orders of timescale separation parameter $\epsilon$. We will also characterize the internal thermodynamic flows at each order of approximation.
\subsection{Adiabatic approximation}
\label{eqn: TS_adiabatic}
When the dynamics is fully time-scale separated (adiabatic approximation), i.e., $\epsilon \to 0$, we obtain:
\begin{eqnarray}
    d_t P_t^{\mbb{X},0}(\Vec{\Gamma}) &=& 0 \label{eqn: marg_order_0},\\
    d_t \mathbb{P}_t^{(0)}(y|\Vec{\Gamma}) &=&  \sum_{\rho, y'} \lambda_\rho^{y,y'}(x)\mathbb{P}_t^{(0)}(y'|\Vec{\Gamma}). \label{eqn: cond_order_0}
\end{eqnarray}
In this regime, the dynamics in the subsystem $\mbb{X}$ appears frozen, in comparison to the fast dynamics in $\mbb{Y}$. To account for the effect of the dynamics of $\mbb{Y}$ back on \mbb{X}, we need to consider the dynamical equation for the next-order term in the timescale parameter $\mathcal{O}(\epsilon^1)$, giving us:
\begin{eqnarray}
    d_t P_t^{\mbb{X},1}(\Vec{\Gamma}) &=& \Vec{\mathcal{L}}_{\mbb{X}} P_t^{\mbb{X},0}(\Vec{\Gamma})- \partial_v \left[\sum_y \mathbb{P}_t^{(0)}(y|\Vec{\Gamma}) f_{\mbb{XY}} (x,y)\right] P_t^{\mbb{X},0}(\Vec{\Gamma}) .\label{eqn: marg_order_1}
\end{eqnarray}

If the two subsystems are initially uncorrelated, i.e. $\mathbb{P}_0(y|\Vec{\Gamma}) = P^{\mathbb{Y}}_0(y)$, then the adiabatic dynamics (Eq.~\eqref{eqn: cond_order_0}) leads to a velocity-independent distribution $\mathbb{P}_t(y|x)$ at any time $t$. Combining Eq.~\eqref{eqn: cond_order_0} with Eq.~\eqref{eqn: marg_order_0} and Eq.~\eqref{eqn: marg_order_1}, we obtain a closed set of equations capturing the adiabatic dynamics:
\begin{eqnarray}
    d_t \mathbb{P}_t(y|x) &=&  \sum_{\rho, y'} \lambda_\rho^{y,y'}(x)\mathbb{P}_t(y'|x) + \mathcal{O}(\epsilon),\label{eqn: adiabatic_Y}\\
    d_t P_t^{\mbb{X}}(\Vec{\Gamma}) &=& \epsilon\left[-v\partial_x  - [f^{\rm{eff}}_{\mbb{X}} (x)+g^{\rm{eff}}(x)]\partial_v +  \gamma \partial_v(v+ \beta_{\mbb{X}}^{-1}\partial_v) \right] P_t^{\mbb{X}}(\Vec{\Gamma}) + \mathcal{O}(\epsilon^2),\label{eqn: adiabatic_X}
\end{eqnarray}
where $f_{\mbb{X}}^{\rm{eff}}(x, t)$ and $g^{\rm{eff}}(x, t)$ are the effective forces from the conservative potential $U(x,y)$ and non-conservative force $g(x,y)$, given as:
\begin{eqnarray}
    f_{\mbb{X}} ^{\rm{eff}}(x, t) &\equiv& -\left[\partial_x \hat{U}_{\mbb{X}}(x) + \sum_y \mathbb{P}_t(y|x)\partial_x \hat{U}_{\mbb{XY}}(x,y)\right] \label{eqn: f_x_eff},\\
    g^{\rm{eff}}(x, t) &\equiv& \sum_y g(x,y) \mathbb{P}_t(y|x) \label{eqn: g_x_eff}.
\end{eqnarray}
The fast dynamics in the subsystem $\mbb{Y}$ quickly relaxes to its stationary state $\pi_y(x)$, satisfying $d_t \pi_y(x) = \sum_{\rho, y'} \lambda_\rho^{y,y'}(x)\pi_y(x) = 0$. Hence, the slow adiabatic dynamics of $\mbb{X}$ is equivalent to an underdamped Brownian particle with effective conservative potential $\hat{U}_{\mbb{X}}^{\rm{eff}}(x) = - \int\! dx \;f_{\mbb{X}} ^{\rm{eff}}(x)$ and an effective non-conservative force $g^{\rm{eff}}(x)$ obtained using $\pi_y(x)$. The underdamped dynamics of $\mathbb{X}$ thus corresponds to an averaging of the fast dynamics, capturing a quasi-static evolution of the subsystem $\mathbb{X}$.   \par
 \par

For systems with purely conservative interactions, i.e. $g(x,y)=0$, the adiabatic dynamics in $\mbb{X}$ relaxes to equilibrium with a modified Gibbs state $P^{\mbb{X}}_{\rm{ad}}(\Gamma) \propto \exp{\left(-\beta_{\mbb{X}} [(v^2/2)+\hat{U}_{\mbb{X}}^{\rm{eff}}(x)]\right)}$ at the same temperature $T_{\mbb{X}}$. Although the fast dynamics in $\mbb{Y}$ can be far from equilibrium, it is unable to drive the slow underdamped dynamics out of equilibrium under the adiabatic approximation. This is reflected in the vanishing energy (Eq.~\eqref{eqn: E_x_to_y}) and information flows (Eq.~\eqref{eqn: Idot_etom}) in the steady-state, which are nonequilibrium resources capable of driving a system out of equilibrium:
\begin{eqnarray}
    \Dot{\mathcal{I}} = \epsilon  \Dot{\mathcal{I}}^{\mbb{X}} = -\Dot{\mathcal{I}}^{\mbb{Y}}  = 0 + \mathcal{O}(\epsilon^2),\\
    \Dot{\mathcal{E}} = \epsilon  \Dot{\mathcal{E}}^{\mbb{X}} = - \Dot{\mathcal{E}}^{\mbb{Y}} = 0 + \mathcal{O}(\epsilon^2).
\end{eqnarray}
Hence, there is a need to go beyond the adiabatic approximation to capture the flows between the subsystems and the associated nonequilibrium phenomenology. 
\subsection{Beyond Adiabatic approximation}
\label{sec: TS_beyond}
Going to the next order in $\epsilon^2$, we derive the following equations from the perturbative expansion of the master equation Eq.~\eqref{eqn: TS_master_eqn}:
\begin{eqnarray}
    d_t \mathbb{P}_t^{(1)}(y|\Vec{\Gamma}) &=& \sum_{\rho, y'} \lambda_\rho^{y,y'}(x)\mathbb{P}_t^{(1)}(y'|\Vec{\Gamma}) - v\partial_x \mathbb{P}_t^{(0)}(y|\Vec{\Gamma}) + \frac{\mathbb{P}_t^{(0)}(y|\Vec{\Gamma}) \partial_v P_t^{\mbb{X},0}(\Vec{\Gamma})}{P_t^{\mbb{X},0}(\Vec{\Gamma})}\left[\left(\sum_y \mathbb{P}_t^{(0)}(y|\Vec{\Gamma}) f_{\mbb{XY}} (x,y)\right) - f_{\mbb{XY}} (x,y)\right]\label{eqn: cond_order_1},\\
    d_t P_t^{\mbb{X},2}(\Vec{\Gamma}) &=& \Vec{\mathcal{L}}_{\mbb{X}} P_t^{\mbb{X},1}(\Vec{\Gamma}) -\left[\sum_y \mathbb{P}_t^{(0)}(y|\Vec{\Gamma}) f_{\mbb{XY}} (x,y)\right] \partial_v P_t^{\mbb{X},1} (\Vec{\Gamma})  -\partial_v\left[\sum_y f_{\mbb{XY}} (x,y)\mathbb{P}_t^{(1)}(y|\Vec{\Gamma})P_t^{\mbb{X},0}(\Vec{\Gamma})\right].  \label{eqn: marg_order_2}
\end{eqnarray}
Combining Eqs.~\mref{eqn: marg_order_0, eqn: cond_order_0, eqn: marg_order_1} with the above Eqs.~\mref{eqn: cond_order_1, eqn: marg_order_2}, we derive a closed dynamical equation for the marginal probability density $P_t^{\mbb{X}}(\Vec{\Gamma})$ and the conditional distribution $\mathbb{P}_t$: 
\begin{eqnarray}
         d_t \mathbb{P}_t(y|\Vec{\Gamma}) \!\!&=& \!\! \sum_{\rho, y'} \lambda_\rho^{y,y'}(x)\mathbb{P}_t(y'|\Vec{\Gamma}) + \epsilon \!\left\{\!- v\partial_x \mathbb{P}_t(y|\Vec{\Gamma}) + \mathbb{P}_t(y|\Vec{\Gamma})\left[\!\left(\sum_y f_{\mbb{XY}} (x,y) \mathbb{P}_t(y|\Vec{\Gamma}) \right) \! -\! f_{\mbb{XY}} (x,y)\right]\partial_v \log P_t^{\mbb{X}}(\Vec{\Gamma}) \right\}\! +\! \mathcal{O}(\!\epsilon^2\!), \label{eqn: closed_cond_1st order}\\
         d_t P_t^{\mbb{X}}(\Vec{\Gamma}) &=& \epsilon\left[\Vec{\mathcal{L}}_{\mbb{X}} P_t^{\mbb{X}}(\Vec{\Gamma}) -\partial_v \left(\sum_y f_{\mbb{XY}} (x,y)\mathbb{P}_t(y|\Vec{\Gamma}) P_t^{\mbb{X}}(\Vec{\Gamma})\right)  \right] + \mathcal{O}(\epsilon^3). \label{eqn: closed_marg_2nd order}
\end{eqnarray}


Hence, going beyond the adiabatic approximation, we can capture the influence of the slow UD dynamics in $\mbb{X}$ on the fast MJ dynamics in $\mbb{Y}$, leading to coupled dynamical equations of Eq.~\eqref{eqn: closed_cond_1st order} and Eq.~\eqref{eqn: closed_marg_2nd order}. To make analytical progress, we will again make the assumption that the conditional distribution has already relaxed to its steady-state, for any given marginal density $P_t^{\mbb{X}}$ at time $t$. This assumption implies $d_t\mathbb{P}_{t}^{(0+1)} = 0$ in Eq.~\eqref{eqn: closed_cond_1st order}. The steady-state conditional distribution $\mathbb{P}_{t}^{(0+1)}(y|\Vec{\Gamma})$, Applying matrix perturbation methods to Eq.~\eqref{eqn: closed_cond_1st order} around its adiabatic solution $\pi_y$,  is then given as
\begin{eqnarray}
    \mathbb{P}_{t}^{(0+1)}(y|\Vec{\Gamma})  = \pi_y(x) + \epsilon \left[v\;\sum_{y'}\mathcal{A}^{-1}_{y,y'}(x)\partial_x \pi_{y'}(x) + \partial_v \log P_t^{\mbb{X}}(\Vec{\Gamma}) \sum_{y'} \mathcal{A}^{-1}_{y,y'}(x)f_{\mbb{XY}} (x,y')\pi_{y'}(x)\right] + \mathcal{O}(\epsilon^2). \label{eqn: ss_cond_order_1}
\end{eqnarray}
Here, $\mathcal{A}^{-1}_{y,y'}(x)$ represents the generalized inverse of the generator $\mathcal{L}_{\mathbb{Y}} = \sum_\rho \lambda_\rho^{y,y'}\ket{y}\bra{y'}$ of the dynamics in the subsystem $\mbb{Y}$. It is explicitly given as $\mathcal{A}^{-1}_{y,y'}(x) = \sum_{\chi_i/\{0\}} \frac{\ket{R_\chi^i}\bra{L_\chi^i}}{\chi_i\braket{L_\chi^i|R_\chi^i}}$, where $\bra{L_\chi^i}$ and $\ket{R_\chi^i}$ represent the left and right eigenvectors corresponding to the nonzero eigenvalues $\chi_i$ of the generator $\mathcal{L}_{\mathbb{Y}} = \sum_\rho \lambda_\rho^{y,y'}(x)\ket{y}\bra{y'}$ governing the jump dynamics. Inserting it back into Eq.~\eqref{eqn: closed_marg_2nd order} results in a closed Fokker-Planck equation for the marginal dynamics. This modified Fokker-Planck equation in adimensional form, is given as:
\begin{eqnarray}
     d_t P_t^{\mbb{X}}(\Vec{\Gamma}) = \epsilon \Big[-v\partial_x  - \partial_v[f_{\mbb{X}} ^{\rm{eff}}(x) +g^{\rm{eff}}(x)  -\gamma^{\rm{eff}}(x)v] +  D^{\rm{eff}}(x)\partial_v^2) \Big]P_t^{\mbb{X}}(\Vec{\Gamma})  + \mathcal{O}(\epsilon^3), \label{eqn: modified_FP}
\end{eqnarray}
with the same effective conservative $f_{\mbb{X}} ^{\rm{eff}}(x)$ (Eq.~\eqref{eqn: f_x_eff}) and the non-conservative $g^{\rm{eff}}(x)$ (Eq.~\eqref{eqn: g_x_eff}) force. However, the drift $\gamma^{\rm{eff}}(x)$ and diffusion coefficient $D^{\rm{eff}}(x)$ are modified due to the interactive forces, given as:
\begin{eqnarray}
    \gamma^{\rm{eff}}(x) &=& \gamma - \epsilon\sum_{y,y'}f_{\mbb{XY}} (x,y)  \mathcal{A}^{-1}_{y,y'}(x)\partial_x \pi_{y'}(x), \\
    D^{\rm{eff}}(x) &=& \frac{\gamma}{\beta_{\mbb{X}} } + \epsilon\sum_{y,y'} f_{\mbb{XY}} (x,y) \mathcal{A}^{-1}_{y,y'}(x)f_{\mbb{XY}} (x,y') \pi_{y'}(x).
\end{eqnarray}
In general, this effective UD dynamics does not need to satisfy the detailed balance condition $\gamma^{\rm{eff}}/D^{\rm{eff}} \neq \beta_{\mbb{X}}$ \cite{wachtler2019stochastic}. Therefore, it is possible to drive the subsystem $\mbb{X}$ out of equilibrium through the interactive force $f_{\mbb{XY}}$ with $\mbb{Y}$. We also notice that the additional terms in the damping and the diffusion coefficient vanish by taking $\epsilon \to 0$, thereby recovering the adiabatic dynamics of Eq.~\eqref{eqn: adiabatic_X}. It is important to note that the modified Fokker-Planck equation presented above has been previously derived in the context of the 2-level electron shuttle model \cite{wachtler2019stochastic}, which will be discussed in detail in Sec.~\ref{sec: Electron_shuttle}. Here, we extend this modified Fokker-Planck formalism to more general composite systems, where the MJ process occurs over any arbitrary state space.  \par
In the absence of non-conservative force $g^{\rm{eff}}=0$, we derive the lowest-order contribution of energy and information flows in the steady-state. For the information flow $\Dot{\mathcal{I}}$, we obtain:
\begin{equation}
     \Dot{\mathcal{I}} = -\epsilon^2 \left[k_B \int\! d\Vec{\Gamma}\; P_{\rm{ad}}^{\mbb{X}}(\Vec{\Gamma}) \left\{\sum_y  \nu(x,y)\; \partial_x \hat{U}_{\mbb{XY}}(x,y) \right\} \right] + \mathcal{O}(\epsilon^3),\label{eqn: Iflow_order_2}
\end{equation}
where $\nu(x,y) = \left[\sum_{y'}\mathcal{A}^{-1}_{y,y'}(x)\partial_x \pi_{y'}(x) - \beta_{\mbb{X}} \sum_{y'} \mathcal{A}^{-1}_{y,y'}(x)f_{\mbb{XY}} (x,y')\pi_{y'}(x)\right]$ represents the first-order perturbation in the conditional distribution $\mathbb{P}_t(y|\Vec{\Gamma})$ and $P_{\rm{ad}}^{\mbb{X}}$ is the marginal density of $\mbb{X}$ from the adiabatic approximation of Eq.~\eqref{eqn: adiabatic_X}. Hence, the force part $\Dot{\mathcal{I}} = \Dot{\mathcal{I}}^{\mbb{X}}_F$ captures the essential contribution to the information flow up to $\mathcal{O}(\epsilon^2)$, and the entropic contributions $\Dot{\mathcal{I}}^{\mbb{X}}_S$  only appear at higher order. \par
Similarly, for the energy flows $\Dot{\mathcal{E}}$, we obtain: 
\begin{eqnarray}
    \Dot{\mathcal{E}}  = \epsilon^2 \left[ \int \!d\Vec{\Gamma}\; P_{ss}^{\mbb{X},1}(\Vec{\Gamma})\;v^2\right] + \mathcal{O}(\epsilon^3),\label{eqn: Eflow_order_2}
\end{eqnarray}
where $P_{ss}^{\mbb{X},1}= \lim_{\epsilon\to0}\left[P_{ss}^{\mbb{X}}-P_{\rm{ad}}^{\mbb{X}}\right]/\epsilon$ represent the first order perturbation in the marginal density $P^{\mbb{X}}$. This term captures the nonequilibrium dissipation at $\mbb{X}$ due to energy flows from $\mbb{Y} \to \mbb{X}.$ Hence, systematically going beyond the adiabatic approximation, we captured the lowest-order contributions to the energetic and information flows from $\mbb{Y}$ needed to push the dynamics of $\mbb{X}$ out of equilibrium.

\section{Large mass limit}
\label{sec: MF_approx}
In this section, we consider the dynamics and thermodynamics in the large (but finite) mass limit for the underdamped system. To avoid trivial dynamics in a flat potential, in the limit $m\to\infty$, we assume that the conservative potential $\hat{U}_{\mbb{X}}(x)$ scales with the mass. This ensures that the subsystem $\mbb{X}$ experiences a finite conservative acceleration, $\mathcal{O}(\partial_x \hat{U}_{\mbb{X}}/m)=1$, while the contribution from the interaction potential vanishes $\lim_{m\to \infty} (\partial_x \hat{U}_{\mbb{XY}}/m) = 0$. This approximation effectively separates the energy scales of the two subsystems, i.e. the energy $E_\mathbb{X}(t)$ of the underdamped subsystem $\mbb{X}$ is much higher than the energy of the dot $E_{\mathbb{Y}}(t)$ and the interaction energy $U_{\mathbb{XY}}(t)$ at any given time $t$, such that the subsystem $\mbb{Y}$ and the environment of $\mathbb{X}$ only weakly influence the underdamped dynamics of $\mathbb{X}.$ \par

Rewriting the initial master equation in Eq.~\eqref{eqn: master_equation_joint} to identify contributions for different orders of $\mathcal{O}(1/m^n)$, we get the following:
\begin{eqnarray}
    d_t P_t(\Vec{\Gamma},y) = \sum_{\rho,y'}\lambda_\rho^{y,y'}(x) P_t(\Vec{\Gamma},y') + \Big\{\left[-v\partial_x  - a_{\mbb{X}} (x) \partial_v \right] + \frac{1}{m}\left[f_{\mbb{XY}} (x,y) + \gamma \partial_v \;v\right] + \frac{1}{m^2} \gamma\beta_{\mbb{X}}^{-1}\partial_v^2 \Big\}P_t(\Vec{\Gamma},y) ,
    \label{eqn: Mass_master_eqn}
\end{eqnarray}
where we have defined the conservative acceleration $a_{\mbb{X}} = -\partial_x \hat{U}_{\mbb{X}}/m $ and the interactive forces $f_{\mbb{XY}} (x,y) = -\partial_x \hat{U}_{\mbb{XY}}(x,y) + g(x,y)$. In the diverging mass $m\to \infty$ limit, the subsystem $\mbb{X}$ has Hamiltonian dynamics $\Dot{x}_t = v_t$ and $\Dot{v}_t = a_{\mbb{X}}(x_t) = -\partial_x \hat{U}_{\mbb{X}}(x_t)/m$ due to its initial energy $E_{\mbb{X}} = (1/2)mv_0^2+ \hat{U}_{\mbb{X}}(x_0)$ \cite{strasberg2021autonomous}. Here, we go beyond this limit to the next order contribution to the dynamics for systems with large, but finite mass $m$. Similar to the previous section, we perturbatively expand the conditional distribution $\mathbb{P}_t(y|\Vec{\Gamma})$ and marginal probability density $P_t^{\mbb{X}}(\Vec{\Gamma})$, leading to the following dynamical equation up to  $\mathcal{O}(1/m)$:
\begin{eqnarray}
         d_t P_t^{\mbb{X}}(\Vec{\Gamma}) &=& \left\{\left[-v\partial_x  - a_{\mbb{X}} (x) \partial_v \right] + \frac{1}{m}\left[\gamma \partial_v \;v - \sum_y \mathbb{P}_t(y|\Vec{\Gamma})f_{\mbb{XY}}(x,y) \right] \right\} P_t^{\mbb{X}}(\Vec{\Gamma}) + \mathcal{O}(m^{-2}),\label{eqn: MF_closed_marg}\\
          d_t \mathbb{P}_t(y|\Vec{\Gamma}) &=&  \sum_{\rho, y'} \lambda_\rho^{y,y'}(x)\mathbb{P}_t(y'|\Vec{\Gamma}) + \left[-v\partial_x  - a_{\mbb{X}} (x) \partial_v \right]\mathbb{P}_t(y'|\Vec{\Gamma}) + \mathcal{O}(m^{-1}) .\label{eqn: cond_MF_0}
\end{eqnarray}
Due to the lack of a diffusive term up to $\mathcal{O}(m^{-1})$, if we initialize the system in an uncorrelated state such that $P_0(\Vec{\Gamma}, y)= P^{\mbb{Y}}_0(y)\delta(x-x_0)\delta(v-v_0)$, then the marginal dynamics evolves to $P_t^{\mbb{X}}(\Vec{\Gamma}) = \delta(x-x_t)\delta(v-v_t)$ at time $t$, satisfying the deterministic dynamics:
\begin{equation}
\begin{aligned}  
    dx_t &= v_t dt,\\
    dv_t &= a_{\mbb{X}}(x_t) dt + \frac{1}{m}\left[-\gamma v_t +  \sum_y \mathbb{P}_t(y|\Vec{\Gamma}_t)f_{\mbb{XY}}(x_t,y) \right] dt +\mathcal{O}(m^{-2}).
    \label{eqn: Langevin_MF}  
\end{aligned}
\end{equation}
The above dynamical equation goes beyond the constant energy timescales of Hamiltonian dynamics of subsystem $\mbb{X}$ at the infinite mass limit. It captures the lowest order contribution due to damping and interaction with subsystem $\mbb{Y}$ for large but finite mass. However, the above equation Eq.~\eqref{eqn: Langevin_MF} is not closed as it is still dependent on the conditional probability $\mathbb{P}_t(y|\Vec{\Gamma}_t)$. To close the dynamical equation up to $\mathcal{O}(1/m)$, we need the conditional probability $\mathbb{P}_t(y|\Gamma_t)$ along a deterministic trajectory $\Vec{\Gamma}_t$ up to $\mathcal{O}(1)$ at any time $t$. This can be obtained by solving:
 \begin{eqnarray}
     d_t \mathbb{P}_t(y|\Vec{\Gamma}_t) &=&  \sum_{\rho, y'} \lambda_\rho^{y,y'}(x)\mathbb{P}_t(y'|\Vec{\Gamma}_t) + \mathcal{O}(m^{-1}).
     \label{eqn: MF_prob_Y}
 \end{eqnarray}
The term due to the Hamiltonian flow in Eq.~\eqref{eqn: cond_MF_0} drops out when the conditional probability $\mathbb{P}_t(y|\Gamma_t)$ is only along the deterministic trajectory $\mathbb{P}_t(y|\Vec{\Gamma}_t) = \int d\Vec{\Gamma}\; \mathbb{P}_t(y|\Vec{\Gamma}) \delta(\Vec{\Gamma}-\Vec{\Gamma}_t)=  \int d\Vec{\Gamma}\; \mathbb{P}_t(y|\Vec{\Gamma}) P_t^{\mbb{X}}(\Vec{\Gamma})$. In this setting, the conditional distribution $\mathbb{P}_t(y|\Vec{\Gamma}_t)$ is equal to the corresponding marginal probability $P_t^{\mbb{Y}}(y)$, i.e. $P_t^{\mbb{Y}}(y) = \mathbb{P}_t(y|\Vec{\Gamma}_t) +\mathcal{O}(1/m)$. Hence, combining Eq.~\eqref{eqn: Langevin_MF} along with Eq.~\eqref{eqn: MF_prob_Y}, we obtain the closed form dynamics consistent upto $\mathcal{O}(1/m)$, given as:
\begin{equation}
\begin{aligned}  
    dx_t &= v_t dt,\\
    dv_t &= a_{\mbb{X}}(x_t) dt + \frac{1}{m}\left[-\gamma v_t +  \sum_y P_t^{\mbb{Y}}(y)f_{\mbb{XY}}(x_t,y) \right] dt +\mathcal{O}(m^{-2}),\\
    d_t P_t^{\mbb{Y}}(y) &=  \sum_{\rho, y'} \lambda_\rho^{y,y'}(x_t)P_t^{\mbb{Y}}(y) + \mathcal{O}(m^{-1}).
    \label{eqn: Large_mass_general}  
\end{aligned}
\end{equation}
Hence, we effectively obtain a deterministic underdamped dynamics for the subsystem $\mbb{X}$, which drives the stochastic MJ dynamics through its dependence on the Poisson rates $\lambda_\rho^{y,y'}(x)$. This dynamics is equivalent to the meanfield dynamics \cite{gorelik1998shuttle, wachtler2019stochastic, vigneau2022ultrastrong}, obtained by doing an ad-hoc approximation $\lambda_\rho^{y,y'}(x) \approx \lambda_\rho^{y,y'}(\langle x \rangle)$, which leads to a closed form equation for $\langle x \rangle, \langle v \rangle$ and $P_t^{\mbb{Y}}(y)$. It is important to highlight that the dynamics of Eq.~\eqref{eqn: Large_mass_general} is essentially deterministic and need not have a time-independent stationary state. Instead, the nonlinear dynamics (Eq.~\eqref{eqn: Large_mass_general}) may exhibit a time-dependent attractor, such as a limit cycle. In contrast, the full stochastic dynamics Eq.~\eqref{eqn: master_equation_joint} has a unique steady-state in the long time limit \( P_{ss}(\vec{\Gamma}, y) = \lim_{t\to\infty} P_t(\vec{\Gamma}, y) \). This indicates that to accurately capture the stationary state, one must consider higher order terms, beyond \( \mathcal{O}(m^{-1}) \), accounting for the diffusive terms in the UD dynamics. Note that the derivation of the mean-field dynamics relies on the assumption that the energy of the underdamped subsystem $\mbb{X}$ is much larger than the energy scales at any given time, i.e $ E_{\mbb{X}}(\Vec{\Gamma}_t) \gg E_{\mbb{Y}}(y_t), \hat{U}_{\mbb{XY}}(\Vec{\Gamma}_t,y_t)$, which is naturally achieved for $m\gg 1$ and $\Gamma_t \neq (0,0)$.  \par

The deterministic dynamical equations of Eq.~\eqref{eqn: Large_mass_general} leads to an incomplete description of the thermodynamics, as the higher-order diffusion term plays an important role in the thermodynamic quantities. For instance, the heat $\Dot{Q}_{\mbb{X}}$ flow in $\mbb{X}$ contains a
term $-\gamma/\beta m$ arising from the thermal diffusion term
of $\mathcal{O}(m^{-2})$ in the master equation Eq.~\eqref{eqn: Mass_master_eqn}. Due to our truncation to $\mathcal{O}(m^{-1})$ in the derivation of MF dynamics, these diffusive effects are not captured. To rigorously derive thermodynamic quantities, the dynamics must be extended to at least $\mathcal{O}(m^{-2})$. Despite this, thermodynamic quantities can still be estimated from MF dynamics using the localized joint probability density $P_t(\Vec{\Gamma}, y)= P^{\mbb{Y}}_t(y)\delta(x-x_t)\delta(v-v_t)$, as done in \cite{wachtler2019stochastic}. The first law at the average level due to meanfield dynamics satisfies:
\begin{eqnarray}
         d_t \overbar{E}  = \vphantom{ \sum_\rho }\Dot{\overbar{Q}}_{\mbb{X}}  + \Dot{\overbar{W}}_{\mbb{X}}  + \sum_\rho \left[\Dot{\overbar{Q}}_\rho + \Dot{\overbar{W}}_\rho\right],
    \label{eqn: first_law_MF}
\end{eqnarray}
where each of thermodynamic quantities are computed from the meanfield dynamics:
\begin{equation}
\begin{aligned}
    \Dot{\overbar{Q}}_{\mbb{X}} &= - \gamma v_t^2  , \hspace{1cm}     \Dot{\overbar{W}}_{\mbb{X}} = \sum_y P^{\mbb{Y}}_t(y) \:v_t\: g(x_t, y) ,\\
    \Dot{\overbar{Q}}_\rho =  \sum_{y>y'}\overbar{J}^{y,y'}_\rho(\Vec{\Gamma}_t)&\left\{\left[U(x_t, y) - U(x_t, y')\right]- w^{y' \to y}_\rho(x_t)\right\},\hspace{0.5cm}
    \Dot{\overbar{W}}_\rho =  \sum_{y>y'} \overbar{J}^{y,y'}_\rho(\Vec{\Gamma}_t) \: w_\rho^{y'\to y}(x_t),\\
    \Dot{\overbar{\mathcal{E}}}^{\mbb{X}}(t) = \sum_y P^{\mbb{Y}}_t(y) &\,v_t\,\partial_x \hat{U}_{\mbb{XY}}(x_t,y), \hspace{0.5cm}
    \Dot{\overbar{\mathcal{E}}}^{\mbb{Y}}(t) =\sum_{\rho, y>y'} \overbar{J}^{y,y'}_\rho(\Vec{\Gamma}_t)\left[\hat{U}_{\mbb{XY}}(x_t, y) - \hat{U}_{\mbb{XY}}(x_t, y')\right],
\end{aligned}
\end{equation}
where $\overbar{J}_\rho^{y,y'} = \lambda_\rho^{y,y'}(x_t) P^{\mbb{Y}}_t(y') -  \lambda_\rho^{y',y}(x_t) P^{\mbb{Y}}_t(y)$ is the average probability current. In particular, the heat flow $\Dot{\bar{Q}}_{\mbb{X}}$ to its reservoir $\mbb{X}$, as described by the mean-field (MF) dynamics, accounts only for the interaction with the reservoir through the damping force $-\gamma v_t$. \par


Here, the deterministic nature of the subsystem $\mbb{X}$ with the localized probability density $P_t(\Vec{\Gamma}, y)$, the joint entropy $S(t)$ has contributions only from the subsystem $\mbb{Y}$, i.e. $\overbar{S}(t) \approx S_{\mbb{Y}}^{\text{marg}}(t) = -k_B \sum_y P^{\mbb{Y}}_t(y) \log P^{\mbb{Y}}_t(y).$ As a result, it also fails to account for changes in correlations, since the joint distribution is an independent product of the marginal densities: $P_t(\Vec{\Gamma}, y) = P^{\mbb{Y}}_t(y) P^{\mbb{X}}_t(\Vec{\Gamma}) = P^{\mbb{Y}}_t(y)\delta(x-x_t)\delta(v-v_t)$. Hence, the mutual information and information flows vanish in this limit, i.e. $d_tI(t) = \Dot{\mathcal{I}}^{\mbb{X}}(t) + \Dot{\mathcal{I}}^{\mbb{Y}}(t) = 0$ (See Sec.~\ref{subsec: infoflows}). \par

The entropy production rate is the sum of changes in the system entropy $d_t S_{\mbb{Y}}^{\text{marg}}(t)$ and changes in the reservoir entropy, leading to:
\begin{eqnarray}
    \Dot{\overbar{\sigma}}_{\rm{tot}} &=& d_t \overbar{S} -\frac{\Dot{\overbar{Q}}_{\mbb{X}}}{T_{\mbb{X}}} - \sum_\rho \frac{\Dot{\overbar{Q}}_{\rho}}{T_{\mbb{Y}}}\\
    &=& \sum_{\rho} \left[\lambda_\rho^{y,y'}(x_t) P^{\mbb{Y}}_t(y') -  \lambda_\rho^{y',y}(x_t) P^{\mbb{Y}}_t(y) \right]\log\frac{\lambda_\rho^{y,y'}(x_t) P^{\mbb{Y}}_t(y')}{\lambda_\rho^{y',y}(x_t) P^{\mbb{Y}}_t(y)} + \frac{\gamma}{T_{\mbb{X}}}v_t^2  \ge 0.
\end{eqnarray}
Thus, we obtain consistent thermodynamic laws for meanfield dynamics as it corresponds just to the large mass limit $m\gg1$.
\section{Example: Electron shuttle}
\label{sec: Electron_shuttle}
In this section, we consider one of the simplest nanoelectromechanical systems, called the electron shuttle. The system consists of a quantum dot hosted by an underdamped harmonic oscillator, placed under a voltage bias between two leads. Electron shuttling is a form of periodic transport or ``shuttling'' of charges between the leads due to the interplay between the mechanical oscillations and sequential tunneling events. It is the nanoscopic equivalent of the classical bell experiment, in which a metallic ball bounces back and forth between two leads at constant high voltage drops, resulting in periodic charge transfer at the macroscale \cite{tuominen1999stepwise}.  However, with modern nanomechanical setups at ultra-low temperatures, charge shuttling has been observed experimentally at the single-electron level \cite{tuominen1999stepwise,scheible2004silicon, erbe2001nanomechanical,moskalenko2009nanomechanical,konig2012voltage,wen2020coherent}. These setups have been extensively studied theoretically, both in the classical and quantum regimes \cite{gorelik1998shuttle,isacsson1998shuttle,clerk2005quantum,weiss1999accuracy,boese2001influence,braig2003vibrational,sapmaz2003carbon,novotny2003quantum,pistolesi2004full,novotny2004shot,pistolesi2004charge,fedorets2004quantum,bennett2006laser,utami2006quantum,blanter2004single}. Recently, the classical setup has been thermodynamically studied \cite{wachtler2019stochastic}, and various autonomous thermodynamic cycles using the self-oscillations have been proposed \cite{tonekaboni2018autonomous, wachtler2019proposal, strasberg2021autonomous}. 
\par

\begin{figure}[h!]
    \centering
    \includegraphics[trim=10 0 10 0, clip, width=0.85\textwidth]{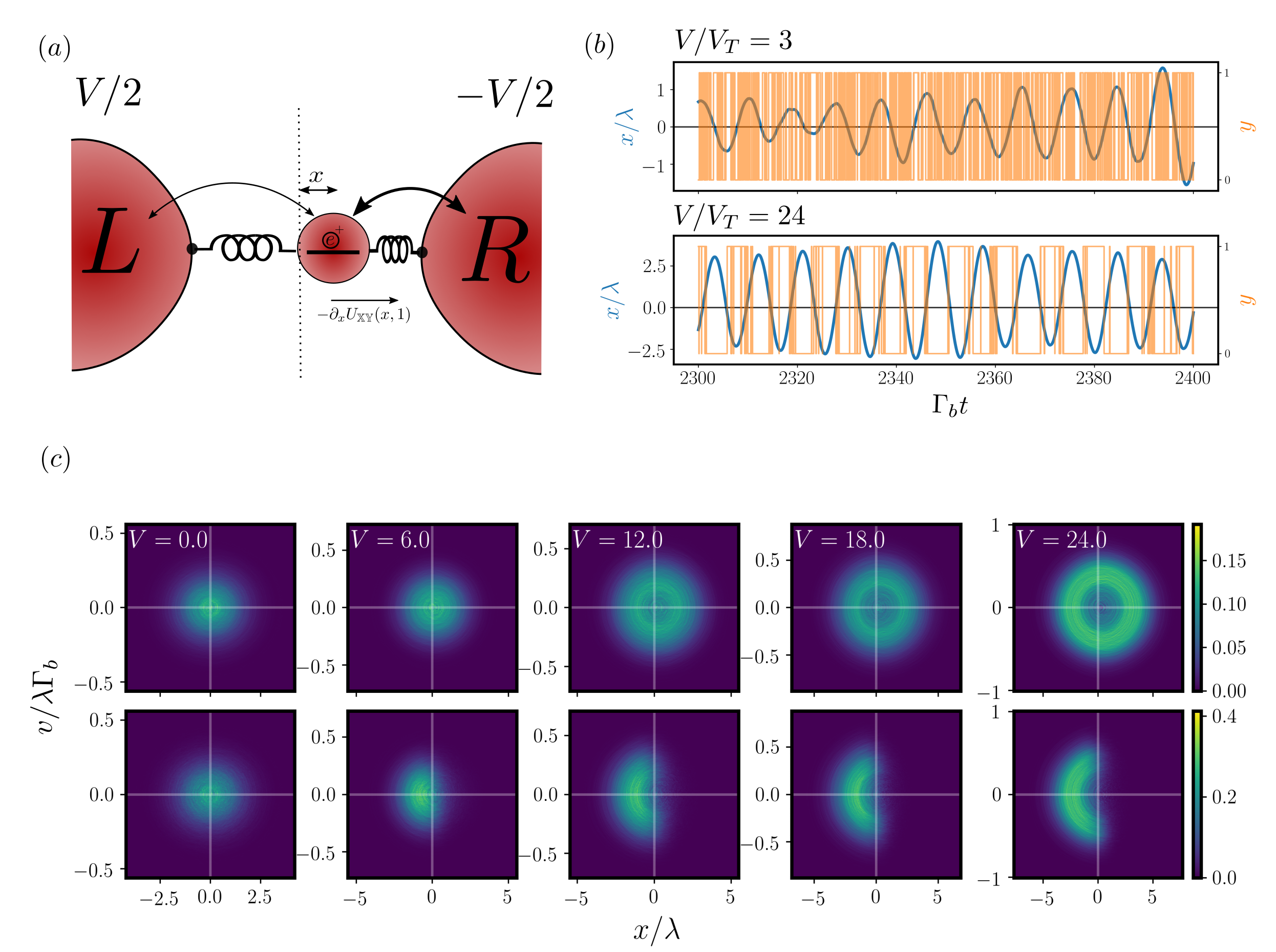}
    \caption{(a): Schematic diagram of the electron shuttle, with an oscillating dot, which experiences a electrostatic force $f_{\mbb{XY}} (x,y) = \alpha V y$ when the dot is occupied ($y=1$). (b): Sample trajectory of the electron shuttle for increasing voltage $V$. The position $x$ of the oscillator is in blue and the state of the quantum dot $y$ is in orange.  The critical voltage, as computed from MF dynamics (Eq.~\eqref{eqn: MF_shuttle}), when the shuttle transitions from incoherent oscillations to coherent oscillations is $V_{\rm{cr}}/V_T \sim 9.2$. (c):  Marginal probability density $P_{ss}^{\mathbb{X}}(x,v)$ (top panel) and the conditional probability density $\mathbb{P}_{ss}(x,v|1)$ (bottom panel) given the occupied dot state for increasing powering voltages $V/V_T$ at the coupling parameter $\alpha \lambda=0.03$.  Parameters: $\omega/\Gamma_b = 0.1292, \gamma/m\omega = 0.01, m\lambda^2\Gamma_b^2\beta=0.08$ with $\lambda=1$, $\Gamma_b=5$ and $\beta=1$.}
    \label{fig: trajectory_shuttle}
\end{figure}

Here, we consider the classical description of the electron shuttling as a result of incoherent quantum tunneling between the leads and the oscillating dot. For simplicity, we assume the dot to be in the ultra-strong Coulomb blockade regime (i.e. ultra-low temperatures and/or small capacitances) such that the dot can accommodate only a single electron. In this context, the dot can be modeled as a single-electron level with addition energy $\mathbb{e}_0$, whose state is described by its occupancy $y\in\{0,1\}$. The dot is further hosted by a nanomechanical oscillator, with mass $m$ and natural frequency $\omega$, that moves in one dimension between the leads. Its instantaneous state $\Vec{\Gamma}\equiv(x,v)$, is characterized by its position $x$ and velocity $v$ (See Fig.~\ref{fig: trajectory_shuttle}(a)). When the dot is occupied ($y=1$), the oscillator is charged and experiences a Coulomb force $f_{\mbb{XY}} = \alpha V y$ due to the electric field between the leads, where $V$ is the voltage drop across the leads, $\alpha$ is the coupling strength with the dimension of the inverse distance. In these nanoelectromechanical (NEM) systems, this force can deform the mechanical component of the system due to electromechanical coupling, which to the lowest order has energy $\hat{U}_{\mbb{XY}} = -\int \!dx \,f_{\mbb{XY}}=  -\alpha V y\, x$. Hence, the total internal energy of the shuttle is given as
\begin{eqnarray}
    \hat{E}(\Vec{\Gamma}, y) = \frac{1}{2}m v^2 + \frac{1}{2}m\omega^2 x^2 + \mathbb{e}_0 y - \alpha V y\, x \;.\label{eqn: energy_shuttle}
\end{eqnarray}
We can identify the energetic contribution due to the isolated electronic subsystem as $\hat{U}_{\mbb{Y}}(y) = \mathbb{e}_0 y$ and mechanical subsystems as $\hat{U}_{\mbb{Y}}(\Vec{\Gamma}) = \frac{1}{2}m v^2 + \frac{1}{2}m\omega^2 x^2$, along with the interaction energy due to electrostatic potential, $\hat{U}_{\mbb{XY}} = -\alpha V y\, x$. \par


\subsection{Stochastic dynamics}
\label{subsec: stochastic dynanmics}

The changes in dot state $y$ are due to tunneling events between the dot and the two leads $\rho\in \{L, R\}$, assumed to be fermionic reservoirs with chemical potentials $\mu_L = \mathbb{e}_0 + V/2$, $\mu_R = \mathbb{e}_0 - V/2$ and inverse temperatures $\beta =(1/k_BT)$. The chemical potential defines the work done by the reservoir $\rho$ to add an electron $(y=0\to1)$ to the dot, i.e., $w_\rho(0\to1) = q_e\mu_\rho$. The oscillator also has a dissipative reservoir with inverse temperature $\beta_{\mbb{X}}$. In this section, we will consider isothermal conditions, i.e $\beta_{\mbb{X}}= \beta_{\mbb{Y}}\equiv \beta$. The jump rates $\lambda_{\rho}^{y,y'}(x)$ for individual tunneling events in the dot are given as
\begin{equation}
\begin{aligned}
 \lambda_{L}^{1,0}(x)  &= \Gamma_b e^{-x/\lambda} f_L(\mathbb{e}_0 - \alpha V x), \hspace{1.7cm}\lambda_{L}^{0,1}(x) = \Gamma_b e^{-x/\lambda}[1-f_L(\mathbb{e}_0 - \alpha V x)];\\
\lambda_{R}^{1,0}(x)  &= \Gamma_b e^{x/\lambda} f_R(\mathbb{e}_0 - \alpha V x), \hspace{1.84cm}\lambda_{R}^{0,1}(x) = \Gamma_b e^{x/\lambda}[1-f_R(\mathbb{e}_0 - \alpha V x)];
\end{aligned}   
\end{equation}
where $f_{\rho}(e) = [\exp(q_e\beta(e - \mu_\rho))+1]^{-1}$ is the Fermi distribution, $\lambda$ is the characteristic tunneling length and $\Gamma_b$ is the bare tunneling rate. Since the tunneling probability exponentially decays with the distance from leads, the tunneling amplitude (or kinetic component) of the jump rates also depends on position $x$ as $\sqrt{\lambda_\rho^{1,0}\lambda_\rho^{0,1}} =\Gamma_b e^{\pm x/\lambda}$. Note that the above rates are thermodynamically consistent, satisfying the local detailed balance condition (Eq.~\eqref{eqn: LDB_condition}). Hence, the stochastic dynamics of the dot and the oscillator are coupled both through their energetic interaction $\hat{U}_{\mbb{XY}}(x,y)=-\alpha V y \,x$ and also through the position-dependent tunneling amplitude $\Gamma_b e^{\pm x/\lambda}$. The master equation (Eq.~\eqref{eqn: master_equation_joint}) for the electron shuttle follows:
\begin{eqnarray}
    d_t P_t(\Vec{\Gamma},y) &=&  (2y-1)[J_L^{1,0}(\Vec{\Gamma}) + J_R^{1,0}(\Vec{\Gamma})] +  \left[ -v \partial_x + \frac{1}{m}\partial_v(m\omega^2 x + \gamma v - \alpha V q) + \frac{\gamma}{m^2\beta_{\mbb{X}} }\,\partial_v^2\right] P_t(\Vec{\Gamma},q).
    \label{eqn: master_equation_shuttle}
\end{eqnarray}
Here, the probability currents $J_{L/R}^{1,0}(\Vec{\Gamma})$ flowing through left $(L)$ and right $(R)$ leads, are given as
\begin{eqnarray}
    J_{L/R}^{1,0}(\Vec{\Gamma}) = \lambda_{L/R}^{1,0}(x) P_t(\Vec{\Gamma},0) -  \lambda_{L/R}^{0,1}(x) P_t(\Vec{\Gamma},1).
    \label{eqn: prob_current_dot}
\end{eqnarray}


The stochastic dynamics of an electron shuttle can be qualitatively understood as follows. Due to tunneling events from the leads, an unoccupied dot becomes occupied and makes the grain charged. Hence, at finite voltage $V$ across the leads, the grain experiences a force $f_{\mbb{XY}}$, pushing the grain towards the right lead with lower potential $\mu_R$. As the grain moves closer to the right lead with $x>0$, the tunneling rates increase exponentially $\lambda_R \propto e^{+x/\lambda}$, facilitating the transfer of charge from the dot to the right lead and leaving the grain temporarily uncharged. Subsequently, the restoring force of the oscillator pulls the grain back beyond its equilibrium position toward the left lead, leading to $x<0$. As the grain moves closer to the left lead, the tunneling amplitude to the left lead, $\lambda_L \propto e^{-x/\lambda}$, increases. This enables the charge to tunnel from the left lead to the dot, thereby recharging the grain and repeating the process. Hence, the applied voltage $V$ induces a self-oscillatory motion of the grain, driven by the electromechanical force $f_{\mbb{XY}}$. \par 

The transition to self-oscillation depends on the interplay between damping, characterized by a damping coefficient $\gamma$, and the strength of the electromechanical force $f_{\mbb{XY}}$, which occurs beyond a critical voltage $V_{\rm{cr}}$ \cite{utami2006quantum, wachtler2019stochastic}. For low voltages $V< V_{\rm{cr}}$, the damping dominates, confining the grain near its equilibrium position $(x,v)=(0,0)$. The tunneling dynamics of the dot thus resembles that of a stationary dot. However, beyond the critical voltage $V > V_{\rm{cr}}$, the electromechanical force $f_{\mbb{XY}}$ overcomes the damping resulting in self-oscillations. As the voltage $V$ increases, the force $f_{\mbb{XY}}$ drives the oscillator to larger amplitudes, and the dot is likely to be occupied, $y=1$, when the grain is closer to the left lead, $x/\lambda \ll0$, and vice versa. For very large voltages $V \gg V_{\rm{cr}}$, the amplitude of the oscillations becomes so large that the dot effectively does not experience the region around $x/\lambda \sim 0$, where tunneling is equally probable for both leads. This prevents short-circuit current flow between the leads. In this high-amplitude regime, the tunneling with the dot predominantly occurs due to a single lead at any time, effectively transferring a single electron from the left to the right lead in a single oscillation. This defines the ``shuttling'' regime of this electromechanical setup. Note that the stochastic dynamics of Eq.~\eqref{eqn: master_equation_shuttle} does not have a sharp transition to self-oscillation mainly due to the thermal fluctuations in the oscillator. But the meanfield dynamics (discussed in the next Sec.~\ref{subsec: MF_shuttle}) provides a reliable estimate of the transition, as previously shown in \cite{wachtler2019stochastic}.

In Fig.~\ref{fig: trajectory_shuttle}(b), we plot a sample trajectory for the position $x$ and the state of the dot $y$ below and above the critical voltage $V_{\rm{cr}}/V_T$. Here, we consider the timescale-separated case, with bare tunneling rates significantly higher than the oscillator’s natural frequency, i.e. $\Gamma_b \gg \omega$. For small applied voltages $V<V_{\rm{cr}}$ (top panel), the oscillator oscillates irregularly due to its thermal fluctuations around $x=0$, similar to the equilibrium scenario ($V=0$). In this case, the dot behaves as in stationary condition with many tunneling events from both leads. However, above a critical voltage $V>V_{\rm{cr}}$ (bottom panel), the amplitude of the oscillations is larger and the oscillations also become more coherent, which characterizes the regime of self-oscillations. Notice also that there are fewer tunneling events as the dot interacts mainly with a single lead. \par

The onset of self-oscillations can clearly be seen in the marginal distribution $P(x,v)$ of the oscillator. In Fig.~\ref{fig: trajectory_shuttle}(c) (top panel), we plot the marginal distribution $P_{\rm{ss}}^{\mbb{X}}(x,v)$ in the steady-state as we increase the applied voltage $V$.  At zero voltage drop $V=0$, the oscillator and the dot relax independently to its equilibrium, with its stationary behavior of the oscillator described by its Gibbs density $P^{\mbb{X}}(x,v) \propto \exp(-\beta(m\omega^2x^2+mv^2)/2)$. As the voltage increases below the critical voltage $0<V<V_{\rm{cr}}$, the probability density still peaks around zero, but with a larger standard deviation. However, above the critical voltage $V>V_{\rm{cr}}$, the probability density peaks around a circular orbit, reflecting the self-oscillating regime. In this regime, further increasing the voltage $V$ leads to increasing amplitude of oscillations. For the timescale separated parameters considered in Fig.~\ref{fig: trajectory_shuttle}(b) and (c), achieving the shuttling behavior requires very large oscillation amplitudes, which occur at voltages far above the critical threshold, $V \gg V_{\rm{cr}}$, beyond the range of our analysis. \par

In the next sections, we will compare the stochastic dynamics Eq.~\eqref{eqn: master_equation_shuttle} of the electron shuttle with the predictions from the dynamical equations derived by taking the large mass limit (Sec.~\ref{sec: MF_approx}) and the slow oscillator/fast tunneling limit (Sec.~\ref{sec: TS}), respectively.

\subsubsection{Large mass limit : Meanfield dynamics}
\label{subsec: MF_shuttle}

In Sec.\ref{sec: MF_approx}, we established that the stochastic dynamics described by Eq.~\eqref{eqn: master_equation_shuttle} is equivalent to a meanfield (MF) approximation in the large mass limit, by keeping terms up to $\mathcal{O}(m^{-1})$ in the full master equation. The MF approximation effectively neglects fluctuations of the oscillator in the jump dynamics, simplifying the tunneling rates, i.e. $\lambda_{L/R}(x) \approx \lambda_{L/R}(\langle x \rangle)$. This leads to closed-form equations for the mean values $(\overbar{x}(t),\overbar{v}(t),\overbar{p}_1(t)) \equiv (\langle x(t) \rangle, \langle v(t) \rangle, P^{\mbb{Y}}_t(1))$ \cite{wachtler2019stochastic}, given by:
\begin{equation}
\begin{aligned}
    \Dot{\overbar{x}}(t) &= \overbar{v}(t),\\
    m \Dot{\overbar{v}}(t) &= -m \omega^2 \overbar{x}(t) - \gamma \overbar{v}(t) + \alpha V {\overbar{p}_1}(t),\\
    \Dot{\overbar{p}_1}(t) &= \overbar{J}_{L}^{1,0}(t) + \overbar{J}_{R}^{1,0}(t),
    \label{eqn: MF_shuttle}
\end{aligned}
\end{equation}
where $\overbar{J}_{L/R}^{1,0}(t) = \lambda_{L/R}^{1,0}(\overbar{x}(t)) \overbar{p}_0(t) -  \lambda_{L/R}^{0,1}(\overbar{x}(t))  \overbar{p}_1(t)$ is the the probability current to the left $(L)$ and right ($R$) leads, respectively. This MF dynamics (Eq.~\eqref{eqn: MF_shuttle}) consists of coupled nonlinear equations that exhibit a Hopf bifurcation at a critical voltage $V_{\rm{cr}}$ \cite{wachtler2019stochastic}. It transitions from its fixed point, satisfying $(\Dot{\overbar{x}}=0,\Dot{\overbar{v}}=0,\Dot{\overbar{p}}_1=0)$, to a stable limit cycle for $V>V_{\rm{cr}}$. The critical voltage $V_{\rm{cr}}$ corresponds to the voltage at which the real part of Jacobian of the MF dynamics Eq.~\eqref{eqn: MF_shuttle} changes from negative to positive in the linear stability analysis around the fixed point. For a large but finite oscillator mass, stochastic dynamics qualitatively follow this bifurcation behavior, though fluctuations smooth out the sharp transition \cite{wachtler2019stochastic}. In the following analysis, we demonstrate that the MF dynamics effectively captures the transient stochastic behavior in the large-mass limit and predicts the limit cycle as the most probable steady-state behavior, even when the oscillator mass is finite with significant thermal fluctuations. However, the MF dynamics (Eq.~\eqref{eqn: MF_shuttle}) captures the limit cycle attractor, it fails to reach a time-independent stationary state, which requires the higher order $\mathcal{O}(m^{-2})$ diffusive term. \par

\begin{figure}[h!]
    \centering
    \includegraphics[trim=10 0 0 0, clip, width=0.85\textwidth]{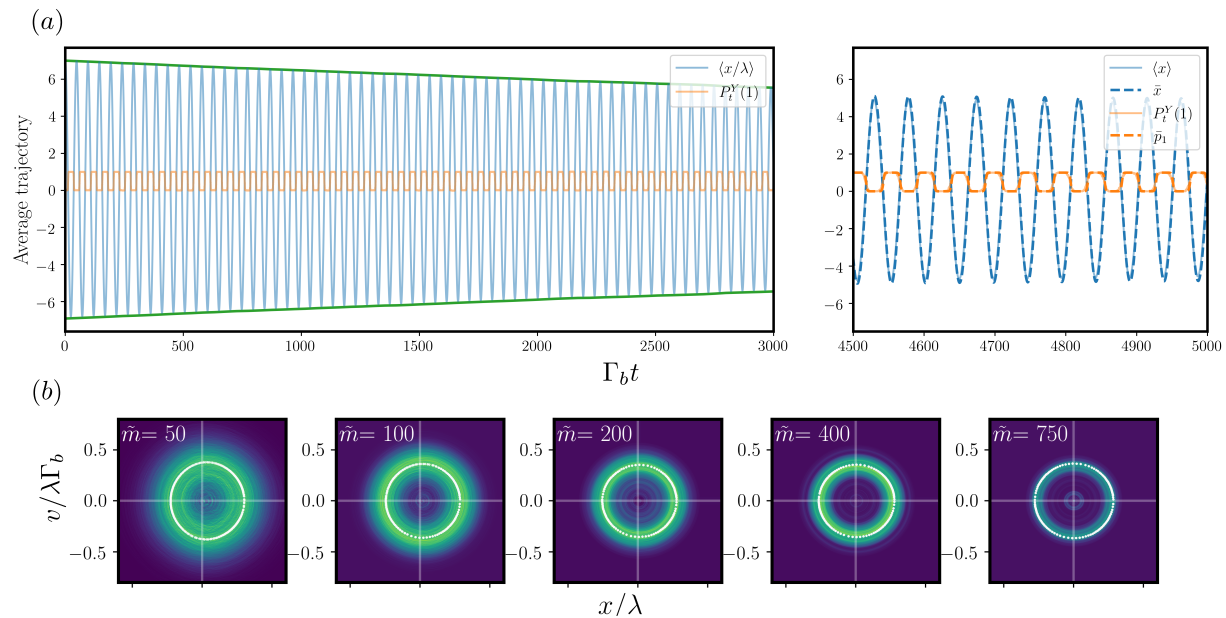}
    \caption{(a): On the left panel, we plot the ensemble averaged trajectory, calculated over $10^3$ trajectories of Eq.~\eqref{eqn: master_equation_shuttle}, of the position $\langle x/\lambda \rangle$ and the average occupancy of the dot $P^{\mbb{Y}}_t(1)$, starting from $x_0/\lambda = 7$, $v_0 =0$ and $P^{\mbb{Y}}_{t=0}(1) =1$. On the right panel, we compare the average trajectory from the stochastic dynamics (Eq.~\eqref{eqn: master_equation_shuttle}) with the meanfield dynamics of Eq.~\eqref{eqn: MF_shuttle} for large, finite mass ($\Tilde{m}\equiv m\beta\lambda^2 \Gamma_b^2 = 250$). (b): The marginal probability density $P^{\mbb{X}}(x,v)$ in the steady-state ($\Gamma_b t\to \infty$) for increasing mass $\Tilde{m}$ of the oscillator. The white curves correspond to limit cycle predicted by the MF dynamics. Parameters: $V/V_T=24$, $\omega/\Gamma_b = 0.1292, \gamma/m\omega = 0.01$ with $\lambda=1$, $\Gamma_b=5$ and $\beta=1$.}
    \label{fig: MF_Large_mass}
\end{figure}

In Fig.~\ref{fig: MF_Large_mass}(a), we plot the ensemble averaged position $\langle x \rangle (t) = \SumInt \;x\; P_t(\Gamma, y)$ and dot occupancy $\langle y \rangle (t) = \SumInt \;y\; P_t(\Gamma, y) = P^{\mbb{Y}}_t(1)$ with corresponding quantities $\overbar{x}(t)$ and $\overbar{p}_1(t)$ from the MF dynamics of Eq.~\eqref{eqn: MF_shuttle} for a large, finite mass $m\gg1$ of the oscillator. The trajectories were initialized with the oscillator at $x_0 =7$ with zero velocity $v_0=0$ and an occupied dot $y_0 =1$. Here, we show that the MF dynamics captures the relaxation of the stochastic system towards the limit-cycle attractor from its initial condition. Thus, the MF dynamics goes beyond the usual infinite mass limit, where the oscillator behaves as an Hamiltonian system with constant amplitude fixed by its initial conditions. In Fig.~\ref{fig: MF_Large_mass}(b), we plot the marginal probability density $P^{\mbb{X}}(x,v)$ in the steady-state ($t\to\infty$) along with the limit cycle predicted by the MF dynamics. As the oscillator mass $m$ increases, the probability density $P^{\mbb{X}}(x,v)$ becomes more sharply peaked around the deterministic limit cycle. The MF dynamics accurately capture the most probable behavior in the steady-state, even for finite oscillator mass. As shown in \cite{wachtler2019stochastic}, the MF dynamics also captures the average thermodynamics of the composite stochastic system in the self-oscillating regime for large voltages $V\gg V_{\rm{cr}}$, when the energy of the limit cycle oscillator $E_{\mbb{X}}=0.5(m\omega^2 x^2 + m v^2)$ is much larger than the rest of the energy (dot + interaction) $\mbb{e}_0 y - \alpha V y x$.

\subsubsection{Fast dot dynamics limit: Timescale separated dynamics}
\label{subsec: TS_shuttle}
In the limit of fast tunneling compared to the oscillation time period, i.e. when $\tau_{\mbb{Y}}=\Gamma_b^{-1}\ll \tau_{\mbb{X}}=(2\pi)/\omega$, the timescale separated (TS) Fokker-Planck equation (Eq.~\eqref{eqn: modified_FP}) for the marginal density of the oscillator in the dimensional form is given as:
\begin{eqnarray}
     d_t P^{\mbb{X}}_t = \Big[-v\partial_x  - \frac{\partial_v[f_{\mbb{X}} ^{\rm{eff}}(x) +g^{\rm{eff}}(x)  -\gamma^{\rm{eff}}(x)v]}{m} +  D^{\rm{eff}}(x)\partial_v^2) \Big]P^{\mbb{X}}_t + \mathcal{O}(\epsilon^2), \label{eqn: modified_FP_shuttle}
\end{eqnarray}
with effective conservative force $f_{\mbb{X}} ^{\rm{eff}}(x)$, non-conservative force $g^{\rm{eff}}(x)$, drift $\Tilde{\Gamma_b}(x)$ and diffusion coefficient $D^{\rm{eff}}(x)$, given as:
\begin{eqnarray}
    f_{\mbb{X}} ^{\rm{eff}}(x) &=& -m\omega^2 x - \alpha V \pi_1(x), \\
    \gamma^{\rm{eff}}(x) &=& \gamma + \frac{\alpha V}{\Gamma_b \chi(x)}\partial_x \pi_1(x), \\
    D^{\rm{eff}}(x) &=& \frac{\gamma}{m^2\beta} + \frac{\alpha^2 V^2  \pi_1(x)}{m^2 \Gamma_b\chi(x)}[1-\pi_1(x)].
\end{eqnarray}
In the above expressions, $\pi_1(x)$ is the stationary conditional probability that the dot is occupied ($y=1$) using the adiabatic approximation, i.e. $\pi_1(x) \equiv \mathbb{P}_{ss}^{(0)}(1|\Vec{\Gamma})= [e^{-x/\lambda} f_L(\mathbb{e}_0 - \alpha V x)+e^{x/\lambda} f_R(\mathbb{e}_0 - \alpha V x)]/\chi(x)$ with $\chi(x)=2\cosh{(x/\lambda)}$ being the escape rate and the normalization condition imposes $\pi_0(x) = 1 - \pi_1(x)$. The effective dynamics (Eq.~\eqref{eqn: modified_FP_shuttle}) does not satisfy the detailed balance condition $\gamma^{\rm{eff}}(x)/(D^{\rm{eff}}(x)m^2)\neq \beta$, which was not possible at the adiabatic level $\Gamma_b/\omega\to0$, and also captures self-oscillations (Fig.~\ref{appfig: prob_comparison}). However, for very large voltages $V\gg V_{\rm{cr}}$, the TS approximation fails, since there is basically one jump per half oscillation. In this regime, the oscillator acts as a shuttle, transferring one electron per period of an oscillation from the left ($L$) lead to the right ($R$) lead. \par 
\begin{figure}[h!]
    \centering
    \includegraphics[trim=0 0 0 0, clip,width=0.85\textwidth]{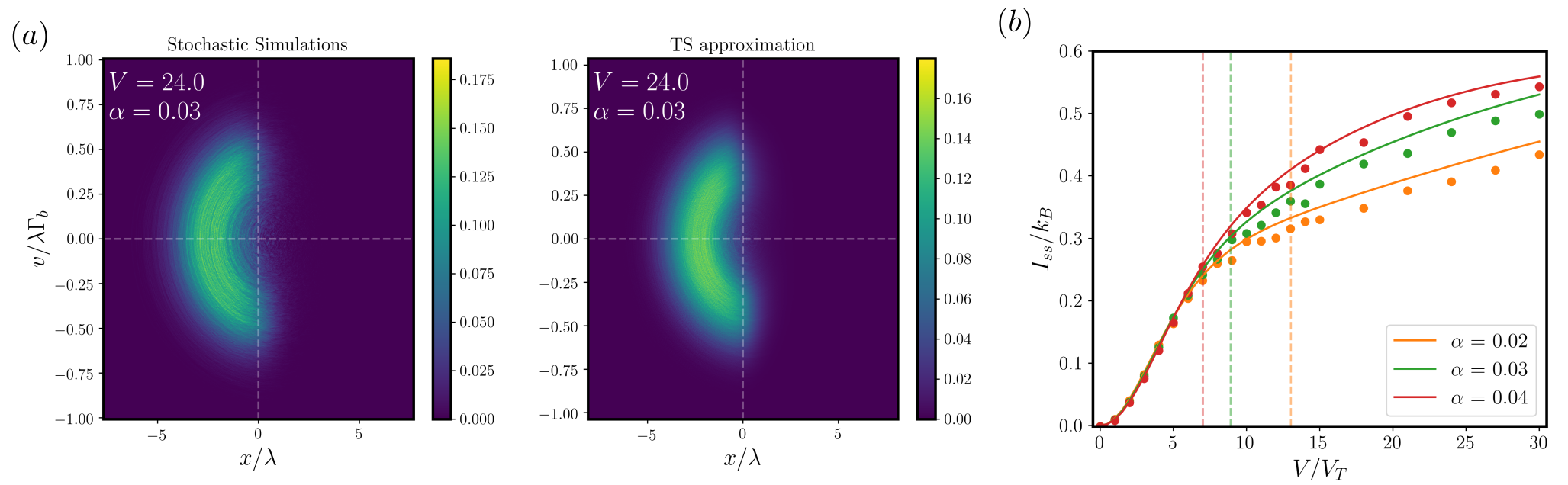}
    \caption{(a): steady-state distribution from the stochastic simulations (left) and the timescale separated dynamics using Eq.~\eqref{eqn: modified_FP_shuttle} (right)
    (b): Mutual information $I_{ss}(\Vec{\Gamma};y)$ between the dot and the oscillator in the steady-state. The markers are obtained from the stochastic simulations and the solid lines are computed using the TS solution (Eq.~\eqref{eqn: mutual_info_TS}). The vertical dashed line depict the critical voltage $V_{\rm{cr}}/V_T$, as computed from the MF dynamics (Eq.~\eqref{eqn: MF_shuttle}), for different coupling strengths $\alpha$. Parameters: $\omega/\Gamma_b = 0.1292, \gamma/m\omega = 0.01, m\lambda^2\Gamma_b^2\beta=50$ with $\Gamma_b=5$, $\lambda=1$ and $\beta=1$.}
    \label{fig: information}
\end{figure}
In the steady-state of the shuttle dynamics, satisfying $d_t P^{\mbb{X}}_{\rm{TS}} =0$ in Eq.~\eqref{eqn: modified_FP_shuttle}, we compute the marginal probability density $P_{\rm{TS}}^{\mbb{X}}(\Vec{\Gamma})$, going beyond the adiabatic approximation. This yields the following joint probability density $P_{\rm{ss}}^{\mbb{X}}(\Vec{\Gamma},y)$ for the shuttle:
\begin{eqnarray}
    P_{\rm{ss}}(\Vec{\Gamma},y) = \mathbb{P}_{\rm{TS}}^{(0+1)}(y|\Vec{\Gamma})P^{\mbb{X}}_{\rm{TS}}(\Vec{\Gamma}) +\mathcal{O}(\epsilon^2). \label{eqn: ss_shuttle_joint}
\end{eqnarray} 
Here, the conditional distribution $\mathbb{P}_{\rm{TS}}^{(0+1)}(y|\Vec{\Gamma})$ from Eq.~\eqref{eqn: ss_cond_order_1} simplifies to:
\begin{eqnarray}
    \mathbb{P}_{\rm{TS}}^{(0+1)}(y|\Vec{\Gamma}) &=& \pi_y(x) + \frac{(2y-1)}{\Gamma_b \chi(x)}\left[-v\partial_x \pi_{1}(x) - \alpha V \pi_{1}(x)\pi_0(x) \partial_v \log P_t^{\mbb{X}}(\Vec{\Gamma}) \right] + \mathcal{O}(\epsilon^{2}) \\
    &=& \pi_y(x) + v\frac{(2y-1)\nu(x)}{\Gamma_b} + \mathcal{O}(\epsilon^{2}), \label{eqn: ss_cond_shuttle_1}
\end{eqnarray}
where $\nu(x) = \left[-\partial_x \pi_{1}(x) + \alpha V \beta \pi_{1}(x)\pi_0(x)\right]/\chi(x)$ is the contribution due to the oscillator dynamics. Since the adiabatic conditional probability $\pi_1(x)$ decreases as one moves from the left lead ($x<0$) to the right lead ($x>0$), this additional contribution is positive, $\nu(x) >0 $. As anticipated from the dynamics, the velocity dependence of $\mathbb{P}_{\rm{TS}}^{(0+1)}(y|\Vec{\Gamma})$ implies that the dot is more likely to be occupied at $x=0$ when approaching from the left lead, i.e. $v>0$.   As shown in \cite{gorelik1998shuttle, wachtler2019stochastic}, the marginal probability density $P_{\rm{\rm{TS}}}^{\mbb{X}}(\Vec{\Gamma})$ for the oscillator $\mbb{X}$ can be transformed into the energy-phase ($E_{\mbb{X}},\phi_{\mbb{X}}$) space, where $E_{\mbb{X}}\equiv(1/2)m\omega^2x^2+(1/2)mv^2$ and $\phi_{\mbb{X}}=\tan^{-1}(\omega x/v)$. For low damping, phase dynamics can be assumed to be fully diffused, simplifying steady-state density to $P_{\rm{TS}}^{\mbb{X}}(E_{\mbb{X}},\phi_{\mbb{X}})\approx (1/2\pi)P_t(E_{\mbb{X}})$ and matches well the stochastic simulations (Appendix.~\ref{appsec: shuttle_energy_space}).

The mutual information $I_{ss}(\Vec{\Gamma};y)$  in the steady-state using Eq.~\eqref{eqn: ss_shuttle_joint} leads to
\begin{eqnarray}
    I_{ss}(\Vec{\Gamma};y) = k_B \int \!\! d\Vec{\Gamma}\:P_{ss}^{\mbb{X}}(\Vec{\Gamma})\left[\pi_0(x)\log\frac{\pi_0(x)}{P^{\mbb{Y}}_{ss}(0)}+\pi_1(x)\log\frac{\pi_1(x)}{P^{\mbb{Y}}_{ss}(1)} + v\frac{\nu(x)}{\Gamma_b}\log\frac{P^{\mbb{Y}}_{ss}(0)\pi_1(x)}{P^{\mbb{Y}}_{ss}(1)\pi_0(x)}\right] + \mathcal{O}(\epsilon^2), \label{eqn: mutual_info_TS}
\end{eqnarray}
where  $P^{\mbb{Y}}_{ss}(y) = \int\!\!d\Vec{\Gamma}\:P_{\rm{ss}}^{\mbb{X}}(\Vec{\Gamma},y)$  is the marginal probability for the dot to be in state $y$. In Fig.~\ref{fig: information}(b), we show that the above expression for mutual information $I_{ss}(\Vec{\Gamma};y)$ matches well with stochastic simulations, as a function of the applied voltage $V$. Since increasing the voltage $V$, the dot is more probable to occupied when closer to the left lead ($x < 0$) and vice versa, leading to higher correlations between the two subsystems. As seen in Fig.~\ref{fig: information}(b), most of the correlations are built for voltages below the critical voltage ($V_{\rm{cr}}/V_T \sim 8.9)$. As the voltage $V > V_{\rm{cr}}$ is further increased in the self-oscillating regime, mutual information continues to grow, though at a slower rate, due to the larger phase space $(x,v)$ for which the correlations are built. \par

In the next section, we will analyze the information thermodynamics of electron shuttle in the steady-state operation. Hence, we will mainly focus on the internal thermodynamic flows, i.e. energy and information flows, between the mechanical (\( \mathbb{X} \)) and electronic (\( \mathbb{Y} \)) subsystems, as the shuttle transitions to self-oscillations. We will also compare these thermodynamic flows from the stochastic simulations of Eq.~\eqref{eqn: master_equation_shuttle} with those derived from the TS dynamics (Sec.\ref{subsec: TS_shuttle}) and the MF dynamics (Sec.~\ref{subsec: MF_shuttle}).

\subsection{Steady-state information thermodynamics}
Given the master equation for the composite system (Eq.~\eqref{eqn: master_equation_shuttle}) and the internal energy $E(\Vec{\Gamma},y)$ of the composite system (Eq.~\eqref{eqn: energy_shuttle}), we can use the formalism discussed in Sec.~\ref{sec: Thermo_composite} to identify the average heat, work, and energy flow in the steady-state in the subsystem $\mbb{X}$ as:
\begin{gather}
    \Dot{Q}_{\mathbb{X}} = - \gamma \left[ \langle v^2\rangle -\frac{1}{m\beta}\right] \hspace{1cm}\text{and}\hspace{1cm} \Dot{W}_{\mathbb{X}} = 0, \label{eqn: heat_shuttle}
\end{gather}
Since there is no additional non-conservative force in the oscillator, i.e. $g(x,y) = 0$, the work contribution is zero $\Dot{W}_{\mbb{X}} = 0$ into the oscillator $\mbb{X}$. However, the oscillator will have non-zero heat flow to its environment as a result of non-zero energy flows through the electromechanical interaction $\hat{U}_{\mbb{XY}}$.\par

Similarly, the steady-state heat and work flows into the dot due to its interaction with the leads ($L/R$) are given as:
\begin{eqnarray}
    \Dot{Q}_L &=&   (-q_eV/2) \int d\Vec{\Gamma}\: J^{1,0}_L(\Vec{\Gamma})  - \alpha q_e V \int d\Vec{\Gamma}\: x\, J^{1,0}_L(\Vec{\Gamma}) ,\\
    \Dot{Q}_R &=&   (q_eV/2) \int d\Vec{\Gamma}\: J^{1,0}_R(\Vec{\Gamma})  - \alpha q_e V \int d\Vec{\Gamma}\: x J^{1,0}_R(\Vec{\Gamma}) ,\\
    \Dot{W}_{\mbb{Y}} &=&  \langle\Dot{W}_{L
}\rangle  + \langle\Dot{W}_{R}\rangle = q_e\sum_\rho\mu_\rho\int d\Vec{\Gamma}\: J^{1,0}_\rho(\Vec{\Gamma})\nonumber\\
 &=&  \frac{q_e V}{2}\int d\Vec{\Gamma}\: [J^{1,0}_L(\Vec{\Gamma}) - J^{1,0}_R(\Vec{\Gamma})].
\end{eqnarray}
Here, the averages and the probability current $J^{1,0}_{L/R}(\Vec{\Gamma}) = \lambda_{L/R}^{1,0}(x) P_{ss}(\Vec{\Gamma},0) -  \lambda_{L/R}^{0,1}(x) P_{ss}(\Vec{\Gamma},1)$ are computed using the steady-state probability density $P_{ss}(\Vec{\Gamma},y)$ of the composite system. Note that the contribution due to the constant energy $\mbb{e}_0$ in the chemical potential $\mu_{L/R}$ vanishes as $\int d\Vec{\Gamma}\: [J^{1,0}_L(\Vec{\Gamma}) + J^{1,0}_R(\Vec{\Gamma})]=0$ in the steady-state. 

Now moving onto the entropic changes at global level, the entropy production $T\Dot{\sigma}_{\rm{tot}}$ in the steady-state is equal to the negative of sum of entropy flows $-\dot{Q}_{\rho}/T$ to the reservoirs (Eq.~\eqref{eqn: sigma_tot_ss_1}), i.e.
\begin{eqnarray}
    T\Dot{\sigma}_{\rm{tot}} &=& -\Dot{Q}_{\mathbb{X}} - \Dot{Q}_{L} -\Dot{Q}_{R} \nonumber\\
    &=& \Dot{W}_{\mbb{Y}} = q_e \langle J^{1,0}_{L} \rangle_{ss} V \ge 0 ,\label{eqn: sigma_tot_shuttle}
\end{eqnarray}
where $q_e\langle J^{1,0}_{L} \rangle_{ss}= q_e\int d\Vec{\Gamma}\: J^{1,0}_L(\Vec{\Gamma})$ (Eq.~\ref{eqn: prob_current_dot}) is the average electric current in the steady-state through the dot. Therefore, the current must flow in the direction of the applied voltage difference $V$ for the isothermal scenario, and work $\dot{W}_{\mbb{Y}} \ge 0$ is inputted to shuttle by the leads. This non-conservative work $\dot{W}_{\mbb{Y}}$ is also the source of nonequilibrium dynamics, leading to self-oscillations. The entropy production can be divided into two partial contributions from the mechanical motion $\Dot{\sigma}_{\mbb{X}}$ and the tunneling in the dot $\Dot{\sigma}_{\mbb{Y}}$ separately (Eq.~\eqref{eqn: sigma_tot_ss_3}). As shown in Sec.~\ref{sec: Thermo_composite}, these partial-entropy production rates are individually positive:
\begin{eqnarray}
    \Dot{\sigma}_{\mbb{Y}} &=& \int d\Vec{\Gamma}  \left[J^{1,0}_L(\Vec{\Gamma},t) \log\frac{ \lambda_L^{1,0}(x)P_t(\Vec{\Gamma}, 0)}{\lambda_L^{0,1}(x) P_t(\Vec{\Gamma}, 1)}+ J^{1,0}_R(\Vec{\Gamma},t) \log\frac{ \lambda_R^{1,0'}(x)P_t(\Vec{\Gamma},0)}{\lambda_R^{0,1}(x) P_t(\Vec{\Gamma}, 1 )}\right]\ge 0 \label{eqn: shuttle_parital_sigma_e},\\
    \Dot{\sigma}_{\mbb{X}} &=&\frac{\gamma}{m T}  \SumInt P_t(\Vec{\Gamma}, y)\left[v+(m\beta)^{-1}\partial_v \log P_t(\Vec{\Gamma}, y)\right]^2 \ge 0.
\end{eqnarray}

In addition to the external thermodynamic flows, the steady-state operation is also mediated by internal energy and information flows between the oscillator $\mbb{X}$ and the dot $\mbb{Y}$. Since the average energies remain constant in the steady-state, i.e., $ d_t E = d_t E_{\mathbb{X}} = d_t E_{\mathbb{Y}} = 0$, this implies that the energy flows are balanced between the subsystems, leading to the relationship $\dot{\mathcal{E}}^{\mathbb{Y}} = -\dot{\mathcal{E}}^{\mathbb{X}}$. To characterize the steady-state energy flow $\dot{\mathcal{E}}$ in the shuttle, we identify the positive direction of the energy flow as $\dot{\mathcal{E}}^{\mathbb{Y}}$ - the flow of energy from the dot $\mathbb{Y}$ to the oscillator $\mathbb{X}$ - such that $\dot{\mathcal{E}} \equiv \dot{\mathcal{E}}^{\mathbb{Y}}$. Using this convention, the steady-state energy flows between the dot and the oscillator are given as:
\begin{eqnarray}  
    \Dot{\mathcal{E}}^{\mathbb{X}} &=&  \Dot{Q}_{\mathbb{X}} + \Dot{W}_{\mathbb{X}} = -\alpha V\SumInt P_t(\Vec{\Gamma},y) \,v\,y , \label{eqn: Elfow_shuttle_X}\\
    \dot{\mathcal{E}} &\equiv& \Dot{\mathcal{E}}^{\mbb{Y}} = \Dot{Q}_L +  \Dot{Q}_R + \Dot{W}_{\mbb{Y}} =- \alpha V \int d\Vec{\Gamma}\: x\, \left[ J^{1,0}_L(\Vec{\Gamma}) + J^{1,0}_R(\Vec{\Gamma}) \right] \nonumber \\
    &=& -\Dot{\mathcal{E}}^{\mathbb{X}} = \gamma \left[ \langle v^2\rangle -\frac{1}{m\beta}\right].
    \label{eqn: E_y_shuttle}
\end{eqnarray}
Hence, the steady-state energy flow $\dot{\mathcal{E}}$ from the dot to the oscillator is fully dissipated as heat at its reservoir $-\Dot{Q}_{\mathbb{X}}$, as there is no work source at the oscillator. \par

The information flow is a dynamical quantity that measures changes in the correlations between the state of the dot and the oscillator (Sec.~\ref{subsec: infoflows}). The steady-state information flows between the oscillator and the dot simplifies to:
\begin{flalign}
      \Dot{\mathcal{I}}^{\mbb{Y}} &= \int d\Vec{\Gamma} \;\left[J_L^{1,0}(\Vec{\Gamma}) + J_R^{1,0}(\Vec{\Gamma})\right]\log\frac{\mathbb{P}_{ss}(\Vec{\Gamma}|1)}{\mathbb{P}_{ss}(\Vec{\Gamma}|0)},\\
      \Dot{\mathcal{I}} &\equiv \Dot{\mathcal{I}}^{\mbb{X}} = -\Dot{\mathcal{I}}^{\mbb{Y}} \nonumber\\
     &= -\frac{\alpha V}{m}\int d\Vec{\Gamma} \; P_{ss}^{\mbb{X}}(\Vec{\Gamma}) \partial_v \mathbb{P}_{ss}(1|\Vec{\Gamma}) -\left(\frac{\gamma}{\beta m^2}\right)\int d\Vec{\Gamma} \; P_{ss}^{\mbb{X}}(\Vec{\Gamma}) \frac{\left[\partial_v \mathbb{P}_{ss}(1|\Vec{\Gamma})\right]^2}{\mathbb{P}_{ss}(1|\Vec{\Gamma})[1-\mathbb{P}_{ss}(1|\Vec{\Gamma})]}. \label{eqn: I_X_shuttle}
\end{flalign}
Here, we assume the positive direction of the steady-state information flow as the one due to tunneling dynamics in the dot, $\dot{\mathcal{I}}^{\mathbb{Y}}$ — the flow of information from the dot $\mathbb{Y}$ to the oscillator $\mathbb{X}$ — such that $\dot{\mathcal{I}} \equiv \dot{\mathcal{I}}^{\mathbb{Y}}$. \par

In the next section, we analyze the internal energy $\Dot{\mathcal{E}}$ and information $\Dot{\mathcal{I}}$ flows to determine its true positive direction and the stronger bounds to these flows due to the local second laws. We will also assess the contributions of $\Dot{\mathcal{E}}$ and $\Dot{\mathcal{I}}$ in maintaining the mechanical oscillations of the shuttle. In doing so, we compare these quantities from the stochastic simulations of the initial master equation (Eq.~\eqref{eqn: master_equation_shuttle}) with the predictions from TS (Sec.~\ref{subsec: TS_shuttle}) and MF (Sec.~\ref{subsec: MF_shuttle}) dynamics. \par

\subsubsection{Energy and information flows : Comparison with TS and MF dynamics}


As shown in Sec.~\ref{sec: TS_beyond}, the dynamics and thermodynamics can be computed using the TS approximation. The steady-state distribution $P_{ss}(\Vec{\Gamma},y)$ is accurate up to $\mathcal{O}(\epsilon)$ and the thermodynamic quantities due to the slow oscillator $\mbb{X}$ dynamics are valid up to the order $\mathcal{O}(\epsilon^2)$. The energy flows $\Dot{\mathcal{E}}$ (Eq.~\eqref{eqn: E_y_shuttle}) and information $\Dot{\mathcal{I}}$ flows (Eq.~\eqref{eqn: I_X_shuttle}) can be computed directly by substituting the steady-state distribution $P_{ss}(\Vec{\Gamma},y) = \mathbb{P}_{\rm{TS}}^{(0+1)}(y|\Vec{\Gamma})P^{\mbb{X}}_{\rm{TS}}(\Vec{\Gamma}) +\mathcal{O}(\epsilon^2)$ (Eq.~\eqref{eqn: ss_shuttle_joint}). Using this TS distribution, the energy flow is given by:
\begin{eqnarray}
    \dot{\mathcal{E}}_{\rm{TS}}  = \int d\Vec{\Gamma} \;  \gamma\left[ v^2 -\frac{1}{m\beta}\right] P_{\rm{TS}}^{\mbb{X}}(\Vec{\Gamma}) + \mathcal{O}(\epsilon^3). \label{eqn: E_dot_TS}
\end{eqnarray}
In the energy-phase $(E_{\mbb{X}}, \phi_{\mbb{X}})$ space (Appendix.~\ref{appsec: shuttle_energy_space}), it can be further simplified to $ \dot{\mathcal{E}} = (\gamma/\beta m) [\beta\langle E_{\mbb{X}}\rangle-1]\ge0$, which is zero at equilibrium ($V=0$) where $P_{\rm{TS}}^{\mbb{X}}(\Vec{\Gamma}) \propto \exp\left[-\beta E_{\mbb{X}} \right]$. As the voltage $V$ increases, the average energy grows, satisfying $\beta\langle E_{\mbb{X}}\rangle \ge 1$ (Fig.~\ref{fig: energy_stats}), and resulting in an increasing energy flow $\dot{\mathcal{E}} \ge 0$. As shown in Sec.~\ref{sec: TS_beyond}, under adiabatic approximation, the oscillator relaxes to an equilibrium state for any positive voltage $V$, which leads to $[\beta\langle E_{\mbb{X}}\rangle-1]=0$ with zero energy flows. Hence, by going beyond the adiabatic approximation, we capture the non-zero contributions to these flows.
\par



Now moving on to the information flow, the non-zero contributions computed using the TS probability density is given as:
\begin{eqnarray}
    \Dot{\mathcal{I}}_{\rm{TS}} &\equiv& \frac{\alpha V k_B}{m \Gamma_b}\int d\Vec{\Gamma} \; P_{\rm{TS}}^{\mbb{X}}(\Vec{\Gamma})\nu(x) -\left(\frac{\gamma k_B}{\beta m^2 \Gamma_b^2}\right)\int d\Vec{\Gamma} \; P_{\rm{TS}}^{\mbb{X}}(\Vec{\Gamma}) \frac{\nu^2(x)}{\pi_0(x)\pi_1(x)} + \mathcal{O}(\epsilon^3) \ge 0. \label{eqn: I_dot_TS}
\end{eqnarray}
where $\pi_{(0)1}(x)$ is the zeroth-order $\mathcal{O}(\epsilon^0)$ (adiabatic) stationary conditional probability for the electronic state $y=(0)1$ and $\nu(x)>0$ is a term arising from the first-order correction $\mathcal{O}(\epsilon)$ to the stationary conditional probability, derived in Eq.~\eqref{eqn: ss_cond_shuttle_1}.\par


\begin{figure}[h!]
    \centering
    \includegraphics[trim=0 0 0 0, clip,width=0.95\textwidth]{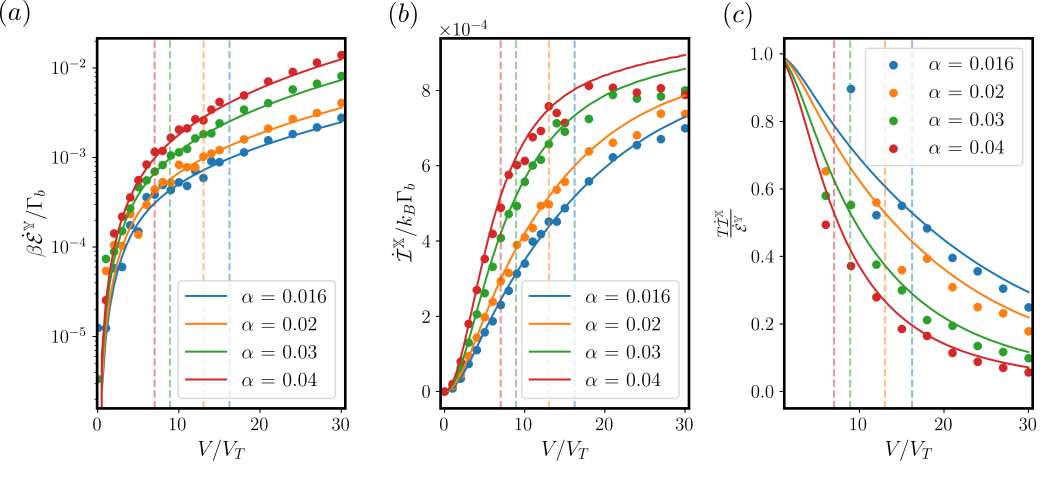}
    \caption{(a): Energetic flow due to the dot dynamics $\Dot{\mathcal{E}}^{\mbb{Y}}$ for increasing interaction strengths $\alpha$. (b): Information flow due to the oscillator dynamics $\Dot{\mathcal{I}}^{\mbb{X}}$ for increasing interaction strengths $\alpha$. (c): Ratio of information flow to the energy flow $T\dot{\mathcal{I}}/\dot{\mathcal{E}}$ between the dot and the oscillator in the steady-state . The markers are obtained from the stochastic simulations and the solid lines are computed using the TS solution (Eq.~\eqref{eqn: I_dot_TS}) and Eq.~\eqref{eqn: E_dot_TS}). The vertical dashed line depict the critical voltage $V_{\rm{cr}}/V_T$, as computed from the MF dynamics (Eq.~\eqref{eqn: MF_shuttle}), for different coupling strengths $\alpha$. Parameters: $\omega/\Gamma_b = 0.1292, \Gamma_b/m\omega = 0.01, m\lambda^2\Gamma_b^2\beta=50$ with $\Gamma_b=5$, $\lambda=1$ and $\beta=1$}
    \label{fig: information_flow}
\end{figure}

When the dot is much faster than the oscillator, i.e. $\epsilon\gg1$, the force part of the information flow is the dominant contribution with $\Dot{I}_F^{\mbb{X}} = \frac{\alpha V k_B}{m \Gamma_b}\int d\Vec{\Gamma} \; P_{\rm{TS}}^{\mbb{X}}(\Vec{\Gamma})\nu(x) \;+\; \mathcal{O}(\epsilon^{-3})$, while the entropic part contributes only to a lower order, i.e. $\Dot{I}_S^{\mbb{X}} = \left(\frac{\gamma k_B}{\beta m^2 \Gamma_b^2}\right)\int d\Vec{\Gamma} \; P_{\rm{TS}}^{\mbb{X}}(\Vec{\Gamma}) \frac{\nu^2(x)}{\pi_0(x)\pi_1(x)}  \;+\; \mathcal{O}(\epsilon^{-4})$. This dominant force contribution $\Dot{I}_F^{\mbb{X}} \ge 0$ is positive, as $\nu(x)>0$ captures the velocity dependence of the conditional distribution $\mbb{P}(y|x,v)$ (see Sect.~\ref{subsec: TS_shuttle}). This results in a net positive information flow due to the oscillator dynamics with $\Dot{\mathcal{I}}_{\rm{TS}}\ge0$. Therefore, in the TS regime, the oscillator behaves as a sensor that gathers information about the state of the dot $y$. The oscillator, unlike the original Maxwell demon, isn't solely exchanging information with the system. Instead, it draws the energy needed for the measurement process from the system itself, highlighting a combination of energetic and informational exchanges. This also implies that the heat dissipated $\dot{\mathcal{E}}$ by the oscillator (sensor) must be larger to account for its information processing $T\Dot{\mathcal{I}}$, which can be clearly seen from the local second laws:
\begin{equation}
    \begin{aligned}
        T\Dot{\sigma}_{\mbb{X}} &= \dot{\mathcal{E}} - T\Dot{\mathcal{I}}\ge 0,   \\
        T\Dot{\sigma}_{\mbb{Y}} &= \Dot{W}_{\mbb{Y}} - \dot{\mathcal{E}} + T\Dot{\mathcal{I}}  \ge 0\;.
    \end{aligned}
    \label{eqn: local_2nd_law_shuttle}
\end{equation}
When we combine both local second laws, we arrive at the global second law: $\Dot{W}_{\mbb{Y}} = q_e  J^{1,0}_{L}(\Vec{\Gamma}) V \ge 0$. Therefore, the electric current $q_e J^{1,0}_{L}(\Vec{\Gamma})$ must flow in the direction of the applied voltage difference $V$ for the isothermal scenario. However, in a model for a suspended nanotube resonator, as demonstrated in \cite{parrondo2023information}, the information flow $\Dot{\mathcal{I}}$ can effectively reverse the current flow against the applied voltage, under the condition of a temperature difference with $T_{\mbb{Y}}\gg T_{\mbb{X}}$.\par

As shown in Fig.~\ref{fig: information_flow}(b), the information flow $\Dot{\mathcal{I}} \equiv \Dot{\mathcal{I}}^{\mbb{X}}\ge 0$ due to the oscillatory dynamics is positive, showing that the oscillator indeed acts like a sensor. Furthermore, the information flows $\Dot{\mathcal{I}}$ from the simulations also saturate to a constant in the self-oscillating regime for large voltages $V\gg V_{\rm{cr}}$, independent of the electromechanical coupling $\alpha$. This can be qualitatively understood, as for large $V\gg V_{\rm{cr}}$, dynamically there is a change in the state of the dot $y$ only for every zero crossing of the position of the oscillator (Fig.~\ref{fig: trajectory_shuttle}(b)) so that the information flow depends only on the frequency of the oscillation. For the simulation parameters with $\Gamma_b/\omega =5$, we find that this saturation of the information flow is not captured by the TS solution (Eq.~\eqref{eqn: I_dot_TS}) in the self-oscillating regime as it goes beyond its regime of applicability. This is mainly due to the limited knowledge of the conditional distribution up to $\mathcal{O}(\epsilon)$. Note that even though the Fokker-Planck equation of Eq.~\eqref{eqn: TS_master_eqn} is exact up to $\mathcal{O}(\epsilon)$, we obtain a solution that includes higher order terms. As demonstrated in Appendix~\ref{appsec: shuttle_energy_space}, retaining these higher order terms yields better agreement with the simulations than the exact solution up to $\mathcal{O}(\epsilon)$, despite the uncontrolled nature of the higher order terms. In Fig.~\ref{fig: information_flow}(c), we plot the ratio of information to energy flows $T\dot{\mathcal{I}}/\dot{\mathcal{E}}$ as the shuttle transitions to self-oscillations for different electromechanical couplings $\alpha$. We find that both the simulations and the TS solution suggest that the self-oscillating regime is maintained mainly by energetic exchanges, as the ratio $T\dot{\mathcal{I}}/\dot{\mathcal{E}}\to 0$ with increasing voltages $V$. \par

\begin{figure}[h!]
    \centering
    \includegraphics[trim=0 0 0 0, clip,width=0.95\textwidth]{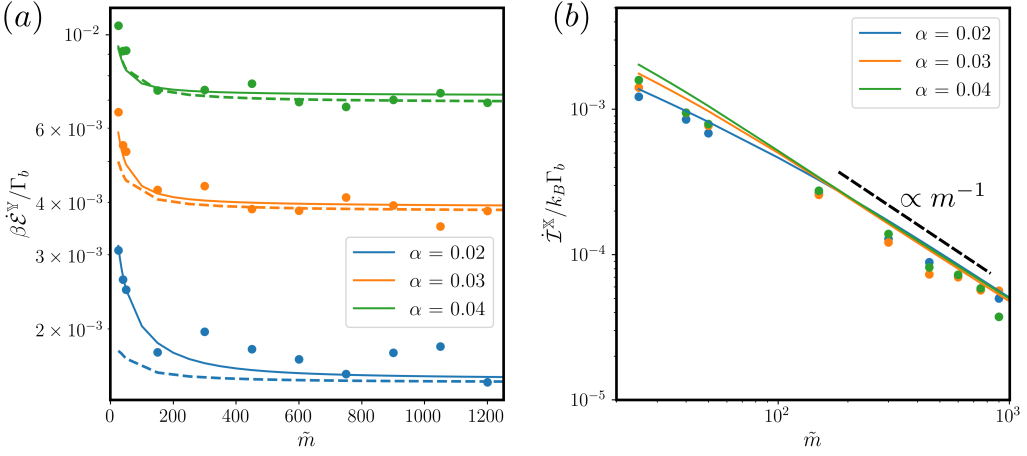}
    \caption{We plot the internal flows as a function of the oscillator mass $m$ in the self-oscillating regime with $V = 24.0$, for different electromechanical coupling $\alpha$. Here, $\tilde{m} = m \lambda^2 \Gamma_b^2\beta$ is a dimensional mass parameter for the shuttle. (a) Energetic flow due to the dot dynamics, $\Dot{\mathcal{E}}^{\mathbb{Y}}_{ss}$, and (b) information flow due to the oscillator dynamics, $\Dot{\mathcal{I}}^{\mathbb{X}}_{ss}$, are shown. Solid lines represent TS solutions (Eqs.~\eqref{eqn: E_dot_TS} and \eqref{eqn: I_dot_TS}), dashed lines depict MF approximation (Eq.~\eqref{eqn: MF_shuttle}), and markers are from the stochastic simulations of initial master equation (Eq.~\eqref{eqn: master_equation_shuttle}). Parameters: $V=24.0, \omega/\Gamma_b = 0.1292, \Gamma_b/m\omega = 0.01$ with $\Gamma_b=5$, $\lambda=1$ and $\beta=1$}
    \label{fig: flows_mass}
\end{figure}

Now keeping all the other parameters fixed, we compare how these internal flows behave as we increase the mass $m$ of the oscillator. In the large mass $\Tilde{m} = m \lambda^2\Gamma_b^2\beta \gg 1$ regime, the stochastic dynamics of the composite system is equivalent to the meanfield dynamics by keeping terms in the master equation up to $\mathcal{O}(m^{-1})$ (Sec.~\ref{sec: MF_approx}). In the electron shuttle, this meanfield dynamics relaxes to a limit cycle in the long time limit (Sec.~\ref{subsec: MF_shuttle}), and hence does not predict the time-independent stationary state of the composite system. However, as shown in Fig.~\ref{fig: MF_Large_mass}(b), the true stationary state also becomes sharply peaked around the limit cycle attractor as mass is increased. This limit cycle is formed from the balance of energy flow through the interaction force $\alpha V {\overbar{p}_1}$ and the heat dissipated due to the damping force $-\gamma \overbar{v}$. Hence, the energy flow $\dot{\mathcal{E}}_{\rm{MF}}$ into the system can be computed from the damping force in the limit cycle attractor by time-averaging over one oscillation, as follows:
\begin{eqnarray}
    \dot{\mathcal{E}}_{\rm{MF}} = \gamma\langle \overbar{v}_t^2 \rangle = \frac{\gamma}{T_p}\int_0^{T_p}d\tau\; \overbar{v}_t^2,
    \label{eqn: E_shuttle_MF}
\end{eqnarray}
where $T_p$ is time-period of the limit cycle oscillation. Note that we can relate the steady-state energy flow to the heat dissipation at the oscillator as the change in state functions, like energy, over an oscillation is zero. However, the mean-field dynamics fails to capture the diffusive aspects of the system, which contribute to the spreading around the limit cycle and to additional heat dissipation due to thermal fluctuations. Although the diffusive term appears on order $\mathcal{O}(m^{-2})$ in the master equation, its cumulative effect on observables such as the heat flow is of order $\mathcal{O}(m^{-1})$ (Eq.~\eqref{eqn: heat_shuttle}), capturing deviations of the velocity distribution from the Maxwell-Boltzmann distribution. From the MF deterministic dynamics, we are only capturing the lowest order contribution to the energy flow due to the damping.\par

Due to the deterministic nature of the meanfield dynamics, the joint distribution $P_{t}(\Vec{\Gamma},1)$ factorizes into the product of the marginal distributions in leading order at all times, i.e. $P_{t}(\Vec{\Gamma},1)= P^{\mbb{X}}_t(\Vec{\Gamma})P^{\mbb{Y}}_t(1)  + \mathcal{O}(m^{-2}) = \delta(x- x_t)\delta(v - v_t)P^{\mbb{Y}}_t(1) + \mathcal{O}(m^{-2})$ and $\mathbb{P}_{t}(\Vec{\Gamma}|1)=P^{\mbb{X}}_t(\Vec{\Gamma})=\delta(x- x_t)\delta(v - v_t) + \mathcal{O}(m^{-2})$. Consequently, the conditional distribution $\mathbb{P}_{t}(\Vec{\Gamma}|1)=P^{\mbb{X}}_t(\Vec{\Gamma})$ is independent of a particular state of the dot $y$ in the deterministic trajectory of the oscillator, which if fully determined by the initial conditions. As a result, the MF dynamics predicts zero information flows $\dot{\mathcal{I}}_{\rm{MF}}=0$, even in the long-time limit, where it reaches the limit cycle attractor. Note that the steady-state distribution becomes more sharply peaked around the limit cycle as mass is increased. This observation aligns with the steady-state information flow computed from the TS regime, with $\epsilon \gg 1$. Specifically, the dominant force contribution to the information flow asymptotically scales with mass as $\mathcal{O}(m^{-1})$ (Eq.~\eqref{eqn: I_dot_TS}). 
Hence, one need to go to higher-order diffusive term in the UD dynamics to capture the non-zero contributions of the information flow.
\par

In Fig.~\ref{fig: flows_mass}(a) and (b), we plot the energy and information flows, respectively, in the self-oscillating regime with $V =24.0 > V_{\rm{cr}}$ as a function of increasing mass of the oscillator. Since we use the same parameters as in Fig.~\ref{fig: information_flow} with $\epsilon = 0.1292$, the internal flows from the TS dynamics (Eq.~\eqref{eqn: E_dot_TS}) captures the stochastic simulations across different masses, with small variations due to broken timescale separation at large voltage $V$. As the mass $m$ increases, the energy flows saturate to a constant value as predicted by the MF dynamics, due to the purely damping contribution along the limit cycle (Eq.~\eqref{eqn: E_shuttle_MF}). For small masses, the energy flow deviates from MF predictions due to the dominant fluctuations in the oscillator. However, the MF dynamics become increasingly accurate with increasing mass, as the stationary state will be exactly peaked at the deterministic limit cycle attractor. In contrast to the energy flows, the information flows decreases as one increases the mass. As shown in Fig.~\ref{fig: flows_mass}(b), the information flow from the TS and the stochastic simulation asymptotically decay with the mass as $m^{-1}$, as in Eq.~\eqref{eqn: I_dot_TS}. \par 

\begin{figure}[h!]
    \centering
    \includegraphics[trim=0 0 0 0, clip,width=0.45\textwidth]{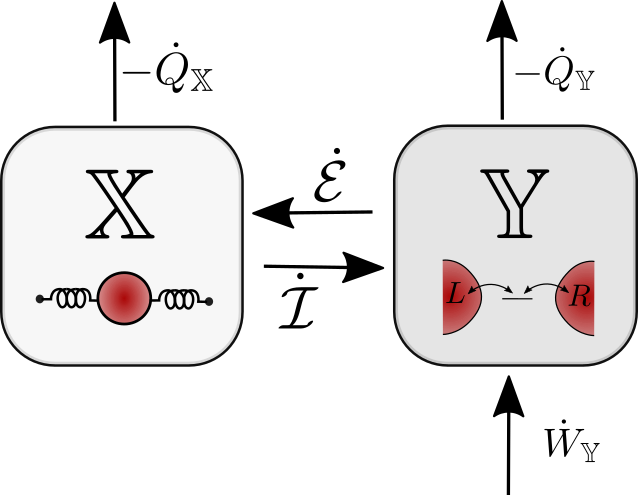}
    \caption{Steady-state thermodynamic flows in the electron shuttle. Arrows indicate the positive flows.}
    \label{fig: Shuttle_SS_engine}
\end{figure}

We now summarize the thermodynamic flows in the steady-state operation of the electron shuttle (Fig.~\ref{fig: Shuttle_SS_engine}). For a positive voltage $V$ across the leads, the input work $\dot{W}_{\mathbb{Y}} \ge 0$ due to a steady current flow through the dot is partially transferred to the oscillator through a positive energy flow $\dot{\mathcal{E}}\equiv \Dot{\mathcal{E}}^{\mbb{Y}} \ge 0$, while the remainder is dissipated at the leads (\( \dot{Q}_{L/R} \leq 0 \)). Since there is no non-conservative force at the oscillator, this energy flow is also fully dissipated in its environment with ($\dot{Q}_{\mathbb{X}} \leq 0 $). In addition to the energy flows, the steady-state of the shuttle also involves information flows between the dot and the oscillator. The oscillator acts as a sensor, measuring the state of the dot by increasing mutual information through its dynamics $\dot{\mathcal{I}}\equiv \Dot{\mathcal{I}}^{\mbb{X}} \ge 0$. As seen from the second law in Eq.~\eqref{eqn: local_2nd_law_shuttle}, this continuous information acquisition about the state of the dot is enabled by the energy flow from the dot itself. In such cases, the information flow cannot reverse the direction of the current flow at the dot, as $\dot{\sigma}_{\rm{tot}} = q_e J_L^{1,0} V \ge 0 $. Moreover, as shown in Fig.~\ref{fig: information_flow}(c), the information flow becomes negligible compared to the energy flows in the self-oscillating regimes for large voltages $V \gg V_{\rm{cr}}$, implying that the partial entropy produced $T\Dot{\sigma}_{\mbb{X}}$ in the oscillator is dominated by the energy flow, which becomes more relevant for increasing the mass of the oscillator. \par

In the next section, we analyze the total entropy production of the shuttle, as it transitions to self-oscillations, and the thermodynamic efficiency of conversion of electric power input to the mechanical oscillations.

\subsubsection{Thermal Engine: From electric to mechanical}

As shown in Eq.~\eqref{eqn: sigma_tot_shuttle}, the inputted power at the dot is equal to the entropy production rate multiplied by the temperature $T\Dot{\sigma}_{\rm{tot}}$, i.e. $\Dot{W}_{\mbb{Y}} = q_e V \langle J^{1,0}_{L}(\Vec{\Gamma})\rangle_{ss}= T\Dot{\sigma}_{\rm{tot}}$. 
From the TS dynamics, we can also compute the total entropy production rate (or input power) in the steady-state as:
\begin{eqnarray}
    T\Dot{\sigma}_{\rm{tot}} &=&  q_e V \langle J^{1,0}_{L}(\Vec{\Gamma})\rangle_{\rm{TS}}  + \mathcal{O}(\epsilon^2) \nonumber\\
    &=& q_e V \int d\Vec{\Gamma}\; P_{\rm{TS}}^{\mbb{X}}(\Vec{\Gamma}) \left[\lambda_{L}^{1,0}(x) \pi_0(x) -  \lambda_{L}^{0,1}(x) \pi_1(x)\right] + \;\mathcal{O}(\epsilon^2) \nonumber \\
    &=& \frac{q_e V}{2}\tanh\left(\frac{V}{4V_T}\right) \int d\Vec{\Gamma}\; P_{\rm{TS}}^{\mbb{X}}(\Vec{\Gamma}) e^{-\frac{x}{\lambda}}\left[1+\tanh\left(\frac{x}{\lambda}\right)\right] + \;\mathcal{O}(\epsilon^2),
    \label{eqn: sigma_TS_shuttle}
\end{eqnarray}
Here, $\mathbb{J}_{L}^{1,0}(x) = \tanh\left(\frac{V}{4V_T}\right) e^{-\frac{x}{\lambda}}\left[1+\tanh\left(\frac{x}{\lambda}\right)\right] + \;\mathcal{O}(\epsilon)$ is the dominant contribution to the conditional steady-state current at any given $(x,v)$. This quantity is maximized when the oscillator is at $x=0$, when the tunneling amplitudes of the both leads are equal. Note that the expression of Eq.~\eqref{eqn: sigma_TS_shuttle} is equivalent to the previous analysis of \cite{wachtler2019stochastic}, as the velocity-dependent contribution of $\mathbb{P}_{\rm{TS}}^{(0+1)}(y|\Vec{\Gamma})$ to entropy production vanishes. This occurs because the marginal density $P_{\rm{TS}}^{\mbb{X},0} \propto e^{-\beta mv^2/2}$ at the adiabatic level is symmetric around $v=0$, whereas the $\mathbb{P}_{\rm{TS}}^{(0+1)}(y|\Vec{\Gamma}) - \pi_y(x) \propto v $. \par

\begin{figure}[h!]
    \centering
    \includegraphics[trim=0 0 0 0, clip,width=0.95\textwidth]{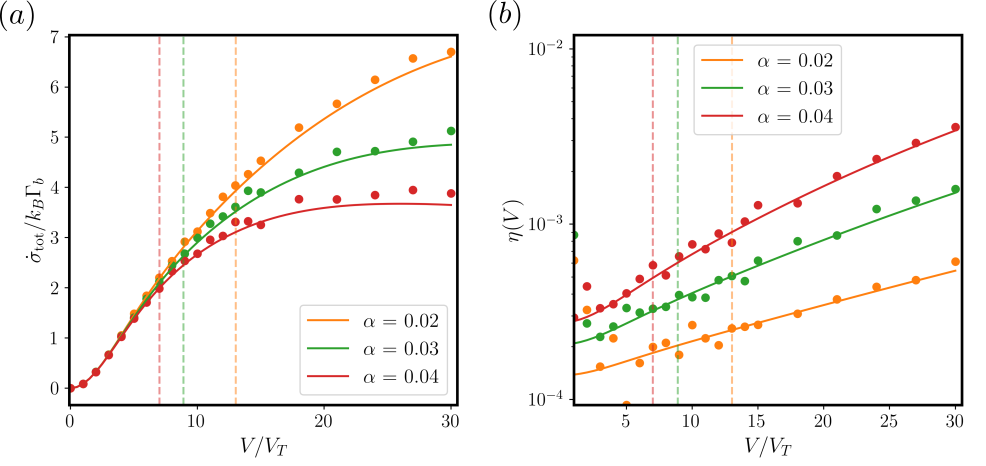}
    \caption{(a): The total entropy production $\Dot{\sigma}_{\rm{tot}}$ as function of applied voltage $V$ for different electromechanical coupling $\alpha$. (b): Transduction efficiency $\eta = \dot{\mathcal{E}}/(T\Dot{\sigma}_{\rm{tot}})$ in the conversion of the applied electric power to mechanical oscillations.  The vertical dashed line depict the critical voltage $V_{\rm{cr}}/V_T$, as computed from the MF dynamics (Eq.~\eqref{eqn: MF_shuttle}), for different coupling strengths $\alpha$. Parameters: $\omega/\Gamma_b = 0.1292, \Gamma_b/m\omega = 0.01, m\lambda^2\Gamma_b^2\beta=50$ with $\Gamma_b=5$ and $\beta=1$}
    \label{fig: efficiency}
\end{figure}

In the limit of $\alpha=0$, there is no interaction force between the dot and the oscillator, implying that the oscillator relaxes to equilibrium with the Gibbs distribution $P^{\mbb{X}}_{ss}(\Vec{\Gamma}) \propto \exp(-\beta m \omega^2 x^2)$, regardless of the applied voltage $V$ across the dot. Averaging over the equilibrium distribution, the entropy production satisfies $ T\Dot{\sigma}_{\rm{tot}} \propto V\tanh\left(V/4V_T\right)$, which is equivalent to a stationary dot at $x=0$. However, for any finite $\alpha$, the shuttle transitions to self-oscillations above a critical voltage $V > V_{\rm{cr}}$, where the amplitude of oscillations increases with voltage $V$. The increase in amplitude implies that the oscillator spends less time around $x=0$, leading to a decrease in current flow $\langle J^{1,0}_{L}(\Vec{\Gamma})\rangle_{\rm{TS}}$. In the extreme case of very large voltages $V\gg V_{\rm{cr}}$, the oscillating dot essentially shuttles a single electron between the leads per oscillation, leading to the electric current $\langle J_{L}^{01}\rangle_{\rm{TS}}$ equal to the oscillation frequency $\omega/(2\pi)$, independent of the jump rates $\Gamma_b$. This decrease in the average current between the leads in the self-oscillating regime is reflected in the entropy production rate with increasing voltages $V$. Furthermore, the amplitude of self-oscillations is also proportional to the electromechanical coupling $\alpha$, due to the larger energy flow $\dot{\mathcal{E)}}$ for a given voltage $V$. This leads to an inverse relationship of entropy production $T\Dot{\sigma}_{\rm{tot}}$ with $\alpha$. Fig.~\ref{fig: efficiency}(a) illustrates the above-mentioned relationship of the total entropy production rate $\Dot{\sigma}_{\rm{tot}}$ with voltage $V$ and electro-mechanical coupling $\alpha$. Similar to the internal flows, the TS estimate of Eq.~\eqref{eqn: sigma_TS_shuttle} agrees well with stochastic simulations, with deviations emerging at large voltages due to the breakdown of the TS approximation.\par

Finally, we analyze the shuttle as a nanomachine, transforming the inputted electrical work into mechanical oscillations. A part of the total power input to the shuttle $\Dot{W}_{\mbb{Y}} = -\Dot{Q}_{\mathbb{X}} - \Dot{Q}_{L} -\Dot{Q}_{R} = T\Dot{\sigma}_{\rm{tot}}$ is consumed by the oscillator through the energy flow $\dot{\mathcal{E}}=-\Dot{Q}_{\mathbb{X}}$, which drives the self-oscillations. We characterize the transduction efficiency $\eta$ as follows:
\begin{eqnarray}
    \eta \equiv \frac{-\Dot{Q}_{\mathbb{X}}}{\Dot{W}_{\mbb{Y}}} = \frac{\dot{\mathcal{E}}}{T\Dot{\sigma}_{\rm{tot}}}.
\end{eqnarray}
The local second laws of Eq.~\eqref{eqn: local_2nd_law_shuttle} imposes restriction on this efficiency, given as $\dot{\mathcal{I}}/\Dot{\sigma}_{\rm{tot}}\le \eta \le 1+ \dot{\mathcal{I}}/\Dot{\sigma}_{\rm{tot}}$. In Fig.~\ref{fig: efficiency}(b), we plot the transduction efficiency $\eta$ as a function of the applied voltage $V$ for different electromechanical strengths $\alpha$.  With stronger coupling between the subsystems, the rate of energy exchange $\dot{\mathcal{E}}$ increases along with a decrease in the entropy production rate $\Dot{\sigma}_{\rm{tot}}$ at any fixed voltage $V$, indicating improved transduction efficiency.


\section{Conclusion and Discussion}
In this article, we extended the framework of classical information thermodynamics to autonomous systems composed of two interacting subsystems: one governed by a Markov jump process and the other by underdamped diffusion. We derived the thermodynamic flows with respect to external heat, work reservoirs, and entropy production. We consistently derived the first and second laws that govern the thermodynamics of the composite system. Next, we derived the internal energetic and informational flows between the subsystems, both at the trajectory and average level. In doing so, we established local second laws for each subsystem, highlighting a measurement-feedback interpretation of their interactions.  At the fluctuating level, we derived integrated fluctuation relations for the partially masked dynamics of the subsystems. Specifically, the partial entropy production, involving the information flows, due to the individual subsystem, satisfies this fluctuation relation. Hence, we showed that the traditional bipartite formalism for purely jump and diffusive dynamics, can be extended to such composite system with mixed dynamics. \par

We also analyzed two practically relevant regimes of operations of these systems: one with slow underdamped dynamics relative to the discrete jump dynamics, and the other in the large mass limit of the underdamped subsystem.  Unlike purely jump or diffusive dynamics, we showed that the nonequilibrium dynamics of the fast (discrete) subsystem cannot drive the slow (underdamped) subsystem out of equilibrium purely through interactions under the adiabatic approximation, resulting in zero energy and information flows.  To address this, we perturbatively derived dynamical equations beyond the adiabatic approximation, capturing the lowest-order contributions to energy and information flows and the associated nonequilibrium effects in slow subsystem. In the large-mass regime, we systematically extended the effective dynamics, capturing the lowest-order effects of damping and interactions beyond traditional Hamiltonian dynamics in the infinite-mass limit. We found that the dynamics in this regime correspond to a mean-field approximation, where non-zero energy flows are present, but information flows vanish due to the deterministic nature of the underdamped system.\par

Finally, we applied our theoretical framework to a paradigmatic model of a nanoelectromechanical system - a single-electron shuttle - in which self-oscillation arises from the coupling between electron tunneling through a quantum dot and the mechanical motion of a nanoresonator. Here, we demonstrated that in the large, but finite-mass regime, the mean-field dynamics effectively capture the transient stochastic behavior at the average level, as the system relaxes into a limit cycle within the self-oscillating regime. Similarly, we showed that the timescale-separated dynamics accurately represent the steady-state probability distribution and the mutual information, even in the self-oscillating regime with small amplitudes where multiple tunneling events occur per oscillation. We analyzed the energy and information flows, as the shuttle transitions towards the self-oscillations above a critical voltage. Under isothermal conditions, we demonstrated that for any applied voltage across the leads, the oscillator acts as a sensor, measuring the state of the dot, but using the energy supplied by the dot due to its tunneling events. We also found that the self-oscillating regime is primarily driven by energetic exchanges, with the information flows saturating as the dot dynamics becomes synchronized with mechanical oscillations for large voltages. Lastly, we analyzed the transduction efficiency of the shuttle in converting electrical work into mechanical oscillations.\par

Our thermodynamic framework opens up the avenue to study hybrid CMOS circuits \cite{freitas2021stochastic}, coupling with inductors \cite{freitas2020stochastic, gopal2024thermodynamic} and/or quartz crystals \cite{matthys1983crystal, siniscalchi2020ultra}. Such electronic systems can be designed to build precise clocks beyond the limitations of the thermodynamic uncertainty relation (TUR) \cite{pietzonka2022classical, gopal2024thermodynamic}. The electron shuttle, capable of precise single-electron transport per oscillation, has previously been explored in the context of clocks \cite{culhane2024powering} and single-electron sources \cite{wachtler2022electron}. However, a detailed analysis is needed to identify the parameter space in which the original TUR for any integrated current observable may be violated in this model. Using timescale separation between jump and diffusive dynamics, we derived a modified Fokker-Planck equation for the underdamped subsystem, which requires the full spectral information of the generator of jump dynamics. Applying this approach to systems with macroscopic state spaces, such as CMOS circuits, remains an area for future exploration, with potential simplifications by considering only the slowest relaxation modes. In this article, we assume that the MJ process in the discrete subsystem depends only on the position of the underdamped subsystem. In contrast, if the jump dynamics were velocity-dependent, the fast Markov jump (MJ) dynamics could drive the slow underdamped (UD) subsystem out of equilibrium, even under the adiabatic approximation. This behavior does not occur when the jump dynamics depend solely on the position, as discussed in Sec.~\ref{sec: TS}. However, a careful evaluation of the local detailed balance condition is necessary to ensure that the correct equilibrium state is reached, due to the odd parity of the velocity under time-reversal \cite{spinney2012nonequilibrium,lee2013fluctuation}. Additionally, further analysis is required to verify the validity of the generalized integrated fluctuation theorem (IFT) under these conditions. \par

In the isothermal case of the electron shuttle, we observed that self-oscillations are primarily driven by energy transduction rather than information flows. Since the minimum energetic cost to generate it and its maximum transduced energy (due to coherent oscillations) are scaled by its corresponding temperatures, the information flow becomes more significant in non-isothermal settings \cite{grelier2024unlocking, parrondo2023information}. Further studies are required to identify the design principles such that the shuttle functions effectively as an information engine and assess its transduction efficiency in these contexts. Additionally, on the theoretical front, anomalous contributions to thermodynamic quantities have previously been observed when considering the overdamped approximation, neglecting inertia \cite{celani2012anomalous,murashita2016overdamped}. The systematic derivation of such anomalous contributions to information thermodynamics in the overdamped limit in such mixed dynamics is a topic of future research. 

\bibliography{main}

\appendix
\counterwithin{figure}{section}

\section{Second Law : Average Description}
\label{appsec: 2nd_law_simplification}
At the average description, the nonequilibrium system entropy at a given time $t$ of the composite system is given as,
\begin{eqnarray}
    S(t) =  \SumInt P_t(\Vec{\Gamma}, y)[s_i(y)-k_B\log P_t(\Vec{\Gamma}, y)].
\end{eqnarray}
Taking the time derivative of the Shannon entropy, we can distinguish the partial contributions from the individual subsystems and identify the total entropy production rate:
\begin{flalign}
    d_t S &= \SumInt d_t P_t(\Vec{\Gamma}, y)[s_i(y)-k_B\log P_t(\Vec{\Gamma}, y)] \label{eqn: 2nd law_average}\\
            &= \SumInt \left(\sum_\rho\sum_{y'} J_\rho^{y,y'}(\Vec{\Gamma}) - \nabla.J_{\Vec{\Gamma}}^y\right)[s_i(y)-k_B\log P_t(\Vec{\Gamma}, y)]\\
            &= \underbrace{\int \!d\Vec{\Gamma}\! \sum_{\rho, y>y'} J^{y,y'}_\rho(\Vec{\Gamma},t)\left[ s_i(y) - s_i(y') - k_B \log\frac{ P_t(\Vec{\Gamma}, y)}{ P_t(\Vec{\Gamma}, y')}\right]}_{\Dot{S}_{\mbb{Y}}}+ \underbrace{\vphantom{\sum_{\rho, y>y'}} k_B\SumInt  \nabla.\left[\Vec{\mathcal{L}}^{\rm{diss}}P_t(\Vec{\Gamma}, y)\right]\log P_t(\Vec{\Gamma}, y) }_{\Dot{S}_{\mbb{X}}}.\nonumber
\end{flalign}
The first term in the expression above $\Dot{S}_{\mbb{Y}}$ denotes changes in the Shannon entropy of the system due to dynamics in subsystem $\mbb{Y}$. It is the sum of the entropy flow to the reservoirs $\rho$, $\Dot{\Sigma}_\rho^r = - \Dot{Q}_\rho/T_{\mbb{Y}} $, and the corresponding partial entropy production rate, $\Dot{\sigma}_{\mbb{Y}}$,
\begin{flalign}
     \Dot{S}_{\mbb{Y}}= \int d\Vec{\Gamma} \sum_{\rho, y>y'} J^{y,y'}_\rho(\Vec{\Gamma},t) \left[ s_i(y) - s_i(y') - k_B \log\frac{ P_t(\Vec{\Gamma}, y)}{ P_t(\Vec{\Gamma}, y')}\right] =  \frac{1}{T_{\mbb{Y}}}\sum_\rho \Dot{Q}_\rho+ \Dot{\sigma}_{\mbb{Y}}, \\
     \Dot{\sigma}_{\mbb{Y}} = \sum_\rho \int d\Vec{\Gamma} \sum_{y>y'} J^{y,y'}_\rho(\Vec{\Gamma},t) \log\frac{ \lambda_\rho^{y,y'}(x)P_t(\Vec{\Gamma}, y')}{\lambda_\rho^{y',y}(x) P_t(\Vec{\Gamma}, y)}\ge 0 .\label{eqn: 2ndlaw_electronic}
\end{flalign}
The second term corresponding to the underdamped dynamics can be further split into the corresponding entropy change in reservoir $\mbb{X}$, $-\langle\Dot{Q}_m\rangle/T_{\mbb{X}}$ and partial entropy production rate $\Dot{\sigma}_{\mbb{X}}$,
\begin{flalign}
    \Dot{S}_{\mbb{X}}= k_B\SumInt  \nabla.\left[\Vec{\mathcal{L}}^{\rm{diss}}P_t(\Vec{\Gamma}, y)\right] \log P_t(\Vec{\Gamma}, y) =  \frac{\langle\Dot{Q}_{\mbb{X}}\rangle}{T_{\mbb{X}}} + \Dot{\sigma}_{\mbb{X}},\\
    \Dot{\sigma}_{\mbb{X}} =\frac{\gamma}{T_{\mbb{X}}}  \SumInt P_t(\Vec{\Gamma}, y)\left[v+(m\beta_{\mbb{X}})^{-1}\partial_v \log P_t(\Vec{\Gamma}, y)\right]^2 \ge 0. \label{appeqn: partial_sigma_X}
\end{flalign}

\section{Marginalized dynamics and thermodynamics}
\label{appsec: marginal}
Marginalization to one of the subsystems can be understood as a coarse-graining procedure that retains only the degrees of freedom specific to that subsystem at any given time. In this section, we explore the effective dynamics and thermodynamics of such coarse-grained states.

\subsection{Marginal UD dynamics}
We will first analyze the case with only continuous degrees of freedom $\Vec{\Gamma}$, while integrating out the discrete electronic state $y$. This situation arises when one can only access the position and velocity of the subsystem $\mbb{X}$, whereas changes in the discrete states are inaccessible. This is equivalent to the marginalization done for a coupled underdamped system in \cite{herpich2020effective}. From the knowledge of the full dynamics, the corresponding Fokker-Planck equation for the marginal probability density, $P_t^{\mbb{X}}(\Vec{\Gamma}) = \sum_y P_t(\Vec{\Gamma},y)$, is given as:
\begin{eqnarray}
    d_t P_t^{\mbb{X}}(\Vec{\Gamma}) = -\nabla.\Big(\left[\Vec{\mathcal{L}}^{\rm{det}}_{\mbb{X}} +\Vec{\mathcal{L}}^{\rm{diss}}_{\mbb{X}} \right] P_t^{\mbb{X}}(\Vec{\Gamma})\Big).
    \label{eqn: marginal_dynamics_underdamped}
\end{eqnarray}
Here, the effective underdamped probability current is further split into deterministic and dissipative contributions:
\begin{eqnarray}
    \Vec{\mathcal{L}}^{\rm{det}}_{\mbb{X}} = \begin{bmatrix}
        v \\
        \frac{1}{m}\left[-\partial_x \hat{U}_{\mbb{X}}(x)+g^{y}(\Vec{\Gamma},t)\right]
    \end{bmatrix} ,\;\;\; \Vec{\mathcal{L}}^{\rm{diss}}_{\mbb{X}} \equiv \begin{bmatrix}
                                        0\\
                                        v+(\beta m)^{-1} \partial_v \log P_t^{\mbb{X}}(\Vec{\Gamma})
                                        \end{bmatrix}.
\end{eqnarray}
Comparing it with Eq.~\eqref{eqn: prob_current_operator_joint}, the key difference is the emergence of an additional time-dependent non-conservative force $g^{y}(\Vec{\Gamma},t)$, given as
\begin{equation}
    g^{y}(\Vec{\Gamma},t)\equiv\sum_y \mathbb{P}_t(y|\Vec{\Gamma})\left[-\partial_x \hat{U}_{\mbb{XY}}(x,y) + g(x,y)\right]= \sum_y \mathbb{P}_t(y|\Vec{\Gamma}) f_{\mbb{XY}}(x,y), \label{eqn: effective_g}
\end{equation}
due to the influence of the discrete state dynamics through the interaction force $f_{\mbb{XY}}(x,y)$. Note that the dynamics in Eq.~\eqref{eqn: marginal_dynamics_underdamped} is not a closed equation, as it still depends on the time-dependent conditional distribution $\mathbb{P}_t(y|\Vec{\Gamma})$. 
\par

The marginalized dynamics of Eq.~\eqref{eqn: marginal_dynamics_underdamped} mimics that of an underdamped particle with internal energy $E_{\mbb{X}} = (1/2)mv^2 + \hat{U}_{\mbb{X}}(x)$ influenced by a non-conservative force $g^y(\Vec{\Gamma}, t)$ in a thermal bath with inverse temperature $\beta_{\mbb{X}}$. The first law for the marginal dynamics is given as,
\begin{flalign}
    \partial_t E_{\mbb{X}}  &=  - \gamma\left[\langle v^2\rangle -\frac{1}{\beta m}\right] +  \int  \!\!d\Vec{\Gamma} \; P^{\mbb{X}}_t(\Vec{\Gamma})\:g^y(\Vec{\Gamma},t)\: v\nonumber\\
    &= \Dot{Q}_{\mbb{X}}  + \Dot{W}_{\mbb{X}} - \Dot{\mathcal{E}}^{\mbb{X}}. 
    \end{flalign}
Similar to the composite system, the marginal entropy production can be identified from the sum of entropy change in the subsystem $\mbb{X}$ along with its reservoir entropy, leading to:
\begin{equation}
    \begin{aligned}
    \Dot{\Sigma}_{\mbb{X}} &= d_t S_{\mbb{X}}^{\text{marg}} - \beta_{\mbb{X}} \Dot{Q}_{\mbb{X}} =  \Dot{\sigma}_{\mbb{X}} + \Dot{\mathcal{I}}^{\mbb{X}}_F + \Dot{\mathcal{I}}^{\mbb{X}}_S.
    \end{aligned}
\label{eqn: entropy_marg_X}
\end{equation}
However, as mentioned in main text, this quantity can take negative values leading to apparent violations of the second law. \par

A similar apparent violation of the second law has previously been observed in the entropy production $\Dot{\Sigma}_{\mbb{X}}$ of molecular refrigeration systems \cite{PhysRevLett.93.120602, PhysRevE.75.022102, munakata2012entropy}, where a velocity-dependent force performs the feedback cooling. The second law is restored in such systems by introducing an "entropy pumping" term. As noted in \cite{herpich2020effective}, this entropy pumping contribution is exactly equal to the force contribution of the information flow $\Dot{\mathcal{I}}^{\mbb{X}}_F= \frac{k_B}{m}\int d\Vec{\Gamma} \; P_t^{\mbb{X}}(\Vec{\Gamma})  \partial_v \;g^y(\Vec{\Gamma},t)$, arising from interactions with the subsystem creating the force. This leads to a positive corrected entropy production $\Dot{\Sigma}^{\rm{pump}}_{\mbb{X}}$ at $\mbb{X}$, given as:
\begin{eqnarray}
    \Dot{\Sigma}^{\rm{pump}}_{\mbb{X}} &=& d_t S_{\mbb{X}}^{\text{marg}} - \beta_{\mbb{X}} \Dot{Q}_{\mbb{X}} - \Dot{\mathcal{I}}^{\mbb{X}}_F\\ 
    &=& \beta_{\mbb{X}} \gamma \SumInt P_t^{\mbb{X}}(\Vec{\Gamma})\left[v+(\beta_{\mbb{X}} m)^{-1}\partial_v \log P_t^{\mbb{X}}(\Vec{\Gamma})\right]^2 \ge 0\\
    &=&  \Dot{\sigma}_{\mbb{X}} + \Dot{\mathcal{I}}^{\mbb{X}}_S \le \Dot{\sigma}_{\mbb{X}}.
\end{eqnarray}
However, this corrected entropy production $\Dot{\Sigma}^{\rm{pump}}_{\mbb{X}}$ only provides a lower bound to the true partial entropy production at $\mbb{X}$, i.e. $\Dot{\Sigma}^{\rm{pump}}_{\mbb{X}} \le \Dot{\sigma}_{\mbb{X}}$, because it neglects the entropic part of the information flow $\Dot{\mathcal{I}}^{\mbb{X}}_S \le 0$. This entropic part of the information flow, $\Dot{\mathcal{I}}^{\mbb{X}}_S \le 0$, is lost through such a marginalization procedure. Thus, while the entropy pumping term corrects the negative entropy production, it does not account for all the entropy exchanged due to subsystem interactions, and the partial entropy production at a subsystem must include both force and entropic contributions of the information flow.\par

Putting it all together, we can reformulate the marginal entropy balance as follows:
\begin{eqnarray}
    d_t S_{\mbb{X}}^{\text{marg}} = \beta_{\mbb{X}}\Dot{Q}_{\mbb{X}} + \Dot{\Sigma}_{\mbb{X}} + \Dot{\mathcal{I}}^{\mbb{X}}_F .
\end{eqnarray}
This is equivalent to the entropy balance done for molecular refrigerator using feedback control of their velocities \cite{PhysRevLett.93.120602}.

\subsection{Marginal MJ dynamics}
Now we will shift our focus to the complementary case where we can only keep track of the electronic degrees of freedom $y$. Here, we will be left with the marginal distribution $P_t^{\mbb{Y}}(y) = \int d\Vec{\Gamma}\:P_t(\Vec{\Gamma},y)$, after integrating out continuous degrees of freedom. Integrating Eq.~\eqref{eqn: master_equation_joint}, one finds that the dynamics in the discrete states can still be described by a Markov jump process, given by
\begin{eqnarray}
    d_t P_t^{\mbb{Y}}(y) = \sum_\rho \sum_{y'} \left[W_\rho^{y,y'}(\Vec{\Gamma},t) P_t^{\mbb{Y}}(y') - W_\rho^{y',y}(\Vec{\Gamma},t) P_t^{\mbb{Y}}(y)\right],
    \label{appeqn: marginal_dynamics_MJP}
\end{eqnarray}
where we have defined marginalized jump rates $W_\rho^{y,y'}(\Vec{\Gamma},t) \equiv \int d\Vec{\Gamma}\,\mathbb{P}_t(\Vec{\Gamma}|y')\lambda_\rho^{y,y'}(x)$. As in the previous case, Eq.~\eqref{appeqn: marginal_dynamics_MJP} cannot be solved independently due to the dependence of $W_\rho$ on the conditional distribution $\mathbb{P}_t(\Vec{\Gamma}|y')$. However, there is no trivial identification of new non-conservative force from the effective dynamics in Eq.~\eqref{appeqn: marginal_dynamics_MJP}. Assuming that the energy of the electronic subsystem is described $\hat{U}_{\mbb{Y}}(y)$, we can write the first law as:
\begin{eqnarray}
    \partial_t E_{\mbb{Y}}(y)  &=& \sum_{\rho, y>y'}\int d\Vec{\Gamma}\: J^{y,y'}_\rho(\Vec{\Gamma})\left[\hat{U}_{\mbb{Y}}(y) - \hat{U}_{\mbb{Y}}(y')\right]\nonumber\\
    &=& \Dot{Q}_{\mbb{Y}}  + \Dot{W}_{\mbb{Y}} - \Dot{\mathcal{E}}^{\mbb{Y}} .
\end{eqnarray}
Assuming that one can measure the changes in the environment entropy $-\beta_{\mbb{Y}} \langle\Dot{Q}_{\rho}\rangle$, the marginal entropy production can be computed as:
\begin{flalign}
    \Dot{\Sigma}_{\mbb{Y}} =  d_t S_{\mbb{X}}^{\text{marg}} - \beta_{\mbb{Y}}\sum_\rho\Dot{Q}_\rho .
\end{flalign}

From the entropy balance for the marginal dynamics in subsystem $\mbb{Y}$, we can connect this to the partial entropy production $\Dot{\sigma}_{\mbb{Y}}$, as:
\begin{eqnarray}
    \Dot{\sigma}_{\mbb{Y}} = d_t S_{\mbb{X}}^{\text{marg}} - \beta_{\mbb{Y}}\sum_\rho\Dot{Q}_\rho - \Dot{\mathcal{I}}^{\mbb{Y}} = \Dot{\Sigma}_{\mbb{Y}} -  \Dot{\mathcal{I}}^{\mbb{Y}} .
\end{eqnarray}
where $\Dot{\mathcal{I}}^{\mbb{Y}}$ is the directional information flow to the electronic subsystem, whose explicit form is given in Eq.~\eqref{appeqn: partial_sigma_X}.

\section{Proof of IFT for partial entropy production}
\label{appsec: IFT_proof}
Consider a trajectory $[\underbar{$ y$},\underbar{$\Vec{\Gamma}$};\{\rho_k\}] \equiv \{\Vec{\Gamma}(\tau), \mathcal{N}_\rho^{y,y'}(\Vec{z_\tau}) \}$, defined for $0\le \tau \le t$, of the stochastic dynamics described by Eq.\mref{eqn: master_equation_joint}. The probability of observing such a trajectory is given as,
\begin{flalign}
    P[\underbar{$y$},\underbar{$\Vec{\Gamma}$};\{\rho_k\}] &\propto P_0(y_0.\Vec{\Gamma}_0) \prod_{i=0}^{N}e^{-\int_{t_i}^{t_{i+1}}d\tau \: \left[\mathcal{G}(\Vec{\Gamma}_\tau, y_i) + \mathcal{R}(y_i,x_\tau)\right]} \prod_{i=1}^{N} \lambda_{\rho_i}^{y_{i},y_{i-1}}(x_{t_{i}})\\
    &=  P_0(y_0,\Vec{\Gamma}_0) e^{-\int_{0}^{t} d\tau \: \left[\mathcal{G}(\Vec{\Gamma}_\tau, y_\tau) + \mathcal{R}(y_\tau,x_\tau)\right]} \prod_{i=1}^{N} \lambda_{\rho_i}^{y_{i},y_{i-1}}(x_{t_{i}})
\end{flalign}
 where $\mathcal{G}(\Vec{\Gamma}_\tau, y_i) = \frac{\beta_{\mbb{X}}}{4\gamma}\left[ m\Dot{v}_\tau + \partial_x  U(x_t,y_i) + \gamma v_\tau - g(\Vec{\Gamma}_t, y_i) \right]^2 - \frac{\gamma}{2 m} + \frac{1}{2m} \partial_v g(\Vec{\Gamma}_t, y_i)$ is the Onsager-Machlup action for underdamped dynamics in Stratonovich notation, and using $y_\tau = y_i$ when $\tau \in [t_{i-1}, t_i)$. The contribution to path probability when no jumps occur has an exponential distribution with the escape rate $\mathcal{R}(y, x) = \sum_\rho \sum_{y'} \lambda_\rho^{y', y} (x) $. We will also consider the initial state to be drawn from the initial probability density $P_0(y_0,\Vec{\Gamma}_0)$, without loss of generality.\par
 Hence the path probability can be split as a product due to contributions from the dynamics in subsystem $\mbb{X}$ $\mathcal{P}_{\mbb{X}}[\underbar{$y$},\,\underbar{$\Vec{\Gamma}$}| y_0, \Vec{\Gamma}_0 ]$ and the dynamics in subsystem $\mbb{Y}$ $\mathcal{P}_{\mbb{Y}}[\underbar{$y$},\,\underbar{$\Vec{\Gamma}$}| y_0, \Vec{\Gamma}_0 ]$, as follows $\mathcal{P}[\underbar{$y$},\underbar{$\Vec{\Gamma}$};\{\rho_k\}] = P_0(y_0.\Vec{\Gamma}_0)\mathcal{P}_{\mbb{X}}[\underbar{$y$},\,\underbar{$\Vec{\Gamma}$}| y_0, \Vec{\Gamma}_0 ]\mathcal{P}_{\mbb{Y}}[\underbar{$y$},\,\underbar{$\Vec{\Gamma}$}| y_0, \Vec{\Gamma}_0 ]$  which we will define as,
 \begin{flalign}
     \mathcal{P}_{\mbb{X}}[\underbar{$y$},\,\underbar{$\Vec{\Gamma}$}| y_0, \Vec{\Gamma}_0 ] &\propto e^{-\int_{0}^{t}d\tau \: \mathcal{G}(\Vec{\Gamma}_\tau, y_\tau)}, \label{eqn: path_prob_mech}\\
     \mathcal{P}_{\mbb{Y}}[\underbar{$y$},\,\underbar{$\Vec{\Gamma}$}| y_0, \Vec{\Gamma}_0 ] &\propto e^{-\int_{0}^{t}d\tau  \: \mathcal{R}(y_i,x_\tau)}\prod_{i=1}^{N} \lambda_{\rho_i}^{y_{i},y_{i-1}}(\Vec{\Gamma}_{t_{i}}) .\label{eqn: path_prob_electronic}
 \end{flalign}
With access to only a subsystem and its reservoir, the estimated (marginal) entropy production can lead to violation of the second law. Marginal entropy production due to dynamics in subsystem $\mbb{Y}$ is calculated using its change in entropy and the corresponding entropy flow to its reservoir, 
$\Hat{\Sigma}_{\mbb{Y}}(t) = -\beta_{\mbb{Y}} \sum_\rho \Hat{Q}_\rho(t) + \Delta s_e(t)$. Similarly, the marginal entropy production due to $\mbb{X}$ is given as $ \Hat{\Sigma}_{\mbb{X}}(t)=  -\beta_{\mbb{X}}  \Hat{Q}_{\mbb{X}}(t) + \Delta s_{\mbb{X}}(t)$. In the main text, it was shown that the partial entropy production at the average level, which also accounts for the information flows, restores the second law, i.e. $\sigma_{\mbb{Y}(\mbb{X})}(t) \equiv \langle \Hat{\sigma}_{\mbb{Y}(\mbb{X})}(t) \rangle = \langle \Hat{\Sigma}_{\mbb{Y}(\mbb{X})}(t) - \Hat{I}^{\mbb{Y}(\mbb{X})}(t)\rangle \ge 0$. We will demonstrate below the existence of a generalized integrated fluctuation theorem (IFT) for the partial entropy production, $\Hat{\sigma}_{e(m)}(t)$, given as:
\begin{flalign}
    \langle e^{-\Hat{\sigma}_{\mbb{Y}}(t)/k_B} \rangle =1\hspace{1.5cm}\text{and}\hspace{1.5cm}  \langle e^{-\Hat{\sigma}_{\mbb{X}}(t)/k_B} \rangle =1.
\end{flalign}
where the ensemble average is over all the trajectories up to time t. The IFT naturally recovers to second-law-like inequality for the average partial entropy production, $\sigma_e(t)  \ge 0$ (Eq.~\eqref{eqn: 2ndlaw_electronic}), by the application of Jensen's inequality.\par
To prove its existence, we need to show that there exists modified processes ($1^{\ast}/2^{\ast})$ such that: 
\begin{flalign}
    \Hat{\sigma}_{\mbb{Y}}(t) = k_B \log\frac{\mathcal{P}[\underbar{$y$},\,\underbar{$\Vec{\Gamma}$}]}{\mathcal{P}^*_1[\underbar{$y$}^{\dagg},\,\underbar{$\Vec{\Gamma}$}^{\dagg}]}   \hspace{1.5cm}\text{and}\hspace{1.5cm}    \Hat{\sigma}_{\mbb{X}}(t) = k_B\log\frac{\mathcal{P}[\underbar{$y$},\,\underbar{$\Vec{\Gamma}$}]}{\mathcal{P}^*_2[\underbar{$y$}^{\dagg},\,\underbar{$\Vec{\Gamma}$}^{\dagg}]},
\end{flalign}
where $\mathcal{P}^*_{1}[\underbar{$y$}^{\dagg},\,\underbar{$\Vec{\Gamma}$}^{\dagg}] = P_t(y_t,\Vec{\Gamma}_t)\mathcal{P}_{\mbb{X}^{\ast}}[\underbar{$y$}^{\dagg},\,\underbar{$\Vec{\Gamma}$}^{\dagg}| y_t, \Vec{\Gamma}_t ]\mathcal{P}_{\mbb{Y}}[\underbar{$y$}^{\dagg},\,\underbar{$\Vec{\Gamma}$}^{\dagg}| y_t, \Vec{\Gamma}_t ]$ and $\mathcal{P}^*_{2}[\underbar{$y$}^{\dagg},\,\underbar{$\Vec{\Gamma}$}^{\dagg}] = P_t(y_t,\Vec{\Gamma}_t)\mathcal{P}_{\mbb{X}}[\underbar{$y$}^{\dagg},\,\underbar{$\Vec{\Gamma}$}^{\dagg}| y_t, \Vec{\Gamma}_t ]\mathcal{P}_{\mbb{Y}^{\ast}}[\underbar{$y$}^{\dagg},\,\underbar{$\Vec{\Gamma}$}^{\dagg}| y_t, \Vec{\Gamma}_t ]$ are the probabilities of observing a time-reversed trajectory $\{\underbar{$y$}^{\dagg},\,\underbar{$\Vec{\Gamma}$}^{\dagg}\} = \{\underbar{$x$}_{t-\tau}, -\underbar{$v$}_{t-\tau}, \underbar{$y$}_{t-\tau}\}$ due to a modified dynamics, denoted by $1^{\ast}(2^{\ast}$) which exclusively affects the underdamped ($\mbb{X}^{\ast}$) and jump dynamics ($\mbb{Y}^{\ast}$) respectively \cite{shiraishi2015fluctuation, rosinberg2016continuous}.
\subsection{Electronic partial entropy production}
For the electronic partial entropy production, this requirement can be further simplified to
\begin{flalign}
    \Hat{\sigma}_{\mbb{Y}}(t) &= k_B\log\frac{\mathcal{P}[\underbar{$y$},\,\underbar{$\Vec{\Gamma}$}]}{\mathcal{P}^*_1[\underbar{$y$}^{\dagg},\,\underbar{$\Vec{\Gamma}$}^{\dagg}]}\\
    &=-\sum_\rho \frac{\Hat{Q}_\rho(t)}{T_{\mbb{Y}}}  + \log\frac{\mathcal{P}_{\mbb{X}}[\underbar{$y$},\,\underbar{$\Vec{\Gamma}$}| y_0, \Vec{\Gamma}_0] P_0(y_0.\Vec{\Gamma}_0)}{\mathcal{P}_{\mbb{X}^{\ast}}[\underbar{$y$}^{\dagg},\,\underbar{$\Vec{\Gamma}$}^{\dagg}| y_t, \Vec{\Gamma}_t] P_t(y_t,\Vec{\Gamma}_t)}\\
    &=-\sum_\rho \frac{\Hat{Q}_\rho(t)}{T_{\mbb{Y}}} + \Delta \hat{s}(t) + k_B\log\frac{\mathcal{P}_{\mbb{X}}[\underbar{$y$},\,\underbar{$\Vec{\Gamma}$}| y_0, \Vec{\Gamma}_0]}{\mathcal{P}_{\mbb{X}^{\ast}}[\underbar{$y$}^{\dagg},\,\underbar{$\Vec{\Gamma}$}^{\dagg}| y_t, \Vec{\Gamma}_t]}.
\end{flalign}
In the above sequence of algebraic simplification, we used the local detailed balance condition which relates the entropy change in the reservoirs $\rho$ or the corresponding heat flows, $\Hat{Q}_\rho(t) = [s_i(y_0)-s_i(y_t)]+ \sum_{i=0}^N \delta_{\rho_i, \rho}\log\frac{\lambda_{\rho_i}^{y_{i},y_{i-1}}(\Vec{\Gamma}_{t_{i}})}{\lambda_{\rho_i}^{y_{i-1},y_{i}}(\Vec{\Gamma}_{t_{i}})}$, and $\Delta \hat{s} = [s_i(y_t)-s_i(y_0)]- k_B\log [P_t(y_t,\Vec{\Gamma}_t)/P_0(y_0,\Vec{\Gamma}_0)] = \Delta \hat{s}_{\mbb{Y}} + \Delta \hat{s}_{\mbb{X}}$. Hence, we are left to identify the modified underdamped process $\mbb{X}^{\ast}$ satisfying:
\begin{flalign}
        k_B\log\frac{\mathcal{P}_{\mbb{X}^{\ast}}[\underbar{$y$}^{\dagg},\,\underbar{$\Vec{\Gamma}$}^{\dagg}| y_t, \Vec{\Gamma}_t]}{\mathcal{P}_{\mbb{X}}[\underbar{$y$},\,\underbar{$\Vec{\Gamma}$}| y_0, \Vec{\Gamma}_0]} &= \Delta \hat{s}_{\mbb{X}}(t).
\end{flalign}
The modified underdamped dynamics ($\mbb{X}^{\ast}$) satisfying the above condition is:
\begin{equation}
    \begin{rcases}  
        \begin{aligned}
            dx_t &= v_t dt,\\
            m\: dv_t &=\left[ -\partial_x  U(x_t,y_t) + \gamma v_t + g(x_t, y_t) + \frac{2\gamma}{\beta m}\partial_v \log P_t(y, x_t, -v_t) \right] dt+ dB_t .\label{appeqn: modified_Langevin}  
        \end{aligned}
    \end{rcases} \mbb{X}^{\ast}
\end{equation}
The above dynamics give us the following probability for the time-reversed trajectory, and can be written as $\mathcal{P}^*_m[\underbar{$y$}^{\dagg},\,\underbar{$\Vec{\Gamma}$}^{\dagg}| y_t, \Vec{\Gamma}_t] \propto e^{-\int_t^0 d\tau\: \mathcal{G}^*(\Vec{\Gamma}_\tau^{\dagg}, y_\tau^{\dagg})}$. Here, ${G}^*(\Vec{\Gamma}_\tau^{\dagg}, y_\tau^{\dagg})$ is the action of the modified dynamics, which can be further simplified as follows:
\begin{flalign}
     \mathcal{G}^*(\Vec{\Gamma}_\tau^{\dagg}, y_\tau^{\dagg}) &= \frac{\beta}{4\gamma}\Big[m\Dot{v}_\tau^{\dagg} + \partial_{x^{\dagg}} U(x^{\dagg}_\tau, y_\tau^{\dagg}) - \gamma v_\tau^{\dagg} \!-\! g(x_\tau^{\dagg},\! -v_\tau^{\dagg}, y_\tau^{\dagg}) \!-\! \frac{2\gamma}{\beta m}\partial_{v^\dagger} \!\log P_\tau(y_\tau^{\dagg}, x_\tau^{\dagg},\! -v_\tau^{\dagg}) \Big]^2 \nonumber\\
     &\hspace{1cm}+\! \frac{\gamma}{2 m} + \frac{1}{2m} \partial_{v^\dagger} g(\Vec{\Gamma}_\tau^{\dagg}, y_\tau^{\dagg})+  \frac{\gamma}{\beta m}\partial^2_{v^\dagger} \!\log P_\tau(y_\tau^{\dagg},x_\tau^{\dagg},\! -v_\tau^{\dagg}) \nonumber\\
     &= \frac{\beta}{4\gamma}\Big[m\Dot{v}_\tau + \partial_{x} U(x_\tau, y_\tau) + \gamma v_\tau \!-\! g(x_\tau,\! v_\tau, y_\tau) \!+\! \frac{2\gamma}{\beta m}\partial_{v} \log P_\tau(y_\tau, x_\tau,\! -v_\tau) \Big]^2 \nonumber\\
     &\hspace{1cm}+\! \frac{\gamma}{2 m} - \frac{1}{2m} \partial_{v} g(\Vec{\Gamma}_\tau, y_\tau)+  \frac{\gamma}{\beta m}\partial^2_{v} \log P_\tau(y_\tau, x_\tau,\! -v_\tau).
\end{flalign}
In the last step, we have used that the time reversal operation is odd for the velocity $v$ and also changed $t - \tau \to \tau$. Finally expressing it in terms of the action for the forward trajectory of the original process $\mathcal{G}(\Vec{\Gamma}_\tau, y_\tau)$, we get the required result,
\begin{flalign}
     \mathcal{G}^*(\Vec{\Gamma}_\tau^{\dagg}, y_\tau^{\dagg}) &= \mathcal{G}(\Vec{\Gamma}_\tau, y_\tau) - \Dot{\hat{s}}_{\mbb{X}}.
\end{flalign}
The partial entropy change, $\Delta \hat{s}_{\mbb{X}} = k_B\int_0^t d\tau\, \Dot{\hat{s}}_{\mbb{X}}(\tau)$, up to time $t$ (Sec.~\ref{sec: 2ndlaw_composite}) can be expanded in terms of its rate $\Dot{\hat{s}}_{\mbb{X}}$, given as:
\begin{flalign}
    \Dot{\hat{s}}_{\mbb{X}}(\tau) &= \Vec{\Gamma}_\tau\:.\:\nabla\log P_\tau(\Vec{\Gamma}_\tau,y_\tau) + \frac{\nabla\:.\:\Vec{J}^y_{\Vec{\Gamma}}(\tau)}{P_\tau(\Vec{\Gamma}_\tau,y_\tau)}\nonumber\\
    &= \frac{-1}{m}\left[m\Dot{v}_\tau  +\partial_x U(x_\tau,y_\tau) +\gamma v_\tau - g(x_\tau, y_\tau) \right]  -\frac{\gamma}{m} - \frac{\gamma}{\beta m^2}\frac{\partial^2_{v} \log P_\tau(y_\tau, x_\tau,\! -v_\tau)}{P_\tau(y_\tau, x_\tau,\! -v_\tau)}.
\end{flalign}

\subsection{Partial entropy production : UD dynamics}
For the partial entropy production due to UD dynamics, the requirement of Eq.~\eqref{} implies that the above equation can be further simplified as,
\begin{flalign}
    \Hat{\sigma}_{\mbb{X}}(t) &= k_B\log\frac{\mathcal{P}[\underbar{$y$},\,\underbar{$\Vec{\Gamma}$}]}{\mathcal{P}^*_2[\underbar{$y$}^{\dagg},\,\underbar{$\Vec{\Gamma}$}^{\dagg}]}\\
    &= -\frac{\Hat{Q}_{\mbb{X}}(t)}{T_{\mbb{X}}}  + k_B\log\frac{\mathcal{P}_{\mbb{Y}}[\underbar{$y$},\,\underbar{$\Vec{\Gamma}$}| y_0, \Vec{\Gamma}_0] P_0(y_0.\Vec{\Gamma}_0)}{\mathcal{P}^*_{\mbb{Y}}[\underbar{$y$}^{\dagg},\,\underbar{$\Vec{\Gamma}$}^{\dagg}| y_t, \Vec{\Gamma}_t] P_t(y_t,\Vec{\Gamma}_t)}\\
    &= -\frac{\Hat{Q}_{\mbb{X}}(t)}{T_{\mbb{X}}}  + \Delta s(t) + k_B\log\frac{\mathcal{P}_{\mbb{Y}}[\underbar{$y$},\,\underbar{$\Vec{\Gamma}$}| y_0, \Vec{\Gamma}_0]}{\mathcal{P}^*_{\mbb{Y}}[\underbar{$y$}^{\dagg},\,\underbar{$\Vec{\Gamma}$}^{\dagg}| y_t, \Vec{\Gamma}_t]}\\
    \implies\Delta \hat{s}_{\mbb{Y}}(t) &= k_B\log\frac{\mathcal{P}^*_{\mbb{Y}}[\underbar{$y$}^{\dagg},\,\underbar{$\Vec{\Gamma}$}^{\dagg}| y_t, \Vec{\Gamma}_t]}{\mathcal{P}_{\mbb{Y}}[\underbar{$y$},\,\underbar{$\Vec{\Gamma}$}| y_0, \Vec{\Gamma}_0]}+ [s_i(y_t)-s_i(y_0)].\label{eqn: IFT_cond_m}
\end{flalign}
The modified jump process ($\mbb{Y}^{\ast}$) satisfying the above condition has rates, given as:
\begin{equation}
    \begin{rcases}
        \begin{aligned}
            W_\rho^{y,y'}(x_t,v_t,t) = \frac{\lambda_\rho^{y,y'}(x_t)P_t(y,x_t,-v)}{P_t(y',x_t,-v_t)} \label{appeqn: modified_rate}        
        \end{aligned}   
    \end{rcases} \mbb{Y}^{\ast}
\end{equation}
The above rate leads to a modified escape rate $\mathcal{R^*}(y, x, v, t)$,
\begin{eqnarray}
    \mathcal{R^*}(y, x, v, t) = \mathcal{R}(y, x) + \sum_{\rho, y'} \frac{J_\rho^{y,y'}(x, -v)}{P_t(y, x, -v)}. \label{eqn: modified_escape}
\end{eqnarray}
Combining Eq.~\eqref{appeqn: modified_rate} and Eq.~\eqref{eqn: modified_escape} with Eq.~\eqref{eqn: path_prob_electronic}, we get the necessary condition:
\begin{eqnarray}
    k_B\log\frac{\mathcal{P}^*_{\mbb{Y}}[\underbar{$y$}^{\dagg},\,\underbar{$\Vec{\Gamma}$}^{\dagg}| y_t, \Vec{\Gamma}_t]}{\mathcal{P}_{\mbb{Y}}[\underbar{$y$},\,\underbar{$\Vec{\Gamma}$}| y_0, \Vec{\Gamma}_0]} = -k_B\sum_{k=1}^{N}\log\frac{P_{t_{k}}(\Vec{\Gamma}_{t_{k}},y_{t_{k}})}{P_{t_k}(\Vec{\Gamma}_{t_k},y_{t_{k-1}})}   +k_B \int_0^\mathcal{T}\left[\sum_{\rho,y'}\frac{J^{y_t,y'}_\rho(\Vec{\Gamma}_t,t)}{P_t(\Vec{\Gamma}_t,y_t)}\right]dt
\end{eqnarray}

\section{Electron Shuttle: Beyond adiabatic approximation in the energy space}
\label{appsec: shuttle_energy_space}
In this section, we show the derivation of the modified Fokker-Planck equation for the electron shuttle using the general framework of Sec.~\ref{sec: TS_beyond}. The generator $\mathcal{L}_{\mbb{Y}}$ of the jump dynamics due to tunneling events is given as, 
\begin{eqnarray}
\mathcal{L}_{\mbb{Y}} = 
\begin{bmatrix}
-[\lambda_L^{0,1}(x) + \lambda_R^{0,1}(x)] & \lambda_L^{1,0}(x) + \lambda_R^{1,0} (x)\vspace{0.2cm}\\ 
\lambda_L^{0,1}(x) + \lambda_R^{0,1}(x)  & -[\lambda_L^{1,0}(x) + \lambda_R^{1,0}(x)]
\end{bmatrix}.
\end{eqnarray}
The eigenvalues of the above generator are $0$ and $-\Gamma_b\chi(x)$, where  $\Gamma_b\chi(x)= \sum_{\rho,y,y'}\lambda_\rho^{y,y'}(x)= 2\Gamma_b\cosh(x/\lambda)$ is the escape rate. The left $\bra{L_0}$ and right $\ket{R_0}$ eigenvectors to the $0$ eigenvalue are: $\bra{L_0}=(1,1)$  and $\ket{R_0}=(\pi_0(x), \pi_1(x))^{\dagg}$ respectively. Note that the right eigenvector $\ket{R_0}$ is the steady-state probability at the adiabatic limit. Similarly, the left $\bra{L_{\chi}}$ and right $\ket{R_{\chi}}$ eigenvectors corresponding to the remaining eigenvalue $-\Gamma\chi(x)$ are: $\bra{L_{\chi}}=[\pi_1(x),\, \pi_0(x)] $ and $\ket{R_{\chi}}=[1,-1]^{\dagg}$. The generalized inverse $\mathcal{A}^{-1}_{y,y'}(x)$ is then given as, 
\begin{eqnarray}
\mathcal{A}^{-1}_{y,y'}(x) = \frac{1}{\Gamma\chi(x)}
\begin{bmatrix}
-\pi_1(x) & \pi_0(x) \\ 
\pi_1(x)  & -\pi_0(x)
\end{bmatrix}.
\end{eqnarray}
Substituting the $\mathcal{A}^{-1}_{y,y'}(x)$ to Eq.~\eqref{eqn: cond_order_1}, the conditional distribution $\mathbb{P}_{ss}^{(0+1)}(y|\Vec{\Gamma})$ for the electron shuttle simplifies to:
\begin{eqnarray}
    \mathbb{P}_{\rm{TS}}^{(0+1)}(y|\Vec{\Gamma})  = \pi_y(x) + v\frac{(2y-1)\nu(x)}{\Gamma_b} + \mathcal{O}(\epsilon^{2}), \label{appeqn: ss_cond_shuttle_1}
\end{eqnarray}
where $\nu(x) = \left[-\partial_x \pi_{1}(x) + \alpha V \beta_{\mbb{X}} \pi_{1}(x)\pi_0(x)\right]/\chi(x)$ is the contribution due to the oscillator dynamics. In this expression, we have used the equilibrium adiabatic solution for the marginal density, satisfying $\partial_v \log P_{\rm{ad}} = -\beta m v + \mathcal{O}(\epsilon)$. As expected from the dynamics, the velocity dependence of $\mathbb{P}_{\rm{TS}}^{(0+1)}(y|\Vec{\Gamma})$ implies that the dot is more probable to be occupied at $x=0$ only when coming from the left lead, i.e. $v>0$. We also derive the modified FP equation for the shuttle (Eq.~\eqref{eqn: modified_FP_shuttle}) using $\mathcal{A}^{-1}_{y,y'}(x)$ to Eq.~\eqref{eqn: modified_FP}.\par
\begin{figure}[h!]
    \centering
    \includegraphics[trim=0 0 0 40, clip,width=\textwidth]{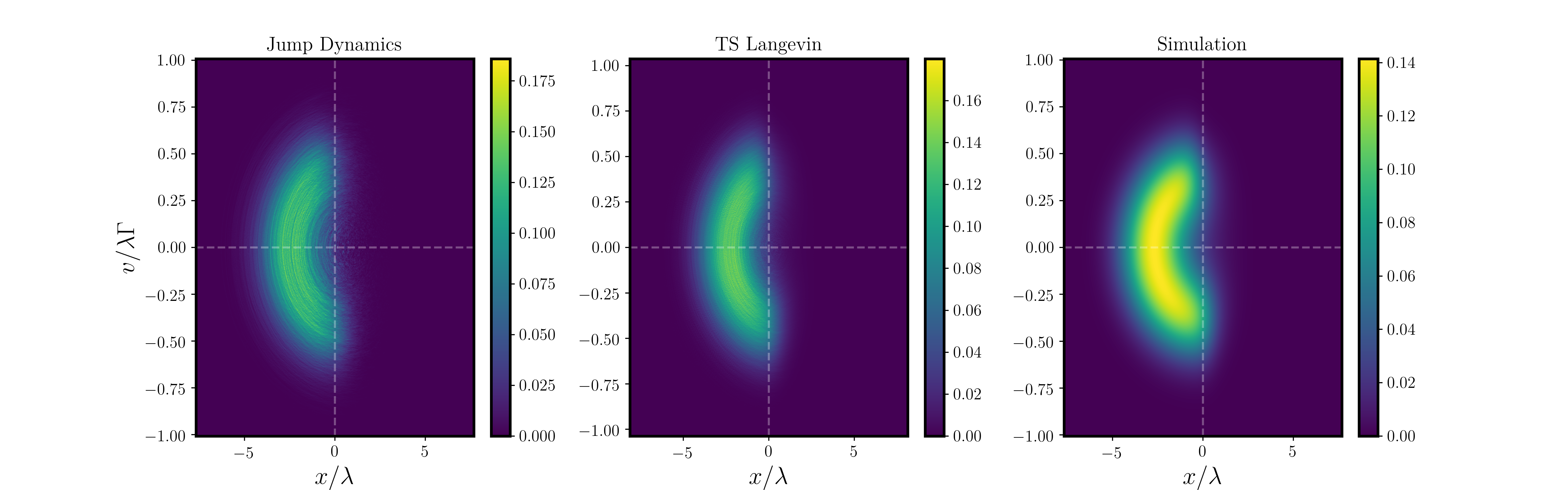}
    \caption{Comparison of the steady-state probability density for the occupied dot $P_{ss}(x,v,1)$ computed using stochastic simulations for initial master equation of Eq.~\eqref{eqn: master_equation_shuttle} (left), modified Fokker-Planck equation of Eq.~\eqref{eqn: modified_FP_shuttle} (center) and the semi-analytical solution in the energy $E$ space of Eq.~\eqref{appeqn: Pss_energy} (right).  Parameters: $V/V_T=24, \omega/\Gamma_b = 0.1292, \gamma/m\omega = 0.01, \alpha\lambda=0.04, m\lambda^2\Gamma_b^2\beta=50$ with $\lambda=1$, $\Gamma_b=5$ and $\beta=1$}
    \label{appfig: prob_comparison}
\end{figure}

To further gain insight on the oscillatory dynamics, one can equivalently work in energy-phase $(E_{\mbb{X}},\phi_{\mbb{X}})$ space. For the electron shuttle, the transformation is the following: $\hat{E}_{\mbb{X}} = (1/2)m\omega^2 x^2 +(1/2)mv^2$ and phase $\tan\phi_{\mbb{X}}=\omega x/v$. In this section, we will drop the subscript ${\mbb{X}}$ for ease of notation here, i.e. $(E, \phi) \equiv (\hat{E}_{\mbb{X}}, \phi_{\mbb{X}})$ and note that $E$ used here is not the total energy of the composite system. This allows us to parametrize the position $x$ and the velocity $v$ in terms of $(E, \phi)$, given as:
\begin{eqnarray}
    x = \sqrt{\frac{2E}{m\omega^2}}\sin{\phi}, \hspace{1cm} v = \sqrt{\frac{2E}{m}}\cos{\phi}
\end{eqnarray}
In the low damping $\gamma/m\omega\ll1$ regime, we can effectively work in the energy space as the timescales associated with the energy $\tau_A \propto m\omega/\gamma$ become larger than the timescale associated with phase changes $\tau_\phi \propto 2\pi/\omega$, i.e. $\tau_E\gg\tau_\phi$. At the lowest order of this time scale separation, the joint probability density can be assumed to satisfy $P_t(E,\phi)\approx (1/2\pi)P_t(E)$. Averaging Eq.~\eqref{eqn: modified_FP_shuttle} over $\phi$ in the $(E,\phi)$ space, one obtains the following master equation \cite{wachtler2019stochastic},
\begin{eqnarray}
    \partial_t P_t(E) =\partial_E\left[2Em(\hat{\gamma}(E) +\hat{D}(E)m^2\partial_{E})P_t(E)\right] +\mathcal{O}(\epsilon^3)
    \label{appeqn: shuttle_master_energy}
\end{eqnarray}
with the effective damping $\hat{\gamma}(E)$ and diffusion $\hat{D}(E)$ coefficient computed as
\begin{eqnarray}
    \hat{\gamma}(E) &=\frac{1}{2\pi}\int_0^{2\pi} d\phi\: \gamma^{\rm{eff}}(E, \phi)\cos^2\phi\\
    \hat{D}(E) &=\frac{1}{2\pi}\int_0^{2\pi} d\phi\: D^{\rm{eff}}(E, \phi)\cos^2\phi.
    \label{eqn: TS_E_FP}
\end{eqnarray}
Solving for the steady-state probability density $\partial_t P_{\rm{TS}}(E) = 0$ in the above Fokker-Planck equation, we get
\begin{eqnarray}
    P_{\rm{TS}}^{\mathbb{X}}(E) \propto \exp\left(-\int_0^{E} dE' \frac{\hat{\gamma}(E')}{m^2\hat{D}(E')}\right).
    \label{appeqn: Pss_energy}
\end{eqnarray}
As seen in Fig.~\ref{appfig: prob_comparison}, the above solution $P_{\rm{TS}}(E)$ matches well with steady-state density $P_{ss}^{\mathbb{X}}(x,v)$ by directly solving  Eq.~\eqref{eqn: modified_FP_shuttle}, which also matches well with the stochastic simulations of full system. Since $P_{\rm{TS}}(E)$ is an even function in $x$ and $v$, it fails to capture the small shift in the equilibrium position of the modified potential of the oscillator $U_{\rm{eff}}(x)$. \par

Hence, the joint probability $P(E,\phi,y)$ in the energy-phase coordinates can again be expressed in terms of the conditional probability $\mathbb{P}(y|E,\phi)$ as:
\begin{eqnarray}
    P_{ss}(E,\phi,y) &=& \mathbb{P}_{ss}(y|E,\phi)P_{ss}^{\mathbb{X}}(E),\\
    P_{ss}^{\mathbb{X}}(E,\phi) &\approx& \frac{1}{2\pi}P_{\rm{TS}}^{\mathbb{X}}(E),\\
    \mathbb{P}_{ss}(y|E,\phi) = \pi_1(E,\phi) - \frac{(2y-1)}{\Gamma_b \chi(E,\phi)}&\Bigg[&\frac{\sin \phi \cos \phi}{m\omega} \partial_{E} \pi_1(E,\phi) + \frac{\cos^2 \phi}{m} \partial_{\phi} \pi_1(E,\phi) \nonumber\\
    &+& \sqrt{2Em}\; \alpha V \cos \phi\; \pi_1(E,\phi)\pi_0(E,\phi) \partial_{E} \log P_{\rm{TS}}^{\mathbb{X}}(E)  \Bigg] + \mathcal{O}(\epsilon^2)   . 
\end{eqnarray}

Even though the TS Fokker-Planck equation in Eq.~\eqref{eqn: modified_FP_shuttle} is exact up to $\mathcal{O}(\epsilon^2)$, the steady-state solution $P_{\rm{TS}}^{\mathbb{X}}(E)$ of Eq.~\eqref{appeqn: Pss_energy}, directly computed from the Fokker-Planck equation, contain higher order terms in $\epsilon$. To keep consistency, we will expand it up to first order in $\epsilon$, leading to: 
\begin{eqnarray}
    P_{\rm{TS}}^{\mathbb{X},(0+1)}(E) = \frac{1}{N}\exp(-\beta E)\left[ 1 - \beta\int_0^{E}\! dE'\left(\frac{2\hat{\gamma}(E')}{\gamma} - 1\right) \right] +\mathcal{O}(\epsilon^2),
\end{eqnarray}
where $N$ is the normalization constant and $\left(\frac{2\hat{\gamma}(E')}{\gamma} - 1\right) \propto \Gamma_b^{-1}$ is the first order correction. The first order term captures the nonequilibrium features of oscillator due to the coupling with the dot. The lowest order contributions of the energy and information flows ($\mathcal{O}(\epsilon^2)$) in the low damping regime, is explicitly given as:
\begin{equation}
    \begin{aligned}
    \dot{\mathcal{E}}^{(2)} &=  -\frac{\beta}{m}\int_0^{\infty} \!dE\; \frac{e^{-\beta E}}{N}E\left[\int_0^{E}dE'\left(\frac{2\hat{\gamma}(E')}{\gamma} - 1\right)\right] + \mathcal{O}(\epsilon^3), \\
    \dot{\mathcal{I}}^{(2)} &=  \frac{\alpha V}{m \Gamma_b}\int_0^{\infty} \!dE\! \int_0^{2\pi}\!d\phi\;\frac{e^{-\beta E}}{2\pi N}\frac{\nu(E,\phi)}{\chi(E,\phi)} + \mathcal{O}(\epsilon^3). 
    \end{aligned}
    \label{eqn: internal_flow_shuttle_2}
\end{equation}
Here, we rewrote the position-dependent functions $\nu(x)=\nu(E,\phi)$ and $\chi(x)=\chi(E,\phi)$ in the energy-phase $(E,\phi)$ space.
\begin{figure}[h!]
    \centering
    \includegraphics[trim=0 0 430 0, clip,width=0.5\textwidth]{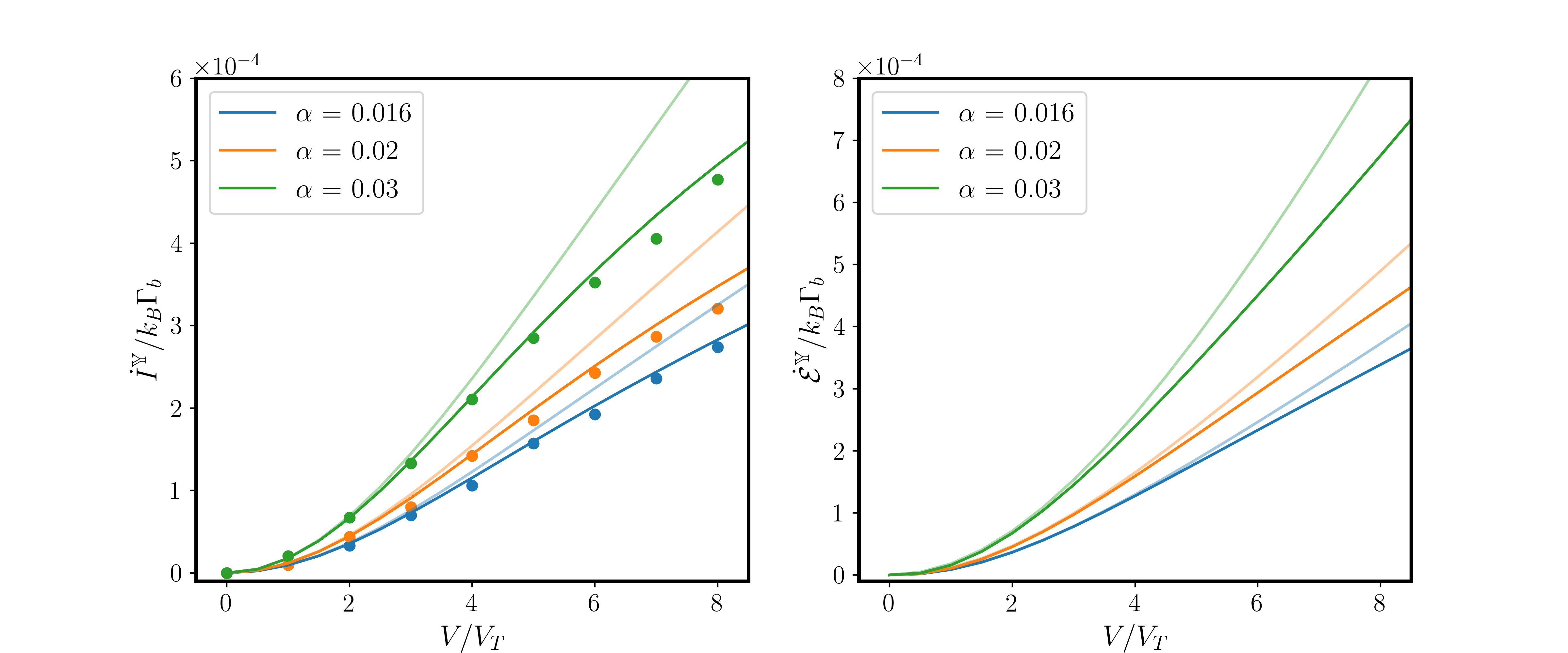}
    \caption{Comparison of the steady-state information exactly up to $\mathcal{O}(\epsilon^2)$ (light) of Eq.~\eqref{eqn: internal_flow_shuttle_2} with the flows computed using the full solution $P_{ss}^{\mathbb{X}}(E)$ (dark) of Eq.~\eqref{eqn: I_dot_TS} and Eq.~\eqref{eqn: E_dot_TS}. The markers are from the stochastic simulations of the full system. Parameters: $\omega/\Gamma = 0.1292, \gamma/m\omega = 0.01, m\lambda^2\Gamma^2\beta=50$ with $\Gamma=5$ and $\beta=1$}
    \label{appfig: flow_comp}
\end{figure}
In Fig.~\ref{appfig: flow_comp}, we compare the exact information flows up to $\mathcal{O}(\epsilon^2)$ with the flows computed using the full solution $P_{ss}^{\mathbb{X}}(E)$ keeping higher order terms, as done in Eq.~\eqref{eqn: I_dot_TS} and Eq.~\eqref{eqn: E_dot_TS}. We find that the exact thermodynamic flows up to $\mathcal{O}(\epsilon^2)$ only work for small voltages $V$ whose range of validity further decreases with increased electromechanical coupling $\alpha$ (with lower critical voltage $V_{\rm{cr}})$. This is expected in the electron shuttle, as with increasing voltage, the oscillation amplitude grows, causing the dot to interact predominantly with a single lead at a time. In such regimes, the number of lower tunneling events decreases as the dot spends less time near the center $x \sim 0$, which leads to failure of the TS assumption. But as shown in the Fig.~\ref{fig: information_flow}, keeping higher order terms of the full solution $P_{ss}^{\mathbb{X}}(E)$ captures the stochastic simulations even for larger voltages $V > V_{\rm{cr}}$ in the self-oscillating regime.

\begin{figure}[h!]
    \centering
    \includegraphics[trim=0 0 0 0, clip,width=0.5\textwidth]{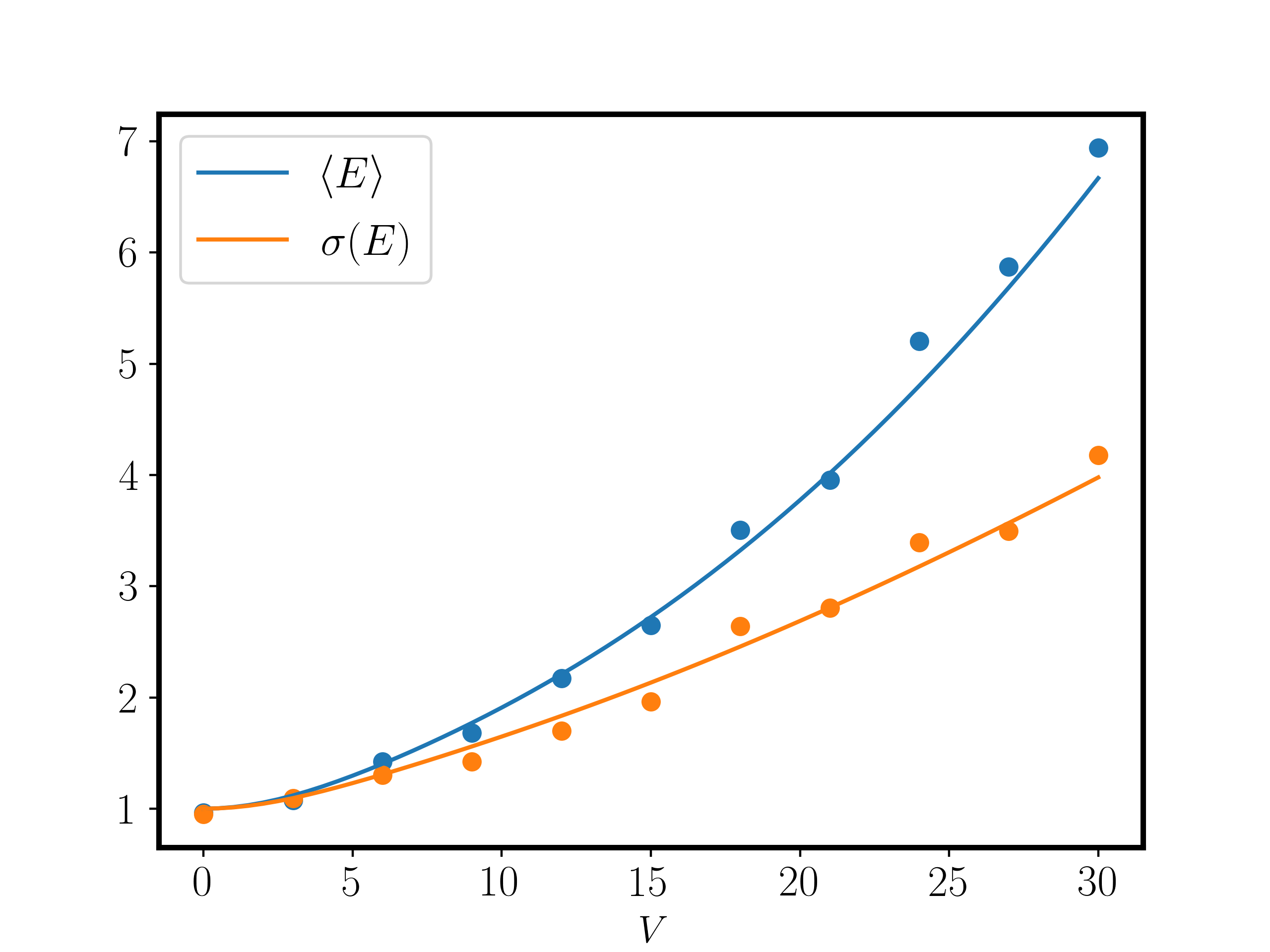}
    \caption{Average and the standard deviation of the energy $E \equiv \hat{E}_{\mbb{X}}$ of the oscillator for increasing powering voltages $V$. The solid lines are obtained from the TS solution of Eq.~\eqref{appeqn: Pss_energy} and the markers are from the simulations. Parameters: $\omega/\Gamma = 0.1292, \alpha\lambda = 0.04, \gamma/m\omega = 0.01, m\lambda^2\Gamma^2\beta=0.08$ with $\Gamma=5$ and $\beta=1$}
    \label{fig: energy_stats}
\end{figure}

\end{document}